\documentclass[journal]{IEEEtran}
\usepackage{amsmath,amsfonts}
\usepackage{algorithmic}
\usepackage{algorithm}
\usepackage{array}
\usepackage{textcomp}
\usepackage{stfloats}
\usepackage{url}
\usepackage{verbatim}
\usepackage{graphicx}
\usepackage{cite}
\usepackage{amsthm}
\usepackage{amssymb}
\usepackage{epsfig}
\usepackage{graphicx}
\usepackage{balance}
\usepackage{epstopdf}
\usepackage{multicol}
\usepackage{multirow}
\usepackage{threeparttable}
\usepackage{tabularx}
\usepackage{booktabs}
\usepackage{cite}
\usepackage{cases}
\usepackage{cuted}
\usepackage[usenames,dvipsnames]{color}
\usepackage[caption=false,font=footnotesize]{subfig}
\usepackage{makecell}
\usepackage{bm}
\allowdisplaybreaks[2]

\begin{document}

\title{MPFSR-Enhanced GNNs: Spectral Graph Neural Networks Enhancement Through Learnable Multiple-Parameter Graph Fractional Fourier Transforms}

\author{Manjun~Cui, Xiaopeng~Cheng, Yangfan~He, and Zhichao~Zhang,~\IEEEmembership{Member,~IEEE}
	\thanks{This work was supported in part by the Open Foundation of Hubei Key Laboratory of Applied Mathematics (Hubei University) under Grant HBAM202404; in part by the Foundation of Key Laboratory of System Control and Information Processing, Ministry of Education under Grant Scip20240121; and in part by the Startup Foundation for Introducing Talent of Nanjing Institute of Technology under Grant YKJ202214. \emph{(Manjun~Cui and Xiaopeng~Cheng are co-first authors.)} \emph{(Corresponding author: Zhichao~Zhang.)}}
	\thanks{Manjun~Cui and Xiaopeng~Cheng are with the School of Mathematics and Statistics, Nanjing University of Information Science and Technology, Nanjing 210044, China (e-mail: cmj1109@163.com; 18355938105@163.com).}
	\thanks{Yangfan~He is with the School of Communication and Artificial Intelligence, School of Integrated Circuits, Nanjing Institute of Technology, Nanjing 211167, China, and also with the Jiangsu Province Engineering Research Center of IntelliSense Technology and System, Nanjing 211167, China (e-mail: Yangfan.He@njit.edu.cn).}
	\thanks{Zhichao~Zhang is with the School of Mathematics and Statistics, Nanjing University of Information Science and Technology, Nanjing 210044, China, with the Hubei Key Laboratory of Applied Mathematics, Hubei University, Wuhan 430062, China, and also with the Key Laboratory of System Control and Information Processing, Ministry of Education, Shanghai Jiao Tong University, Shanghai 200240, China (e-mail: zzc910731@163.com).}}

\markboth{IEEE TRANSACTIONS ON PATTERN ANALYSIS AND MACHINE INTELLIGENCE}
{Shell \MakeLowercase{\textit{et al.}}: Bare Demo of IEEEtran.cls for Journals}


\maketitle

\begin{abstract}
	Graph neural networks (GNNs) excel in processing non-Euclidean data, but traditional spectral GNNs rely on static bases and fundamentally lack active spectral regulation. Although the graph fractional Fourier transform (GFRFT) introduces cross-domain modulation, it applies a uniform fractional parameter across all frequencies. This ignores frequency heterogeneity and restricts the models' adaptive capacity in graph node classification tasks. In the paper, we propose two novel types of multiple-parameter GFRFTs (MPGFRFTs) and establish their corresponding theoretical frameworks, including essential properties, computational complexity, and parameters differentiability. By assigning independent, learnable fractional parameters to distinct frequency bands, MPGFRFTs enable fine-grained spectral regulation. Then, we operationalize this mathematical framework by designing the adaptive multiple-parameter fractional spectral regulation (MPFSR) module, a plug-and-play component for mainstream spectral models. We also establish rigorous theoretical bounds on the spectral stability of this module, guaranteeing a stable and reliable convergence during the end-to-end parameters optimization. Experiments demonstrate that integrating the proposed MPFSR module alleviates the constraints of static bases and yields performance gains in node classification on complex graphs, advancing a novel paradigm for active spectral modulation in graph representation learning.
\end{abstract}

\begin{IEEEkeywords}
Fractional spectral regulation module, graph fractional Fourier transform, graph node classification, multiple-parameter graph fractional Fourier transform, spectral graph neural network.
\end{IEEEkeywords}

\section{Introduction}
\indent Graph neural networks (GNNs) \cite{Scarselli08Graph,wu2020comprehensive,corso2024graph} have emerged as a powerful tool for processing non-Euclidean structured data, demonstrating exceptional modeling capabilities in fundamental tasks such as graph node classification, link prediction, and graph representation learning across domains like social network analysis, bioinformatics, and recommendation systems. Based on their feature aggregation mechanisms, GNNs are typically categorized into spatial-based and spectral-based approaches. Spatial GNNs achieve feature aggregation through message passing between nodes \cite{gilmer17a,li2021spatial}. While they have achieved impressive results regarding computational efficiency and scalability, they often lack the capacity to analytically resolve the frequency components of a graph's global topological structure. In contrast, spectral GNNs are rooted in graph signal processing (GSP) theory \cite{Ortega08,ortega22introduction,Leus23,Song22} and characterize the underlying structural features of graphs from a frequency-domain perspective \cite{wang22am,mo2025autosgnn}. Despite their solid mathematical foundation, the depth of research and the adaptive regulatory capabilities of current spectral GNNs still require further exploration, largely constrained by the computational complexity of graph eigendecomposition and the flexibility of transformation operators.

\indent Looking at the development trajectory of spectral GNNs, the core challenge has consistently been the efficient and flexible design of graph filters. Early spectral models directly performed full eigendecomposition on the graph Laplacian matrix \cite{Ortega18GSP,wu2019simplifying,bo2021beyond}. While this laid the theoretical groundwork for spectral convolution, its extremely high computational complexity severely limited its application on large-scale graphs. To overcome this bottleneck, researchers introduced polynomial approximation strategies, such as Chebyshev \cite{he2022convolutional}, Jacobi \cite{wang2022powerful} and Bernstein \cite{he2021bernnet}, which bypassed explicit decomposition and endowed the models with spatial localization. GCN simplified the filter to a first-order linear approximation \cite{xu2019graph,dong2021adagnn}, which drastically improved computational efficiency but at the cost of weakening the model's expressive power to approximate arbitrary complex filters. In recent years, with the rise of self-attention mechanisms, cutting-edge models like Specformer have begun utilizing Transformer architectures in the spectral domain to achieve global, set-to-set associative learning at the eigenvalue level. However, most of these methods still rely on static orthogonal bases or fixed spectral transformations. They fundamentally lack a mechanism for active, deep-level spectral regulation of complex graph signals.

\indent A cornerstone of GSP is the graph Fourier transform (GFT) \cite{Sandryhaila13,Gavili17,Lu19,Domingos20,Patane23,Qi22}, which extends the classical Fourier transform (FT) \cite{bochner1949fourier,nussbaumer1982fast} to graph-structured data by projecting signals onto the eigenbasis of a graph shift operator (GSO) \cite{Girault15,Gavili17,Ji23,Isufi22,Liu25,Xia21,Li24}, such as the graph Laplacian or adjacency matrix. While the GFT enables spectral analysis on graphs, it lacks the ability to explore intermediate representations between the vertex and frequency domains. 
To address this limitation, Wang et al. \cite{Wang18} first proposed the graph fractional Fourier transform (GFRFT), which introduces a fractional order parameter to interpolate between the identity transform and the full GFT. In this formulation, the GFRFT is defined by applying a fractional power of the GFT matrix derived from a given GSO, thereby extending the core idea of the classical fractional Fourier transform (FRFT) \cite{Pan09,Zhao24} to graph domains via spectral-domain modulation. Subsequent works have further developed the GFRFT framework by extending it along multiple directions \cite{Ozturk21,yan2021windowed,gan2025windowed,Alik24Wiener,alikacsifouglu2025joint}. More recently, a hyper-differential operator-based GFRFT has been introduced in \cite{Alikasifoglu24}, which generalizes the notion of fractional powers and ensures differentiability with respect to the order parameter. This advancement is significant as it enables gradient-based learning of the fractional order parameter within task-specific optimization pipelines. As a result, the transform becomes not only flexible but also learnable, allowing it to adapt to data-driven objectives in applications.

\indent Although the GFRFT provides a highly promising cross-domain analytical tool for graph signal processing, current deep spectral models still face significant limitations regarding fine-grained spectral regulation.  On the one hand, the core paradigm of traditional spectral GNNs relies heavily on the classical GFT. These methods project signals into the spectral domain and employ various filters to amplify or attenuate specific frequency components. However, such approaches depend on the static eigendecomposition of the Laplacian matrix. Because their transformation bases remain fixed throughout the training process, they fundamentally lack active spectral regulation mechanisms, severely restricting the models' expressive capacity for complex, non-stationary graph signals. On the other hand, although recent research has introduced the GFRFT, which enables cross-domain processing of signals between the spatial and frequency domains by introducing a fractional parameter, providing a new perspective for spectral modulation, existing GFRFT schemes typically use a uniform fractional parameter for all graph frequencies. This globally consistent transformation ignores the heterogeneous characteristics of signals in different frequency bands, resulting in a significant lack of flexibility. Specifically, in real-world graph data, low-frequency components typically encode global community structures and smooth feature distributions, whereas high-frequency components often capture local variations, noise, or intricate edge details. This frequency heterogeneity is particularly crucial for graph node classification tasks, where the model's discriminative power heavily relies on effectively exploiting both homophilic (low-frequency) and heterophilic (high-frequency) neighborhood patterns. Applying a single fractional parameter fails to provide fine-grained regulation across the entire spectrum. Consequently, this uniform transformation paradigm leads to a severe lack of adaptive capacity when processing complex, non-stationary graph signals, thereby creating a bottleneck for further improving the discriminative power of spectral GNNs.

\indent To overcome the bottleneck of the aforementioned single-parameter spectral transformation, this paper proposes the multiple-parameter graph fractional Fourier transform (MPGFRFT). Abandoning the globally consistent transformation paradigm, this method assigns independent fractional order parameters to different frequency bands, providing a purely mathematical framework that supports multi-dimensional, differentiated, and non-linear spectral regulation. To validate the effectiveness of MPGFRFT and endow it with data-driven learning capabilities, we encapsulate this mathematical framework into an adaptive multiple-parameter fractional spectral regulation (MPFSR) module. By deeply integrating this plug-and-play MPFSR module into mainstream deep spectral models (including SpectralCNN, LanczosNet, Specformer, and GrokFormer), we evolve their traditional static transformations into dynamically learnable fractional bases. Coupled with a differential learning rate optimization strategy, this functional module empowers the enhanced networks to adaptively extract optimal features within the continuous intermediate spatial-spectral domain. This mechanism enables the models to effectively preserve macroscopic homophilic structures within low-frequency components while precisely capturing heterogeneous details within high-frequency components. Consequently, it significantly boosts node classification accuracy on complex heterogeneous graph data, paving a novel technical pathway for exploring advanced spectral regulation mechanisms in graph representation learning.

\indent  In summary, this paper drives a fundamental paradigm shift for spectral GNNs. By transitioning from the GFT, which inherently lacks active spectral regulation mechanisms to a dynamic, multiple-parameter fractional framework, we pioneer the introduction of active MPFSR module into deep graph learning. The main contributions of this paper are summarized as follows:
\begin{itemize}

	\item We propose two types of MPGFRFTs and establish their corresponding theoretical frameworks, providing a mathematical foundation for fine-grained graph signal analysis.
	
	\item  We pioneer the integration of fractional-order transformations with deep neural networks. This novel mechanism breaks the bottleneck of fixed filter designs and achieves active spectral regulation.
	
	\item We establish rigorous theoretical bounds on the spectral perturbation of the proposed modules, which theoretically guarantees the stability of the end-to-end parameter optimization.
	
	\item We design the adaptive MPFSR module as a plug-and-play component to operationalize our mathematical framework. Integrating it into mainstream spectral models alleviates the constraints of static bases, enhances representational flexibility, and yields performance gains in graph node classification.
	
\end{itemize}

\begin{figure*}[htbp]
	\centering
	\includegraphics[width=5in]{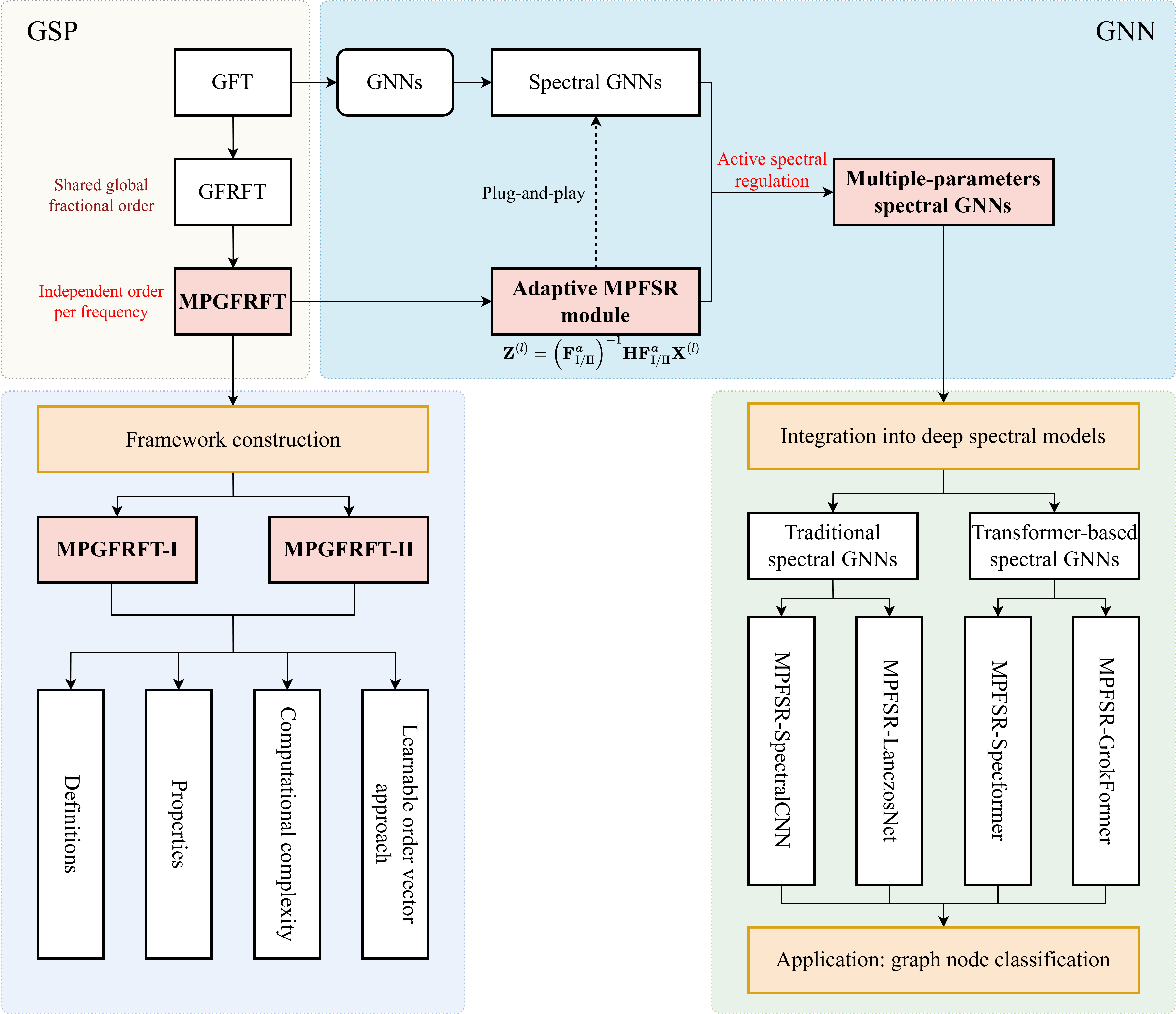}
	\caption{The overall technical framework, illustrating the theoretical construction of MPGFRFT and its plug-and-play integration into deep spectral GNNs.}
	\label{fig:MPGFRFT}
\end{figure*}

\indent  The remainder of this paper is organized as follows. Section~\ref{sec2} provides the necessary preliminaries. Section~\ref{sec3} introduces the definition and theoretical properties of the proposed MPGFRFTs, provides a detailed computational complexity analysis, and develops a learnable order vector scheme under this framework. Section~\ref{sec4} presents the proposed MPFSR module and details its plug-and-play integration into deep spectral GNN architectures. Section~\ref{sec5} evaluates the effectiveness of the enhanced models through graph node classification experiments. Finally, Section~\ref{sec6} concludes the paper. To provide a clear overview, Fig.~\ref{fig:MPGFRFT} visually summarizes the overall technical framework, spanning from the theoretical construction of MPGFRFT to its integration into deep spectral GNNs. All the technical proofs of our theoretical results are relegated to the Appendix parts.

\section{Preliminaries} \label{sec2}
\indent Let $\mathcal{G}=\left\{\mathcal{N},\mathcal{E},\mathbf{A}\right\}$ be a graph or network with node set $\mathcal{N}$, edge set $\mathcal{E}$ and adjacency matrix $\mathbf{A}\in\mathbb{R}^{N\times N}$.  The weighted adjacency matrix $\mathbf{W}$ encodes edge weights, where $W_{mn}$ quantifies the strength of the connection between nodes $m$ and $n$, and all diagonal elements satisfy $W_{nn}=0$. The Laplacian is defined as $\mathbf{L=D-W}$, where $\mathbf{D}$ is the diagonal degree matrix with $D_{nn}=\sum_{m=1}^{N}W_{nm}$. In GSP, the general GSO $\mathbf{Z}$ plays a central role in defining spectral transforms. Typical choices for the GSO include the adjacency matrix $\mathbf{A}$, the weighted adjacency matrix $\mathbf{W}$, the Laplacian $\mathbf{L}$, or their normalized variants. Let the Jordan decomposition of the GSO be written as $\mathbf{Z} = \mathbf{P}_{Z} \mathbf{J_{Z}} \mathbf{P}_{Z}^{-1}$, where $\mathbf{J_{Z}}$ is the Jordan block matrix. In particular, if $\mathbf{Z}$ is diagonalizable, it admits the eigendecomposition $\mathbf{Z} = \mathbf{U} \boldsymbol{\Lambda} \mathbf{U}^{-1}$, where $\mathbf{U}$ consists of the eigenvectors of $\mathbf{Z}$ and $\boldsymbol{\Lambda}$ is a diagonal matrix of eigenvalues. Unless otherwise stated, we assume throughout this paper that the GSO is diagonalizable. A graph signal is defined as a function $\mathbf{x}: \mathcal{N}\rightarrow \mathbb{C}$, which can be represented as a vector $\mathbf{x}=[x(0),x(1),\cdots,x(N-1)]^{\mathrm{T}}\in \mathbb{C}^{N}$. 

\indent The GFT of a signal $\mathbf{x}$ is defined as
\begin{align}
	\widehat{\mathbf{x}} = \mathbf{F}\mathbf{x} = \mathbf{U}^{-1}\mathbf{x},
\end{align}
where $\mathbf{F}=\mathbf{U}^{-1}$ is the GFT matrix. 

\indent Suppose that the GFT matrix $\mathbf{F}$ is diagonalizable, the eigendecomposition of $\mathbf{F}$ is	
\begin{align}  \label{GFT matrix}
	\mathbf{F} = \mathbf{V} \boldsymbol{\Lambda}_F \mathbf{V}^{-1},
\end{align}
where $\boldsymbol{\Lambda}_F={\rm diag}(\lambda_0,\lambda_1,\cdots,\lambda_{N-1})$ is a diagonal matrix of eigenvalues, and $\mathbf{V}=[\mathbf{v}_0,\mathbf{v}_1,\cdots,\mathbf{v}_{N-1}]$ is the matrix whose columns are the corresponding eigenvectors of $\mathbf{F}$. In this paper, we assume that the eigenvalues are distinct.
Then, the GFRFT matrix of order $a$ is defined as
\begin{align}
	\mathbf{F}^a = \mathbf{V} \boldsymbol{\Lambda}_F^a \mathbf{V}^{-1},
\end{align}
where $\boldsymbol{\Lambda}_F^a={\rm diag}(\lambda_0^a,\lambda_1^a,\cdots,\lambda_{N-1}^a)$.
Thus, the GFRFT of a signal $\mathbf{x}$ is given by \cite{Wang17}
\begin{align}
	\widehat{\mathbf{x}}_a = \mathbf{F}^a \mathbf{x}.
\end{align}

\indent Notably, the GFRFT reduces to the identity matrix when $a = 0$, i.e., $\mathbf{F}^0 = \mathbf{I}_N$, and reduces to the GFT when $a = 1$, i.e., $\mathbf{F}^1 = \mathbf{F}$.

\section{Multiple-parameter graph fractional Fourier transform}\label{sec3}
\indent In this section, we propose the MPGFRFTs and formally establish their mathematical framework by introducing two definitions. We systematically detail their respective mathematical properties and analyze the computational complexity associated with each formulation. Furthermore, we present a learnable order vector approach designed to enable the dynamic and continuous optimization of the fractional orders. This theoretical foundation directly motivates and enables the construction of the adaptive fractional spectral modules discussed in the following chapters.

\subsection{Definitions}
\indent The GFRFT extends the GFT by introducing a fractional order parameter $a$, allowing spectral analysis to be performed in intermediate domains between the vertex and frequency representations. While GFRFT provides additional flexibility over GFT, it remains limited in that all spectral components are modulated using the same fractional order. However, different frequency components may exhibit varying degrees of importance or structure. 

\indent In SP, the MPDFRFT \cite{MPDFRFT06,MPDFRFT16} has been proposed to address similar limitations of single-parameter FRFT. Inspired by this, we extend the scalar order $a$ to an order vector $\boldsymbol{a}=(a_0,a_1,\cdots,a_{N-1})\in \mathbb{R}^{N}$ with $N$ parameters, where each graph frequency component is assigned a distinct fractional order. This leads to the definition of the first type of MPGFRFT (MPGFRFT-I).

\indent \emph{Definition 1:} Given an order vector $\boldsymbol{a}=(a_0,a_1,\cdots,a_{N-1})$, the MPGFRFT-I of signal $\mathbf{x}$ is defined as
\begin{align}
	\widehat{\mathbf{x}}_{\boldsymbol{a}}^{\rm I} = \mathbf{F}_{\rm I}^{\boldsymbol{a}} \mathbf{x},
\end{align}
where $\mathbf{F}_{\rm I}^{\boldsymbol{a}}$ denotes the MPGFRFT-I matrix, given by
\begin{align} \label{MPGFRFT-I matrix1}
	\mathbf{F}_{\rm I}^{\boldsymbol{a}} = \mathbf{V} \boldsymbol{\Lambda}_F^{\boldsymbol{a}} \mathbf{V}^{-1},
\end{align}
with $\boldsymbol{\Lambda}_F^{\boldsymbol{a}}\triangleq \mathrm{diag}(\lambda_0^{a_0},\lambda_1^{a_1},\cdots,\lambda_{N-1}^{a_{N-1}})$.

\indent \emph{Theorem 1:} Let $\boldsymbol{a} = (a_0, a_1, \cdots, a_{N-1})$ be any order vector, and let the eigenvalue matrix $\boldsymbol{\Lambda}_F={\rm diag}(\lambda_0,\lambda_1,\cdots,\lambda_{N-1})$ consist of pairwise distinct eigenvalues. The MPGFRFT-I matrix defined in~\eqref{MPGFRFT-I matrix1} admits the following polynomial representation.
\begin{align}  \label{MPGFRFT-I matrix2}
	\mathbf{F}_{\rm I}^{\boldsymbol{a}} = \sum\limits_{n=0}^{N-1}C_{n,a_n}^{\rm I}\mathbf{F}^{n},
\end{align}
where the coefficient $C_{n, a_{n}}^{\mathrm{I}}$ is given by
\begin{align}
	C_{n, a_{n}}^{\mathrm{I}}=\sum\limits_{j=0}^{N-1} p_{n+1,j+1} \lambda_{j}^{a_{j}},
\end{align}
and the matrix $\mathbf{P} = (p_{ij})$ is defined as
\begin{align}  \label{eq:P}
	\mathbf{P}=\left(p_{i j}\right) \triangleq\left(\begin{array}{cccc}
		1 & \lambda_{0} & \cdots & \lambda_{0}^{N-1} \\
		1 & \lambda_{1} & \cdots & \lambda_{1}^{N-1} \\
		\vdots & \vdots & \ddots & \vdots \\
		1 & \lambda_{N-1} & \cdots & \lambda_{N-1}^{N-1}
	\end{array}\right)^{-1}.
\end{align}
\indent \emph{Proof:}
\indent According to \eqref{MPGFRFT-I matrix1} and \eqref{MPGFRFT-I matrix2}, we have
\begin{align}
	\boldsymbol{\Lambda}_F^{\boldsymbol{a}} = \sum\limits_{n=0}^{N-1}C_{n,a_n}^{\rm I}\boldsymbol{\Lambda}_F^{n},
\end{align}
which can be equivalently represented as a matrix-vector product
\begin{align}
	\left(\begin{array}{cccc}
		1 & \lambda_{0} & \cdots & \lambda_{0}^{N-1} \\
		1 & \lambda_{1} & \cdots & \lambda_{1}^{N-1} \\
		\vdots & \vdots & \ddots & \vdots \\
		1 & \lambda_{N-1} & \cdots & \lambda_{N-1}^{N-1}
	\end{array}\right)\left(\begin{array}{c}
		C_{0, a_0}^{\rm I} \\
		C_{0, a_1}^{\rm I} \\
		\vdots \\
		C_{N-1, a_{N-1}}^{\rm I}
	\end{array}\right)=\left(\begin{array}{c}
		\lambda_{0}^{a_0} \\
		\lambda_{1}^{a_1} \\
		\vdots \\
		\lambda_{N-1}^{a_{N-1}}
	\end{array}\right).
\end{align} 

\indent If the eigenvalues are distinct, the Vandermonde matrix is invertible, and we have
\begin{align}
	\left(\begin{array}{c}
		C_{0, a_0}^{\rm I} \\
		C_{0, a_1}^{\rm I} \\
		\vdots \\
		C_{N-1, a_{N-1}}^{\rm I}
	\end{array}\right)=\mathbf{P}\left(\begin{array}{c}
		\lambda_{0}^{a_0} \\
		\lambda_{1}^{a_1} \\
		\vdots \\
		\lambda_{N-1}^{a_{N-1}}
	\end{array}\right),
\end{align} 
and therefore, we arrive the required result.
\qed

\indent Correspondingly, the second type of MPGFRFT (MPGFRFT-II) can be formulated by applying the fractional orders within the coefficient generation process of the matrix polynomial expansion. 

\indent \emph{Definition 2:} Given an order vector $\boldsymbol{a}=(a_0,a_1,\cdots,a_{N-1})$, the MPGFRFT-II of signal $\mathbf{x}$ is defined as
\begin{align}
	\widehat{\mathbf{x}}_{\boldsymbol{a}}^{\rm II} = \mathbf{F}_{\rm II}^{\boldsymbol{a}} \mathbf{x},
\end{align}
where the MPGFRFT-II matrix $\mathbf{F}_{\rm II}$ is expressed as
\begin{align}  \label{MPGFRFT-II matrix}
	\mathbf{F}_{\rm II}^{\boldsymbol{a}} = \sum\limits_{n=0}^{N-1}C_{n,a_n}^{\rm II}\mathbf{F}^{n},
\end{align}
with coefficients
\begin{align} \label{coefficients-II}
	C_{n, a_{n}}^{\mathrm{II}}=\sum\limits_{j=0}^{N-1} p_{n+1,j+1} \lambda_{j}^{a_{n}}.
\end{align}

\indent \emph{Remark 1:} When the underlying graph is a cycle graph, the MPGFRFT-I matrix degenerates to the MPDFRFT matrix proposed in \cite{MPDFRFT06}. Furthermore, if the matrix exhibits periodicity,  MPGFRFT-I reduces to the type I MPDFRFT defined in \cite{MPDFRFT16}. Similarly, the MPGFRFT-II matrix reduces to the type II MPDFRFT in \cite{MPDFRFT16}, since the GFT matrix of a cycle graph coincides with the DFT matrix. In this case, the coefficients $C_{n, a_{n}}^{\mathrm{II}}$ in \eqref{coefficients-II} correspond to the $(n+1)$-th entry of the inverse DFT of $\boldsymbol{\lambda}^{a_n}=\left(\lambda_{0}^{a_n},\lambda_{1}^{a_n},\cdots,\lambda_{N-1}^{a_n} \right)^{\rm T} $, which matches the weight coefficients used in the type II MPDFRFT \cite{MPDFRFT16}.

\indent \emph{Theorem 2:} Suppose the GSO is real and symmetric, the MPGFRFT-I matrix $\mathbf{F}_{\rm I}^{\boldsymbol{a}}$ is unitary for any order vector $\boldsymbol{a}$, that is,
\begin{align}
	\mathbf{F}_{\rm I}^{\boldsymbol{a}} \left( \mathbf{F}_{\rm I}^{\boldsymbol{a}}\right) ^{\rm H} = \left( \mathbf{F}_{\rm I}^{\boldsymbol{a}}\right) ^{\rm H} \mathbf{F}_{\rm I}^{\boldsymbol{a}} = \mathbf{I}_N.
\end{align}
\indent \emph{Proof:} For a real and symmetric GSO, as given in~\eqref{GFT matrix} and \eqref{MPGFRFT-I matrix1}, the eigenvector matrix $\mathbf{V}$ satisfies $\mathbf{V}^{-1}=\mathbf{V}^{\mathrm{H}}$, and the diagonal matrix of fractional eigenvalues is  $\boldsymbol{\Lambda}_F^{\boldsymbol{a}} = \mathrm{diag}(\lambda_0^{a_0}, \lambda_1^{a_1}, \dots, \lambda_{N-1}^{a_{N-1}})$ with $|\lambda_i| = 1$ for all $i$. Thus,
\begin{align}
	\mathbf{F}_{\rm I}^{\boldsymbol{a}}
	\left(\mathbf{F}_{\rm I}^{\boldsymbol{a}}\right)^{\mathrm{H}}
	&= \mathbf{V}\boldsymbol{\Lambda}_F^{\boldsymbol{a}}\mathbf{V}^{-1}
	\left(\mathbf{V}\boldsymbol{\Lambda}_F^{\boldsymbol{a}}\mathbf{V}^{-1}\right)^{\mathrm{H}} \nonumber\\
	&= \mathbf{V}\boldsymbol{\Lambda}_F^{\boldsymbol{a}}
	\left(\boldsymbol{\Lambda}_F^{\boldsymbol{a}}\right)^{\mathrm{H}}\mathbf{V}^{-1} \nonumber\\
	&= \mathbf{V}\mathbf{I}_N\mathbf{V}^{-1} = \mathbf{I}_N,
\end{align}
which completes the proof.  \qed

\indent \emph{Remark 2:} Even when the GSO is real and symmetric, the MPGFRFT-II matrix $\mathbf{F}_{\rm II}^{\boldsymbol{a}}$ is generally not unitary, because
\begin{align}
	\mathbf{F}_{\rm II}^{\boldsymbol{a}} \left( \mathbf{F}_{\rm II}^{\boldsymbol{a}}\right) ^{\rm H}
	=&\sum\limits_{n=0}^{N-1}\sum\limits_{m=0}^{N-1}C_{n,a_n}^{\rm II}\overline{C_{n,a_n}^{\rm II}}\mathbf{F}^{n}\left(\mathbf{F}^{m}\right)^{\rm H} \nonumber\\
	=&\sum\limits_{n=0}^{N-1}\sum\limits_{m=0}^{N-1}C_{n,a_n}^{\rm II}\overline{C_{n,a_n}^{\rm II}}\mathbf{F}^{n-m},
\end{align}
which is generally not equal to the identity matrix $\mathbf{I}_N$.

\subsection{Properties}
\indent Next, we show some properties of the proposed MPGFRFT-I and MPGFRFT-II. \\
\indent \emph{Property 1 (Identity matrix):} If the order vector $\boldsymbol{a}=(0,0,\cdots,0)$, the MPGFRFT-I matrix and MPGFRFT-II matrix can both reduce to an identity matrix, that is
\begin{align}
	\mathbf{F}_{\rm I}^{\boldsymbol{a}} = \mathbf{F}_{\rm II}^{\boldsymbol{a}} = \mathbf{I}_N.
\end{align}

\indent \emph{Property 2 (Reduction to GFRFT):} When the order vector $\boldsymbol{a}=(a,a,\cdots,a)$, the MPGFRFT-I and MPGFRFT-II both reduce to the GFRFT.

\indent \emph{Property 3 (Reduction to GFT):} When the order vector $\boldsymbol{a}=(1,1,\cdots,1)$, the MPGFRFT-I and MPGFRFT-II both reduce to the GFT.

\indent \emph{Proof:} The proofs of properties 1-3 are straightforward and thus omitted.  \qed

\indent \emph{Property 4 (Index additivity):} Let $\boldsymbol{a}$ and $\boldsymbol{b}$ be two order vectors. For the MPGFRFT-I matrix, the index additivity property holds
\begin{align}
	\mathbf{F}_{\rm I}^{\boldsymbol{a}}\mathbf{F}_{\rm I}^{\boldsymbol{b}} = \mathbf{F}_{\rm I}^{\boldsymbol{a+b}}.
\end{align}
\indent In contrast, the MPGFRFT-II matrix does not satisfy the property.

\indent \emph{Proof:} For the MPGFRFT-I matrix, we have
\begin{align}
	\mathbf{F}_{\rm I}^{\boldsymbol{a}}\mathbf{F}_{\rm I}^{\boldsymbol{b}} =&\mathbf{V} \boldsymbol{\Lambda}_F^{\boldsymbol{a}} \mathbf{V}^{-1}\mathbf{V} \boldsymbol{\Lambda}_F^{\boldsymbol{b}} \mathbf{V}^{-1} \nonumber\\
	=&\mathbf{V} \boldsymbol{\Lambda}_F^{\boldsymbol{a+b}} \mathbf{V}^{-1} = \mathbf{F}_{\rm I}^{\boldsymbol{a+b}}.
\end{align}
\indent For the MPGFRFT-II matrix, we have
\begin{align}
	\mathbf{F}_{\rm II}^{\boldsymbol{a}}\mathbf{F}_{\rm II}^{\boldsymbol{b}}= \sum\limits_{n=0}^{N-1}\sum\limits_{m=0}^{N-1}C_{n,a_{n}}^{\rm II}C_{m,b_{m}}^{\rm II}\mathbf{F}^{n}\mathbf{F}^{m},
\end{align}
which, in general, does not reduce to 
\begin{align}
	\mathbf{F}_{\rm II}^{\boldsymbol{a+b}} = \sum\limits_{n=0}^{N-1}C_{n,a_n+b_n}^{\rm II}\mathbf{F}^{n}.
\end{align} 
\indent Thus, we arrive the required results. \qed

\indent \emph{Property 5 (Reversibility):} The MPGFRFT-I matrix is invertible for any order vector $\boldsymbol{a}$, and its inverse is given by
\begin{align}  \label{Reversibility1}
	\left( \mathbf{F}_{\rm I}^{\boldsymbol{a}}\right)^{-1}=	\mathbf{F}_{\rm I}^{-\boldsymbol{a}}.
\end{align}
\indent The MPGFRFT-II matrix is invertible if the following condition holds
\begin{align} \label{Reversibility2}
	\sum\limits_{n=0}^{N-1}\left(\sum\limits_{k=0}^{N-1} p_{n+1, k+1} \lambda_{k}^{a_{n}}\right) \lambda_{j}^{n} \neq 0
\end{align}
for each $\lambda_{j}$ $(j=0,1,\cdots,N-1)$.

\indent \emph{Proof:} According to the index additivity of MPGFRFT-I matrix, we have
\begin{align}
	\mathbf{F}_{\rm I}^{\boldsymbol{a}}\mathbf{F}_{\rm I}^{-\boldsymbol{a}}=\mathbf{F}_{\rm I}^{\boldsymbol{0}}=\mathbf{I}_N.
\end{align}

\indent Inserting \eqref{GFT matrix} into \eqref{MPGFRFT-II matrix}, we have
\begin{align}
	\mathbf{F}_{\rm II}^{\boldsymbol{a}} =& \sum\limits_{n=0}^{N-1}C_{n,a_n}^{\rm II} \mathbf{V}\left(\boldsymbol{\Lambda}_F\right)^{n} \mathbf{V}^{-1} \nonumber\\
	=& \mathbf{V}\left(\sum\limits_{n=0}^{N-1}C_{n,a_n}^{\rm II}\left(\boldsymbol{\Lambda}_F\right)^{n}\right)\mathbf{V}^{-1}.
\end{align}
\indent Therefore, $\mathbf{F}_{\rm II}^{\boldsymbol{a}}$ is invertible if and only if the matrix $\sum_{n=0}^{N-1}C_{n,a_n}^{\rm II}\left(\boldsymbol{\Lambda}_F\right)^{n}$ is invertible, which is equivalent to requiring that all diagonal entries of the matrix are nonzero.  \qed

\indent \emph{Property 6 (Index commutativity):} Both MPGFRFT-I and MPGFRFT-II satisfy the index commutativity property, i.e.
\begin{align}
	\mathbf{F}_{\rm I/II}^{\boldsymbol{a}}\mathbf{F}_{\rm I/II}^{\boldsymbol{b}} = \mathbf{F}_{\rm I/II}^{\boldsymbol{b}}\mathbf{F}_{\rm I/II}^{\boldsymbol{a}}.
\end{align}

\indent \emph{Proof:} The proof is straightforward and thus omitted.  \qed

\indent \emph{Property 7 (Eigenvalues and eigenvectors):} Assume $\mathbf{V}$ is unitary. Then, the eigenvectors $\mathbf{v}_{k}$ of the GFT matrix $\mathbf{F}$ are also eigenvectors of the MPGFRFT-I matrix, with corresponding eigenvalues $\lambda_{k}^{a_{k}}$. 

\indent Similarly, the eigenvectors $\mathbf{v}_{k}$ of $\mathbf{F}$ remain the eigenvectors of the MPGFRFT-II matrix, with corresponding eigenvalues given by $\sum_{n=0}^{N-1} C_{n, a_{n}}^{\mathrm{II}} \lambda_k^{n}$.  

\indent \emph{Proof:} If $\mathbf{V}$ is unitary, we have
\begin{align}
	\mathbf{F}_{\rm I}^{\boldsymbol{a}}\mathbf{v}_k = \left(\sum\limits_{m=0}^{N-1} \lambda_{m}^{a_{m}} \mathbf{v}_{m} \mathbf{v}_{m}^{\mathrm{H}}\right) \mathbf{v}_{k}=\lambda_{k}^{a_{k}} \mathbf{v}_{k},
\end{align}
\begin{align}
	\mathbf{F}_{\rm II}^{\boldsymbol{a}} \mathbf{v}_{k}
	=\left(\sum\limits_{n=0}^{N-1} C_{n, a_{n}}^{\mathrm{II}} \mathbf{F}^{n}\right) \mathbf{v}_{k} 
	=\sum\limits_{n=0}^{N-1} C_{n, a_{n}}^{\mathrm{II}} \lambda_k^{n} \mathbf{v}_{k}.
\end{align}
\indent Thus, we arrive the required results. \qed

\indent \emph{Property 8 (Linearity):} Both MPGFRFT-I and MPGFRFT-II are linear transformations:
\begin{align}
	\mathbf{F}_{\mathrm{I/II}}^{\boldsymbol{a}}\left(c_{1}\mathbf{x}+c_{2}\mathbf{y}\right)
	=c_{1}\mathbf{F}_{\mathrm{I/II}}^{\boldsymbol{a}}\mathbf{x}+c_{2}\mathbf{F}_{\mathrm{I/II}}^{\boldsymbol{a}}\mathbf{y}.
\end{align}

\indent \emph{Proof:} The proof is straightforward and thus omitted.  \qed

\subsection{Computational complexity}
\indent Regarding matrix construction, for the conventional GFRFT, the eigendecompositions of the GSO and the GFT matrix $\mathbf{F}$ both require $\mathcal{O}(N^3)$ operations. For MPGFRFT-I, the transform matrix is defined as $\mathbf{F}_{\rm I}^{\boldsymbol{a}}=\mathbf{V}\boldsymbol{\Lambda}_F^{\boldsymbol{a}}\mathbf{V}^{-1}$. Forming the diagonal matrix $\boldsymbol{\Lambda}_F^{\boldsymbol{a}}$ only requires $\mathcal{O}(N)$ operations via element-wise exponentiation of the eigenvalues. Therefore, the overall computational complexity for constructing the MPGFRFT-I matrix remains $\mathcal{O}(N^3)$, which is identical to that of the conventional GFRFT matrix. For MPGFRFT-II, the computation of the coefficients $C_{n,a_n}^{\rm II}$ requires $\mathcal{O}(N^2)$ operations. The dominant computational cost arises from the calculation of the matrix powers $\mathbf{F}^n$. Since the multiplication of two $N \times N$ dense matrices requires $\mathcal{O}(N^3)$ operations, recursively computing $\mathbf{F}^n = \mathbf{F} \cdot \mathbf{F}^{n-1}$ for $n=2, \dots, N-1$ requires $\mathcal{O}(N^4)$ operations in total. Consequently, constructing the dense matrix $\mathbf{F}_{\rm II}^{\boldsymbol{a}}$ has an overall computational complexity of $\mathcal{O}(N^4)$.

\indent However, in practical applications where the goal is to transform a specific graph signal $\mathbf{x}$, explicitly constructing the full dense matrix for MPGFRFT-II is unnecessary, and the overall complexity for all three transforms aligns at $\mathcal{O}(N^3)$. For the conventional GFRFT and MPGFRFT-I, the dominant computational cost is the eigendecomposition, taking $\mathcal{O}(N^3)$ operations, while the subsequent matrix-vector multiplications only take $\mathcal{O}(N^2)$. For MPGFRFT-II, the transform is evaluated directly via the matrix polynomial, i.e., $\widehat{\mathbf{x}}_{\boldsymbol{a}}^{\rm II} = \sum_{n=0}^{N-1}C_{n,a_n}^{\rm II}\mathbf{F}^{n}\mathbf{x}$. By leveraging the associativity of matrix multiplication, we recursively evaluate $\mathbf{v}_n = \mathbf{F}\mathbf{v}_{n-1}$ with $\mathbf{v}_0 = \mathbf{x}$. This bypasses the $\mathcal{O}(N^4)$ matrix power computation. Because each of the $N-1$ recursive steps is a matrix-vector multiplication requiring $\mathcal{O}(N^2)$ operations, the total cost for this recursive evaluation is bounded by $\mathcal{O}(N^3)$. Table \ref{tab:complexity} summarizes the computational complexities. 

\begin{table}[htbp]
	\centering
	\caption{Computational Complexity Comparison}
	\label{tab:complexity}
	\renewcommand{\arraystretch}{1.3}
	\resizebox{\linewidth}{!}{
		\begin{tabular}{l c c}
			\toprule
			\textbf{Transform} & \textbf{Matrix construction} & \textbf{Signal transformation} \\
			\midrule
			GFRFT & $\mathcal{O}(N^3)$ & $\mathcal{O}(N^3)$ \\
			MPGFRFT-I & $\mathcal{O}(N^3)$ & $\mathcal{O}(N^3)$ \\
			MPGFRFT-II & $\mathcal{O}(N^4)$ & $\mathcal{O}(N^3)$ \\
			\bottomrule
		\end{tabular}
	}
\end{table}

\subsection{Learnable order vector approach} \label{sec:Learnable Order Vector Approach}
\indent The MPGFRFT-I and MPGFRFT-II matrices are both constructed as polynomial functions of the fractional order vector $\boldsymbol{a}=[a_0, a_1, \dots, a_{N-1}]\in \mathbb{R}^N$, as defined in \eqref{MPGFRFT-I matrix2} and \eqref{MPGFRFT-II matrix}. Thus, these transformations are differentiable with respect to $\boldsymbol{a}$. This property enables their integration into deep learning architectures as trainable layers with learnable fractional orders. Motivated by this observation, we treat $\boldsymbol{a}$ as a set of trainable parameters rather than fixed constants, and study the analytical form of the gradients required for backpropagation.

\indent \emph{Theorem 3:} The gradient of the MPGFRFT matrix $\mathbf{F}_{\rm I/II}^{\boldsymbol{a}}$ with respect to the order vector $\boldsymbol{a} = [a_0, a_1, \dots, a_{N-1}] \in \mathbb{R}^N$ is a third-order tensor defined as
\begin{align} 
	\dot{\mathbf{F}}_{\rm I/II}^{\boldsymbol{a}}
	=\frac{\partial\left(\mathbf{F}_{\rm I/II}^{\boldsymbol{a}}\right)}{\partial \boldsymbol{a}}
	=\left[\frac{\partial \mathbf{F}_{\rm I/II}^{\boldsymbol{a}}}{\partial a_0}, \frac{\partial \mathbf{F}_{\rm I/II}^{\boldsymbol{a}}}{\partial a_1}, \dots, \frac{\partial \mathbf{F}_{\rm I/II}^{\boldsymbol{a}}}{\partial a_{N-1}} \right],
\end{align}
where the $k$-th slice of the tensor corresponds to the partial derivative with respect to $a_k$. Specifically, we have
\begin{align}  \label{MPGFRFT-I slice}
	\frac{\partial \mathbf{F}_{\rm I}^{\boldsymbol{a}}}{\partial a_k} = 
	\sum\limits_{n=0}^{N-1} \left( p_{n+1,k+1} \lambda_k^{a_k} \ln \lambda_k \right) \mathbf{F}^n ,
\end{align}
for the MPGFRFT-I case, and
\begin{align}
	\frac{\partial \mathbf{F}_{\rm II}^{\boldsymbol{a}}}{\partial a_k} = \left( \sum\limits_{j=0}^{N-1} p_{k+1,j+1} \lambda_j^{a_k} \ln \lambda_j \right) \mathbf{F}^k,
\end{align}
for the MPGFRFT-II case.

\indent \emph{Proof:} It is obvious that
\begin{align}
	\frac{\partial \left(\mathbf{F}_{\rm I/II}^{\boldsymbol{a}}\right) }{\partial a_k} = 
	\sum\limits_{n=0}^{N-1} \frac{\partial C_{n,a_n}^{\rm I/II}}{\partial a_k} \mathbf{F}^n.
\end{align}
\indent Differentiating the MPGFRFT-I coefficient $C_{n,a_n}^{\rm I}$ with respect to $a_k$ gives
\begin{align}
	\frac{\partial C_{n,a_n}^{\rm I}}{\partial a_k} = p_{n+1,k+1}\,\lambda_k^{a_k} \ln \lambda_k.
\end{align}
\indent Differentiating the MPGFRFT-II coefficient $C_{n,a_n}^{\rm II}$ with respect to $a_k$ gives 
\begin{align}
	\frac{\partial C_{n,a_n}^{\rm II}}{\partial a_k} = \left\{\begin{matrix}
		\sum\limits_{j=0}^{N-1} p_{k+1,j+1} \lambda_j^{a_k} \ln \lambda_j & k=n\\
		0  & k\neq n
	\end{matrix}\right.
\end{align}
\indent Therefore, we arrive the required results. \qed

\indent We construct a differentiable MPGFRFT layer, where the forward pass at $l$-th layer is
\begin{align}
	\mathbf{x}^{(l)}=\varphi\left(\mathbf{F}_{\rm I/II}^{\boldsymbol{a}} \mathbf{x}^{(l-1)}\right)
\end{align}
with $\varphi$ denoting any differentiable activation function. The learning process involves minimizing a task-specific loss function $\mathcal{L}$, with gradients of the form
\begin{align}
	\frac{\partial \mathcal{L}}{\partial \boldsymbol{a}}=\left(\nabla_{\mathbf{x}^{(l)}} \mathcal{L}\right)^{\top} \frac{\partial \mathbf{x}^{(l)}}{\partial \boldsymbol{a}}
\end{align}
with
\begin{align}
	\frac{\partial \mathbf{x}^{(l)}}{\partial \boldsymbol{a}}
	=\dot{\varphi}\left(\mathbf{F}_{\rm I/II}^{\boldsymbol{a}} \mathbf{x}^{(l-1)}\right) \odot\left(\dot{\mathbf{F}}_{\rm I/II}^{\boldsymbol{a}}\mathbf{x}^{(l-1)}\right),
\end{align}
where $\odot$ denotes element-wise multiplication and $\dot{\mathbf{F}}_{\rm I/II}^{\boldsymbol{a}}$ is given by Theorem 3.  The order vector is then updated via standard stochastic gradient descent
\begin{align}
	\boldsymbol{a}^{(t+1)}=\boldsymbol{a}^{(t)}-\gamma \nabla_{\boldsymbol{a}} \mathcal{L},
\end{align} 
where $\gamma$ is the learning rate. 

\indent While our approach is rooted in the multiple-parameter setting of MPGFRFT, it also generalizes prior scalar-order learning schemes. Recent work on the GFRFT proposed learning a scalar fractional order $a$ to improve transform adaptability \cite{Alikasifoglu24}. This scalar-order approach imposes a uniform spectral transformation and may not be sufficient for graph signals with complex spectral diversity. In contrast, our learnable order vector approach allows each spectral component to evolve independently, enabling stronger task-specific adaptation. Notably, when the order vector is constrained such that $a_0=a_1=\cdots=a_{N-1}$, our method naturally reduces to the scalar-order GFRFT as a special case. 

\section{Adaptive Fractional Spectral Graph Neural Networks}\label{sec4}
\indent In this section, we utilize the mathematical framework of the MPGFRFT to design the adaptive MPFSR module. By transforming static bases into dynamically learnable fractional bases, this module provides deep graph models with the crucial benefit of active and fine-grained spectral modulation. Furthermore, we detail the plug-and-play integration of the MPFSR module into mainstream spectral neural networks to effectively perform the graph node classification task.

\subsection{Adaptive MPFSR module}
\indent Traditional spectral GNNs fundamentally rely on the eigendecomposition of the GSO, such as the Laplacian matrix $\mathbf{L} = \mathbf{U} \boldsymbol{\Lambda} \mathbf{U}^{-1}$. In this paradigm, the eigenvector matrix $\mathbf{U}$ and its inverse $\mathbf{U}^{-1}$ are utilized as static bases. Because these transformation bases remain strictly fixed throughout the training process, the networks lack the capacity to actively regulate their spectral responses to accommodate the heterogeneous characteristics of complex graph signals.

\indent To break the limitation of static spectral filters, we propose the adaptive MPFSR module. Leveraging the theoretical MPGFRFT framework established in Section~\ref{sec3}, this module is instantiated as an active, parameterized neural network component. Rather than utilizing the fixed $\mathbf{U}$ and $\mathbf{U}^{-1}$, the module encapsulates the learnable multiple-parameter fractional order vector $\boldsymbol{a}$ to dynamically generate the fractional transformation bases. Let $\mathcal{M}_{\mathrm{I/II}}(\cdot)$ denote the MPGFRFT transformation operator, the module internally constructs the forward and inverse bases as
\begin{align}
	\left[\mathbf{F}_{\mathrm{I/II}}^{\boldsymbol{a}},\left( \mathbf{F}_{\mathrm{I/II}}^{\boldsymbol{a}}\right)^{-1}\right] = \mathcal{M}_{\mathrm{I/II}}(\mathbf{U}, \boldsymbol{a}).
\end{align}

\begin{figure*}[htbp]
	\centering
	\includegraphics[width=6in]{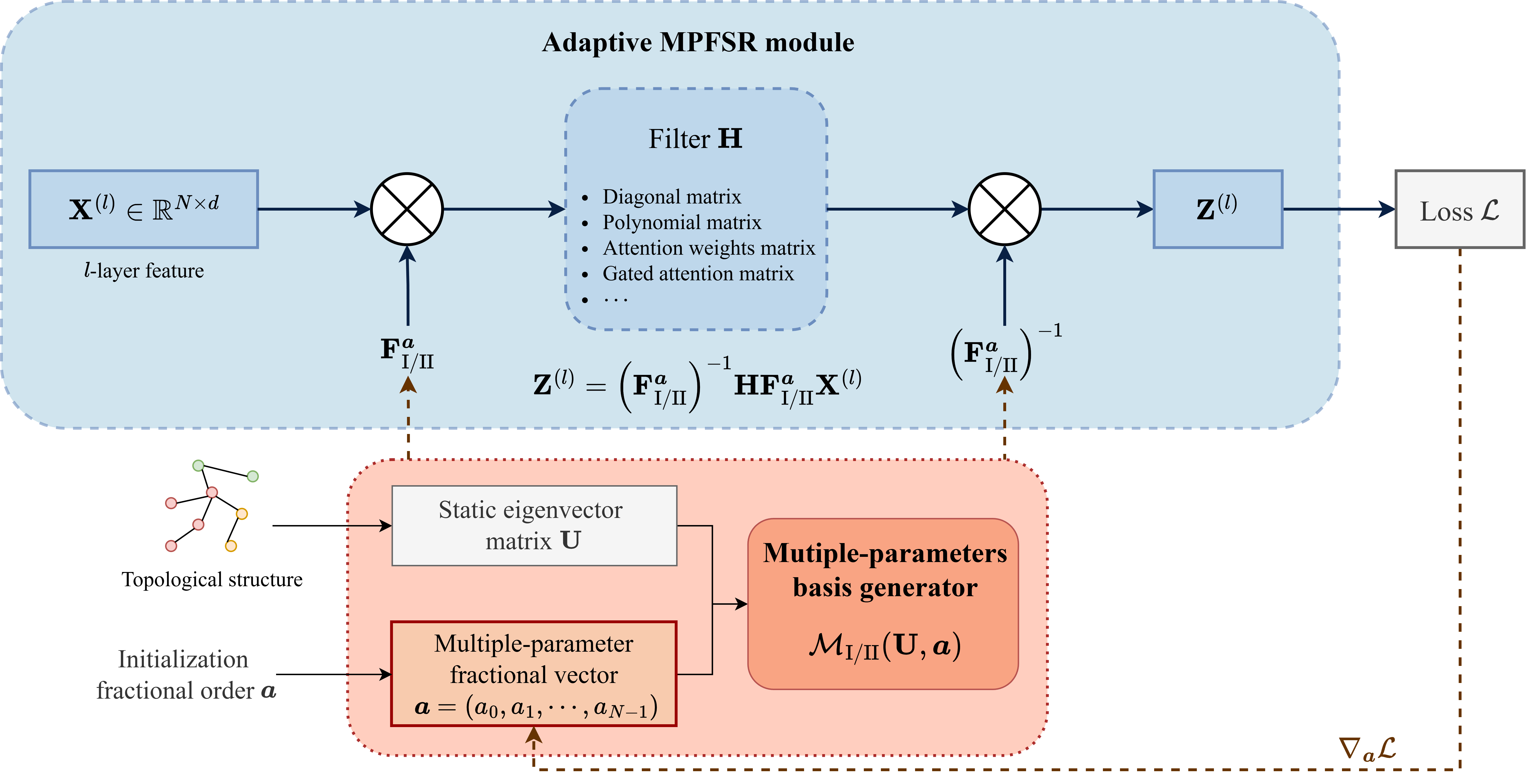}
	\caption{The internal architecture of the adaptive MPFSR module.}
	\label{fig:module}
\end{figure*}

\indent Crucially, the MPFSR module redefines the entire spectral filtering paradigm to enable fine-grained active regulation. The operational pipeline of the module consists of three sequential stages. The internal architecture of this MPFSR module is detailed in Fig.~\ref{fig:module}. First, the signal is projected into the continuous intermediate fractional spectral domain via the basis $\mathbf{F}_{\mathrm{I/II}}^{\boldsymbol{a}}$. Second, a learnable diagonal spectral filter $\mathbf{H}$ modulates the signal in the multiple-parameter fractional domain. Finally, the modulated signal is projected back to the vertex domain via the basis $\left( \mathbf{F}_{\mathrm{I/II}}^{\boldsymbol{a}}\right)^{-1}$. Therefore, for an input node representation matrix $\mathbf{X}^{(l)}$ at the $l$-th layer, the core fractional operation of the MPFSR module is elegantly formulated as
\begin{align} \label{MPFSR module}
	\mathbf{Z}^{(l)} = \left( \mathbf{F}_{\mathrm{I/II}}^{\boldsymbol{a}}\right)^{-1} \mathbf{H} \mathbf{F}_{\mathrm{I/II}}^{\boldsymbol{a}} \mathbf{X}^{(l)},
\end{align}
where $\mathbf{Z}^{(l)}$ denotes the modulated intermediate representation. This intermediate output is subsequently fed into architecture-specific non-linear transformations to compute the next layer's representation $\mathbf{X}^{(l+1)}$. 

\indent By registering the fractional order vector $\boldsymbol{a}$ as explicit, optimizable parameters within the computational graph, the entire basis generation and filtering pipeline becomes end-to-end differentiable. Rather than remaining constrained by the static graph topology, the MPFSR module adaptively updates these spatial-spectral bases driven by the downstream task loss. This mechanism enables the model to dynamically search for the optimal continuous intermediate representation space, fundamentally facilitating task-driven feature extraction for accurate graph node classification.

\indent While this active spectral modulation significantly enhances the expressive power of GNNs, introducing parameterized dynamic bases inherently raises concerns regarding numerical stability and forward-propagation robustness. To mathematically guarantee that the MPFSR module behaves smoothly during the end-to-end optimization of $\boldsymbol{a}$, we establish rigorous theoretical bounds on its spectral stability. Specifically, we evaluate the representation perturbation introduced by our module relative to the standard GFT-based paradigm.

\indent \emph{Theorem 4:} Suppose the GSO is real and symmetric, for any input graph signal $\mathbf{X}^{(l)}\in \mathbb{R}^{N\times d}$, the perturbation in the output representation caused by the MPGFRFT-I based MPFSR module relative to the GFT based module is bounded by
\begin{align}
	&\left\| \left( \mathbf{F}_{\mathrm{I}}^{\boldsymbol{a}}\right)^{-1} \mathbf{H} \mathbf{F}_{\mathrm{I}}^{\boldsymbol{a}} \mathbf{X}^{(l)} - \mathbf{F}^{-1} \mathbf{H} \mathbf{F} \mathbf{X}^{(l)} \right\|_2 \nonumber\\
	\le & 2\pi \|\mathbf{H}\|_2 \|\mathbf{X}^{(l)}\|_2 \|\boldsymbol{a} - \mathbf{1}_N\|_\infty,
\end{align}
where $\mathbf{1}_N=(1,\cdots,1)$ is the $N$-dimensional all-one vector, $\|\cdot\|_\infty$ denotes the $L_\infty$ norm for vectors,  and $\|\cdot\|_2$ denotes the spectral norm for matrices.

\indent \emph{Proof:}  Let $\mathcal{M}_{\mathrm{I}}(\boldsymbol{a}) = \left(\mathbf{F}_{\mathrm{I}}^{\boldsymbol{a}}\right)^{-1} \mathbf{H} \mathbf{F}_{\mathrm{I}}^{\boldsymbol{a}}$. We aim to bound the perturbation $\|\left(\mathcal{M}_{\mathrm{I}}(\boldsymbol{a}) - \mathcal{M}_{\mathrm{I}}(\mathbf{1}_N)\right)\mathbf{X}^{(l)}\|_2$.

\indent According to Theorem 2, applying the submultiplicativity of the spectral norm and the unitary property, we have
\begin{align}
	&\left\|\mathcal{M}_{\mathrm{I}}(\boldsymbol{a}) - \mathcal{M}_{\mathrm{I}}(\mathbf{1}_N)\right\|_2 \nonumber\\
	\le& \left\|\left( \mathbf{F}_{\mathrm{I}}^{\boldsymbol{a}}\right)^{-1} \mathbf{H} \left(\mathbf{F}_{\mathrm{I}}^{\boldsymbol{a}}-\mathbf{F} \right)\right\|_2 + \left\|\left(\left( \mathbf{F}_{\mathrm{I}}^{\boldsymbol{a}}\right)^{-1}-\mathbf{F}^{-1}\right) \mathbf{H} \mathbf{F}\right\|_2 \nonumber\\
	\le& 2 \left\|\mathbf{H}\right\|_2 \left\|\mathbf{F}_{\mathrm{I}}^{\boldsymbol{a}} - \mathbf{F}\right\|_2.
\end{align}

\indent Since $\mathbf{F}_{\mathrm{I}}^{\boldsymbol{a}} = \mathbf{V} \mathbf{\Lambda}_F^{\boldsymbol{a}} \mathbf{V}^H$ and $\mathbf{F} = \mathbf{V} \mathbf{\Lambda}_F \mathbf{V}^H$ share the same unitary eigenvector matrix $\mathbf{V}$, the spectral distance between the bases is determined by their eigenvalues
\begin{align}
	\left\|\mathbf{F}_{\mathrm{I}}^{\boldsymbol{a}} - \mathbf{F}\right\|_2 = \left\|\mathbf{\Lambda}_F^{\boldsymbol{a}} - \mathbf{\Lambda}_F\right\|_2 = \max_{k} \left|\lambda_k^{a_k} - \lambda_k\right|.
\end{align}

\indent Given that the eigenvalues of the unitary matrix $\mathbf{F}$ lie on the unit circle, we denote $\lambda_k = e^{i\omega_k}$ with $\omega_k \in (-\pi, \pi]$. Applying the chord length inequality $|e^{ix} - 1| \le |x|$ for $x \in \mathbb{R}$, we have
\begin{align}
	 \max_{k}\left|\lambda_k^{a_k} - \lambda_k\right| =&  \max_{k} \left|e^{i(a_k - 1)\omega_k} - 1\right| \nonumber\\
	\le& \max_{k}\left|(a_k - 1)\omega_k\right| \nonumber\\
	\le& \pi \left\|\boldsymbol{a} - \mathbf{1}_N\right\|_\infty.
\end{align}
Substituting this uniform upper bound back into the operator difference and multiplying by the input signal norm $\left\|\mathbf{X}^{(l)}\right\|_2$ yields the desired bound.  \qed

\indent \emph{Theorem 5:} Suppose the GSO is real and symmetric, and let $C_P = \sum_{n=0}^{N-1} \sum_{j=0}^{N-1} |p_{n+1, j+1}|$. If the fractional order vector satisfies $\|\boldsymbol{a} - \mathbf{1}_N\|_\infty < \frac{1}{\pi C_P}$, the perturbation in the output representation caused by the MPGFRFT-II based MPFSR module relative to the GFT based module is bounded by
\begin{align}  \label{th5 eq}
	&\left\| \left( \mathbf{F}_{\mathrm{II}}^{\boldsymbol{a}}\right)^{-1} \mathbf{H} \mathbf{F}_{\mathrm{II}}^{\boldsymbol{a}} \mathbf{X}^{(l)} - \mathbf{F}^{-1} \mathbf{H} \mathbf{F} \mathbf{X}^{(l)} \right\|_2 \nonumber\\
	\le & 2\pi\|\mathbf{H}\|_2\|\mathbf{X}^{(l)}\|_2\frac{C_P\|\boldsymbol{a} - \mathbf{1}_N\|_\infty}{1-\pi C_P\|\boldsymbol{a} - \mathbf{1}_N\|_\infty}.
\end{align}

\indent \emph{Proof:}  Let $\mathcal{M}_{\mathrm{II}}(\boldsymbol{a}) = \left(\mathbf{F}_{\mathrm{II}}^{\boldsymbol{a}}\right)^{-1} \mathbf{H} \mathbf{F}_{\mathrm{II}}^{\boldsymbol{a}}$. We aim to bound the perturbation $\left\|\left(\mathcal{M}_{\mathrm{II}}(\boldsymbol{a}) - \mathcal{M}_{\mathrm{II}}(\mathbf{1}_N)\right)\mathbf{X}^{(l)}\right\|_2$.

\indent Applying the triangle inequality and the unitarity of the GFT matrix $\mathbf{F}$, we obtain
\begin{align} \label{th5 eq1}
	&\left\|\mathcal{M}_{\mathrm{II}}(\boldsymbol{a}) - \mathcal{M}_{\mathrm{II}}(\mathbf{1}_N)\right\|_2 \nonumber\\
	\le & \left\| \left(\mathbf{F}_{\mathrm{II}}^{\boldsymbol{a}}\right)^{-1} \mathbf{H} \left(\mathbf{F}_{\mathrm{II}}^{\boldsymbol{a}} - \mathbf{F}\right) \right\|_2 +
	\left\| \left(\mathbf{F}_{\mathrm{II}}^{\boldsymbol{a}}\right)^{-1} \left(\mathbf{F} - \mathbf{F}_{\mathrm{II}}^{\boldsymbol{a}}\right) \mathbf{F}^{-1} \mathbf{H} \mathbf{F} \right\|_2 \nonumber\\
	\le & 2\left\|\mathbf{H} \right\|_2 \left\| \left(\mathbf{F}_{\mathrm{II}}^{\boldsymbol{a}}\right)^{-1}  \right\|_2  \left\|  \mathbf{F}_{\mathrm{II}}^{\boldsymbol{a}} - \mathbf{F} \right\|_2 .
\end{align}

\indent By Property 7, $\mathbf{F}_{\mathrm{II}}^{\boldsymbol{a}}$ shares the unitary eigenvector matrix $\mathbf{V}$ with $\mathbf{F}$. Consequently, $\mathbf{F}_{\mathrm{II}}^{\boldsymbol{a}}$ is a normal matrix. Denoting the eigenvalues of $\mathbf{F}_{\mathrm{II}}^{\boldsymbol{a}}$ as $\mu_k(\boldsymbol{a}) = \sum_{n=0}^{N-1} \sum_{j=0}^{N-1} p_{n+1, j+1} \lambda_j^{a_n} \lambda_k^n$, we have
\begin{align}  \label{th5 eq2}
	\left\|  \mathbf{F}_{\mathrm{II}}^{\boldsymbol{a}} - \mathbf{F} \right\|_2
	=&\max_{k} \left|\mu_k(\boldsymbol{a}) - \lambda_k\right| \nonumber\\
	\le& \max_k \sum_{n=0}^{N-1} \left|\lambda_k\right|^n \sum_{j=0}^{N-1} \left|p_{n+1, j+1}\right| \left|\lambda_j^{a_n} - \lambda_j\right| \nonumber\\
	\le& \pi \max_m \left|a_m-1\right| C_P \nonumber\\
	=& \pi C_P \left\|\boldsymbol{a} - \mathbf{1}_N\right\|_\infty.
\end{align}

\indent Furthermore, as a normal matrix, the inverse spectral norm of $\mathbf{F}_{\mathrm{II}}^{\boldsymbol{a}}$ is strictly determined by the reciprocal of its minimum eigenvalue magnitude. Utilizing the reverse triangle inequality yields
\begin{align}
	\left|\mu_k(\boldsymbol{a})\right| 
	\ge & \left|\lambda_k\right| - \left|\mu_k(\boldsymbol{a}) - \lambda_k\right| \nonumber\\
	\ge & 1 - \pi C_P \left\|\boldsymbol{a} - \mathbf{1}_N\right\|_\infty.
\end{align}
The condition $\|\boldsymbol{a} - \mathbf{1}_N\|_\infty < \frac{1}{\pi C_P}$ guarantees $\min_k |\mu_k(\boldsymbol{a})| > 0$, structurally ensuring invertibility. Thus, 
\begin{align}  \label{th5 eq3}
	 \left\| \left(\mathbf{F}_{\mathrm{II}}^{\boldsymbol{a}}\right)^{-1}  \right\|_2
	 =& \left\|\mathbf{V}\left(\sum\limits_{n=0}^{N-1}C_{n,a_n}^{\rm II}\boldsymbol{\Lambda}_F^{n}\right)^{-1} \mathbf{V}^{\mathrm{H}}\right\|_2 \nonumber\\
	 =& \frac{1}{\min_k\left|\mu_k(\boldsymbol{a})\right|} \nonumber\\
	 \le& \frac{1}{1 - \pi C_P \left\|\boldsymbol{a} - \mathbf{1}_N\right\|_\infty}.
\end{align}

\indent Substituting Eqs.~\eqref{th5 eq2} and \eqref{th5 eq3} into \eqref{th5 eq1} yields the desired result in \eqref{th5 eq}.     \qed

\indent Theorems 4 and 5 establish the rigorous spectral stability of the proposed MPFSR modules, effectively bridging the gap between theoretical design and empirical network optimization. A key insight from both bounds is their linear dependence on the maximum deviation $\|\boldsymbol{a} - \mathbf{1}_N\|_\infty$. This mathematical property directly justifies our training strategy of initializing the fractional parameter vector $\boldsymbol{a}$ to $\mathbf{1}_N$, ensuring that the learning process begins safely within a standard, stable GNN subspace before smoothly expanding its spectral expressivity. Furthermore, these bounded perturbations guarantee Lipschitz continuity with respect to the fractional parameters. Such continuity is crucial for robust model training, as it prevents the forward representations from experiencing numerical explosion during parameter updates, thereby guaranteeing a stable and reliable convergence.

\subsection{Plug-and-play integration into deep spectral models}
\indent A compelling advantage of the proposed MPFSR module is its strict mathematical encapsulation, which endows it with exceptional generalizability and a plug-and-play nature. By generating dynamically updated fractional bases, the module can seamlessly replace the static eigendecomposition operators in existing spectral architectures. To demonstrate its versatility, we integrate the MPFSR module into four representative deep spectral GNNs: SpectralCNN \cite{bruna2014spectral}, LanczosNet \cite{liao2019lanczosnet}, Specformer \cite{Bo2023}, and GrokFormer \cite{Ai2025}. For clarity, the resulting enhanced architectures are denoted as MPFSR-SpectralCNN, MPFSR-LanczosNet, MPFSR-Specformer, and MPFSR-GrokFormer, respectively. The overall end-to-end training procedure for these enhanced architectures is summarized in Algorithm \ref{alg:MPFSR_training}.

\begin{algorithm}[htbp]
	\caption{End-to-end training framework of MPFSR-enhanced GNNs}
	\label{alg:MPFSR_training}
	\begin{algorithmic}[1]
		\REQUIRE Graph $\mathcal{G}=(\mathcal{V}, \mathcal{E})$ with input features $\mathbf{X}$; node labels $\mathbf{Y}$ for the training set $\mathcal{V}_{train}$; static eigenvector matrix $\mathbf{U}$; initialized weights $\mathbf{W}$; fractional order vector $\boldsymbol{a}$; decoupled learning rates $\eta_{\mathbf{W}}, \eta_{\boldsymbol{a}}$ and weight decay $\lambda$.
		\ENSURE Optimized network weights $\mathbf{W}^*$, optimal fractional bases $\mathbf{F}_{\mathrm{I/II}}^{\boldsymbol{a}^*}$, and final node predictions $\hat{\mathbf{Y}}$.
		
		\WHILE{not converged}
		\STATE \textbf{Phase 1: Adaptive multiple-parameter fractional basis generation}
		\STATE Dynamically construct the fractional bases via the MPGFRFT operator: 
		\STATE $\left[\mathbf{F}_{\mathrm{I/II}}^{\boldsymbol{a}},\left( \mathbf{F}_{\mathrm{I/II}}^{\boldsymbol{a}}\right)^{-1}\right] \leftarrow \mathcal{M}_{\mathrm{I/II}}(\mathbf{U}, \boldsymbol{a})$
		
		\STATE \textbf{Phase 2: Forward propagation via MPFSR module integration}
		\STATE Extract the modulated intermediate representation $\mathbf{Z}^{(l)}$ in the fractional domain:
		\STATE $\mathbf{Z}^{(l)} \leftarrow \left( \mathbf{F}_{\mathrm{I/II}}^{\boldsymbol{a}}\right)^{-1} \mathbf{H} \mathbf{F}_{\mathrm{I/II}}^{\boldsymbol{a}} \mathbf{X}^{(l)}$
		\STATE Integrate $\mathbf{Z}^{(l)}$ into the architecture of the selected backbone to yield final node embeddings and predictions $\hat{\mathbf{Y}}$.
		
		\STATE \textbf{Phase 3: Loss computation \& Parameter optimization}
		\STATE Compute the task-specific loss: $\mathcal{L}(\hat{\mathbf{Y}}, \mathbf{Y})$
		\STATE Compute gradients via backpropagation: $\nabla_{\mathbf{W}}\mathcal{L}$ and $\nabla_{\boldsymbol{a}}\mathcal{L}$
		\STATE Update standard network weights of backbone:
		\STATE $\mathbf{W} \leftarrow \mathbf{W} - \eta_{\mathbf{W}} \left( \nabla_{\mathbf{W}}\mathcal{L} + \lambda \mathbf{W} \right)$
		\STATE Update fractional order vector:
		\STATE $\boldsymbol{a} \leftarrow \boldsymbol{a} - \eta_{\boldsymbol{a}} \nabla_{\boldsymbol{a}}\mathcal{L}$
		\ENDWHILE
		
		\RETURN $\mathbf{W}^*$, $\mathbf{F}_{\mathrm{I/II}}^{\boldsymbol{a}^*}$, $\hat{\mathbf{Y}}$
	\end{algorithmic}
\end{algorithm}

\indent \textbf{Integration into traditional spectral GNNs.} For foundational spectral architectures, convolutions are driven by predefined explicit filters or polynomial approximations. In MPFSR-SpectralCNN, the learnable parameters explicitly form the diagonal spectral filter $\mathbf{H} = \mathrm{diag}(\boldsymbol{\theta})$. The layer update subsequently applies a linear transformation and a non-linear activation, formulated as
\begin{align}
	\mathbf{X}^{(l+1)} = \sigma \left( \mathbf{Z}^{(l)} \mathbf{W}^{(l)} \right),
\end{align}
where $\mathbf{W}^{(l)}$ is the layer-specific weight matrix, and $\sigma(\cdot)$ denotes the non-linear activation function.

\indent In MPFSR-LanczosNet, which approximates the Laplacian spectrum for multi-scale diffusions, the filter takes the polynomial form $\mathbf{H} = \sum \theta_s \boldsymbol{\Lambda}^s$. While short-range diffusions continue to capture local topologies via the spatial Laplacian, the long-range diffusions execute within the task-optimized fractional domain to yield $\mathbf{Z}^{(l)}$. The multi-scale representations are then concatenated ($\Vert$) or aggregated as
\begin{align}
	\mathbf{X}^{(l+1)} = \sigma \left( \left[ \mathbf{Z}_{\text{short}}^{(l)} \Vert \mathbf{Z}^{(l)} \right] \mathbf{W}^{(l)} \right),
\end{align}
granting the model unprecedented flexibility to capture complex global topological structures.

\indent \textbf{Integration into transformer-based spectral GNNs.} Recent advancements have introduced transformer architectures to the spectral domain, utilizing self-attention mechanisms to learn global set-to-set frequency interactions. In sophisticated models like Specformer and GrokFormer, the diagonal spectral filter $\mathbf{H}$ is not a directly parameterized matrix. Instead, it is instantiated as a dynamic weight matrix generated on-the-fly by the multi-head attention mechanism operating on the frequency encodings, which we denote as $\mathbf{W}_{\text{attn}}$ for Specformer and $\mathbf{W}_{\text{grok}}$ for GrokFormer. 

\indent When equipped with our module to construct MPFSR-Specformer, the intermediate domain for attention-driven feature scaling shifts from a rigid spectral space to an adaptive fractional one. Specifically, the multi-head attention mechanism generates $H$ distinct frequency weight vectors, yielding $H$ parallel fractional intermediate representations $\{\mathbf{Z}_h^{(l)}\}_{h=1}^H$. Rather than employing a black-box sequence, the layer update explicitly aggregates these multi-head spectral representations with the raw spatial feature $\mathbf{X}^{(l)}$ via a learnable combination layer
\begin{align}
	\mathbf{X}^{(l+1)} = \sigma \left( \theta_0 \mathbf{X}^{(l)} + \sum_{h=1}^{H} \theta_h \mathbf{Z}_h^{(l)} \right),
\end{align}
where $\theta_0$ and $\theta_h$ are the optimizable combination coefficients. This explicit formulation empowers the model to dynamically balance the original spatial topology with multiple task-driven fractional views.

\indent For MPFSR-GrokFormer, which emphasizes robust generalization across homophilic and heterophilic edges, the generation of $\mathbf{Z}^{(l)}$ is driven by gating-coupled attention weights $\mathbf{W}_{\text{grok}}$. Crucially, GrokFormer employs a parallel skip-injection topology. The representation $\mathbf{Z}^{(l)}$ is scaled by a structural gating parameter $g$ and residually injected directly into the attention output stream. By bypassing the spatial self-attention computation, the layer update is concisely formulated as
\begin{align}
	\mathbf{X}^{(l+1)} = \mathrm{FFN} \left( \mathbf{X}^{(l)} + \mathrm{MHA}(\mathbf{X}^{(l)}) + g \cdot \mathbf{Z}^{(l)} \right),
\end{align}
where $\mathrm{MHA}(\cdot)$ and $\mathrm{FFN}(\cdot)$ denote the standard multi-head attention and feed-forward network blocks, respectively. This parallel injection empowers the architecture to independently fuse task-optimized fractional spectral topology without interfering with the spatial attention map.

\indent A comprehensive summary of the differences between the original architectures and our proposed MPFSR-enhanced models is provided in Table~\ref{tab:integration}. By replacing the static graph Fourier basis $\mathbf{U}$ with the task-driven fractional basis $\mathbf{F}_{\mathrm{I/II}}^{\boldsymbol{a}}$, the module introduces active spectral modulation across all paradigms.

\begin{table}[htbp]
	\centering
	\caption{Summary of the proposed MPFSR-enhanced spectral GNN architectures.}
	\label{tab:integration}
	\renewcommand{\arraystretch}{1.3}
	\resizebox{\linewidth}{!}{
		\begin{tabular}{c c c c}
			\toprule
			\textbf{Base model} & \textbf{Enhanced model} & \textbf{Filter matrix form ($\mathbf{H}$)} & \textbf{Transformation basis} \\
			\midrule
			SpectralCNN & MPFSR-SpectralCNN & Diagonal $\mathrm{diag}(\boldsymbol{\theta})$ & \multirow{4}{*}{Static $\mathbf{U} \rightarrow$ Adaptive $\mathbf{F}_{\mathrm{I/II}}^{\boldsymbol{a}}$} \\
			LanczosNet & MPFSR-LanczosNet & Polynomial $\sum \theta_s \boldsymbol{\Lambda}^s$ & \\
			Specformer & MPFSR-Specformer & Attention weights $\mathbf{W}_{\text{attn}}$ & \\
			GrokFormer & MPFSR-GrokFormer & Gated attention $\mathbf{W}_{\text{grok}}$ & \\
			\bottomrule
		\end{tabular}
	}
\end{table}

\section{Experiments} \label{sec5}
This section aims to comprehensively evaluate the effectiveness of the proposed MPFSR framework. Node classification experiments are conducted on real-world graph datasets and the performance gains of the proposed method are systematically analyzed by comparing the original models with their MPFSR-enhanced counterparts.
\subsection{Experimental setup}
\indent \textbf{Datasets description.} Several representative graph node classification datasets are adopted in our experiments. These datasets cover both homophilic and heterophilic graph scenarios, enabling a comprehensive evaluation of the model's adaptability to different graph structural characteristics. In homophilic graphs, adjacent nodes usually share similar class labels, which mainly tests the model's ability to capture low-frequency smooth information. In contrast, in heterophilic graphs, neighboring nodes often belong to different classes, making them more suitable for assessing the model's capability in modeling high-frequency heterogeneous structures and local detailed information.

The datasets used in this study include the homophilic datasets Cora, Citeseer\cite{sen2008collective}, Photo, and Computers\cite{shchur2018pitfalls}, as well as the heterophilic datasets Actor\cite{pei2020geomgcn}, Texas, Cornell, Wisconsin\cite{craven1998learning}, Twitch-PT\cite{rozemberczki2021multi}, and Wiki\cite{yang2020scaling}. Table~\ref{tab:dataset_statistics} summarizes the statistics of the node classification datasets. For each dataset, a publicly available and widely adopted data splitting protocol is employed, where all samples are divided into training, validation, and test sets with a ratio of 60\%/20\%/20\%, so as to ensure the fairness of experimental comparisons. Classification accuracy is used as the evaluation metric. 

\begin{table*}[htbp]
	\centering
	\caption{Statistics of node classification datasets.}
	\label{tab:dataset_statistics}
	\renewcommand{\arraystretch}{1.1}
	\setlength{\tabcolsep}{6pt}
	\resizebox{\textwidth}{!}{
		\begin{tabular}{lcccccccccc}
			\toprule
			Datasets & Cora & Citeseer & Photo & Computers & Twitch-PT & Wiki & Actor & Texas & Cornell & Wisconsin \\
			\midrule
			\#Nodes    & 2,708 & 3,327 & 7,650 & 13,752  & 1,912 & 2,405   & 7,600 & 183 & 183 & 253 \\
			\#Edges    & 5,429 & 4,732 & 238,163 & 245,861 & 31,299 & 17,981 & 33,544 & 295 & 298 & 506 \\
			\#Features & 1,433 & 3,703 & 745 & 767 & 3,169 & 4,973 & 931 & 1,703 & 1,703 & 1,703 \\
			\#Classes  & 7 & 6 & 8 & 10 & 2 & 17 & 5 & 5 & 5 & 5 \\
			$\mathcal{H}$ & 0.81 & 0.74 & 0.83 & 0.78 & 0.57 & 0.46 & 0.22 & 0.06 & 0.12 & 0.14 \\
			\bottomrule
		\end{tabular}
	}
\end{table*}

\indent \textbf{Setup.} To evaluate the effectiveness of the proposed model, several representative spectral graph neural networks are selected as baselines. Specifically, the original backbone models, including SpectralCNN, LanczosNet, Specformer, and GrokFormer, are first adopted as direct comparison methods. To further verify the effectiveness of the proposed MPFSR module, the performance differences before and after introducing this module are systematically compared by constructing the corresponding enhanced variants, namely MPFSR-SpectralCNN, MPFSR-LanczosNet, MPFSR-Specformer, and MPFSR-GrokFormer. All models are trained and tested under the same experimental settings to ensure the fairness and comparability of the results. Moreover, all models are independently run 10 times, and the mean test accuracy together with the corresponding standard deviation is reported to reduce the influence of random initialization and data partitioning.

In addition, to achieve stable optimization of the network parameters and the fractional-order parameter vector, a differential learning rate strategy is adopted. Specifically, the learning rate for the network parameters is set to 0.025, while the fractional order parameter vector is initialized as an all one vector with a learning rate of 0.0001. The hyperparameters are tuned on the validation set, and the search space is defined as follows:
\begin{itemize}
	\item Number of layers: $\{1,2\}$
	\item Number of heads: $\{2,3,4\}$
	\item Hidden dimension: $\{16,32,64,128,256\}$
	\item Number of $K$: $\{1,2,3,4,5,6\}$
	\item Weight decay: $\{10^{-3},5\times10^{-3},10^{-4},5\times10^{-4}\}$
	\item Dropout rate: $\{0.0,0.1,0.2,0.3,0.4,0.5,0.6,0.7,0.8\}$    
\end{itemize}
All experiments are conducted under a unified software and hardware environment. The platform is equipped with an NVIDIA GeForce RTX 5060 GPU. All models are implemented in PyTorch and trained with its automatic differentiation and GPU-accelerated computation mechanisms. A consistent runtime environment is used for all experiments to ensure fairness, reproducibility, and comparability.

\subsection{Main results}
To verify the overall effectiveness of the proposed model, Table~\ref{tab:all_model_node_classification_results} reports the node classification results of the best enhanced model in comparison with several representative graph neural network methods on multiple real-world datasets. To ensure a comprehensive evaluation, the compared methods cover three major categories, namely spatial graph neural networks, spectral graph neural networks, and graph Transformers. Specifically, the spatial graph neural networks include GCN\cite{xu2019graph}, GAT\cite{velivckovic2018graph}, H2GCN\cite{zhu2020beyond}, and HopGNN\cite{chen2023node}; the spectral graph neural networks include ChebyNet\cite{he2022convolutional}, GPRGNN\cite{chien2021adaptive}, BernNet\cite{he2021bernnet}, JacobiConv\cite{wang22am}, and HiGCN\cite{huang2024higher}; and the graph Transformer-based methods include Transformer\cite{dwivedi2020generalization}, GraphGPS\cite{rampavsek2022recipe}, NodeFormer\cite{wu2022nodeformer}, SGFormer\cite{wu2023sgformer}, NAGphormer\cite{chen2023nagphormer}, Specformer, PolyFormer\cite{ma2024polyformer}, and GrokFormer.

The results show that the best enhanced model achieves competitive and state-of-the-art classification performance across multiple datasets, demonstrating strong overall competitiveness  in Table~\ref{tab:all_model_node_classification_results}. The proposed method performs well on both homophilic and heterophilic datasets, indicating strong adaptability to different graph structures. These results show that the proposed enhancement mechanism improves the model’s ability to capture diverse graph structural characteristics and leads to better node classification performance.

To further verify the effectiveness of the proposed enhancement strategy, Table~\ref{tab:node_classification_results} presents the node classification results of the enhanced models and their corresponding original backbone models on multiple datasets. Specifically, the proposed enhancement module is incorporated into different spectral backbone networks, and the classification performance before and after the enhancement is compared under the same experimental settings.
\begin{table*}[htbp]
	\centering
	\caption{Node classification results on real-world node classification tasks: mean accuracy (\%) $\pm$ std. The best results are in bold, while the second-best ones are underlined.}
	\label{tab:all_model_node_classification_results}
	\renewcommand{\arraystretch}{0.95}
	\setlength{\tabcolsep}{5pt}
	\resizebox{0.85\textwidth}{!}{
		\begin{tabular}{lccccc}
			\toprule
			& \multicolumn{3}{c}{Homophilic Datasets} & \multicolumn{2}{c}{Heterophilic Datasets} \\
			\cmidrule(lr){2-4} \cmidrule(lr){5-6}
			& Cora & Citeseer & Photo & Actor & Texas \\
			\midrule
			
			\multicolumn{6}{c}{Spatial-based GNNs} \\
			\midrule
			GCN    & 87.14 $\pm$ 1.01 & 79.86 $\pm$ 0.67 & 88.26 $\pm$ 0.73 & 33.23 $\pm$ 1.16 & 77.38 $\pm$ 3.28 \\
			GAT    & 88.03 $\pm$ 0.79 & 80.52 $\pm$ 0.71 & 90.94 $\pm$ 0.68 & 33.93 $\pm$ 2.47 & 80.82 $\pm$ 2.13 \\
			H2GCN  & 87.96 $\pm$ 0.57 & 80.90 $\pm$ 1.21 & 95.45 $\pm$ 0.67 & 36.31 $\pm$ 2.58 & 91.85 $\pm$ 3.93 \\
			HopGNN & 88.68 $\pm$ 1.06 & 80.38 $\pm$ 0.68 & 94.49 $\pm$ 0.33 & 39.33 $\pm$ 2.79 & 89.15 $\pm$ 4.04 \\
			\midrule
			
			\multicolumn{6}{c}{Spectral-based GNNs} \\
			\midrule
			ChebyNet   & 86.67 $\pm$ 0.82 & 79.11 $\pm$ 0.75 & 93.77 $\pm$ 0.32 & 37.61 $\pm$ 0.89 & 86.22 $\pm$ 2.45 \\
			GPRGNN     & 88.57 $\pm$ 0.69 & 80.12 $\pm$ 0.83 & 93.85 $\pm$ 0.28 & 39.92 $\pm$ 0.67 & 92.95 $\pm$ 1.31 \\
			BernNet    & 88.52 $\pm$ 0.95 & 80.09 $\pm$ 0.79 & 93.63 $\pm$ 0.55 & 41.79 $\pm$ 1.01 & 93.12 $\pm$ 0.65 \\
			JacobiConv & 88.98 $\pm$ 0.46 & 80.78 $\pm$ 0.79 & 95.43 $\pm$ 0.23 & 41.17 $\pm$ 0.64 & 93.44 $\pm$ 2.13 \\
			HiGCN      & 89.23 $\pm$ 0.23 & 81.12 $\pm$ 0.28 & 95.33 $\pm$ 0.37 & 41.81 $\pm$ 0.52 & 92.15 $\pm$ 0.73 \\
			\midrule
			
			\multicolumn{6}{c}{Graph Transformers} \\
			\midrule
			Transformer & 71.83 $\pm$ 1.68 & 70.55 $\pm$ 1.20 & 89.58 $\pm$ 1.05 & 39.95 $\pm$ 0.64 & 88.75 $\pm$ 6.30 \\
			GraphGPS    & 83.42 $\pm$ 1.22 & 75.87 $\pm$ 0.71 & 94.35 $\pm$ 0.25 & 37.68 $\pm$ 0.94 & 83.71 $\pm$ 5.85 \\
			NodeFormer  & 87.32 $\pm$ 0.92 & 79.56 $\pm$ 1.10 & 95.27 $\pm$ 0.22 & 34.62 $\pm$ 1.82 & 84.63 $\pm$ 3.47 \\
			SGFormer    & 87.87 $\pm$ 2.67 & 79.62 $\pm$ 1.63 & 94.34 $\pm$ 0.23 & 41.69 $\pm$ 0.63 & 92.46 $\pm$ 1.48 \\
			NAGphormer  & 88.15 $\pm$ 1.35 & 80.12 $\pm$ 1.24 & 95.49 $\pm$ 0.11 & 40.08 $\pm$ 1.50 & 91.80 $\pm$ 1.85 \\
			PolyFormer  & 87.67 $\pm$ 1.28 & 81.80 $\pm$ 0.76 & 94.08 $\pm$ 1.37 & 41.51 $\pm$ 0.71 & 89.02 $\pm$ 5.44 \\
			\midrule
			
			\multicolumn{6}{c}{Advanced} \\
			\midrule
			Specformer & 88.57 $\pm$ 1.01 & 81.49 $\pm$ 0.94 & 95.48 $\pm$ 0.32 & 41.93 $\pm$ 1.04 & 88.23 $\pm$ 3.84 \\
			GrokFormer & 89.57 $\pm$ 1.43 & 81.92 $\pm$ 1.25 & 95.52 $\pm$ 0.52 & 42.98 $\pm$ 1.48 & \underline{94.59 $\pm$ 2.08} \\
			\midrule
			
			MPFSR-I-GrokFormer & \textbf{89.82 $\pm$ 1.32} & \textbf{82.13 $\pm$ 0.82} & \textbf{95.75 $\pm$ 0.39} & \textbf{44.14 $\pm$ 1.12} & \textbf{96.72 $\pm$ 2.19} \\
			MPFSR-II-GrokFormer & \underline{89.71 $\pm$ 1.56} & \underline{82.09 $\pm$ 1.79} & \underline{95.71 $\pm$ 0.44} & \underline{43.97 $\pm$ 1.37} & 93.44 $\pm$ 1.94 \\
			\bottomrule
		\end{tabular}
	}
\end{table*}

 The enhanced models outperform their original counterparts on most datasets, indicating that the proposed method improves the representation capability of different backbone networks. The results further suggest that the proposed method is not restricted to a specific model architecture and exhibits good generality and transferability across different spectral graph neural networks. Performance gains vary across backbones and datasets, but the overall trend remains clear: the enhanced models generally achieve higher classification accuracy. These results indicate that the proposed multiple-parameter fractional spectral modulation mechanism enables more flexible frequency-domain modeling based on the original static spectral representation and strengthens the joint representation of graph structural characteristics and node attribute information.

\newcolumntype{C}[1]{>{\centering\arraybackslash}m{#1}}

\newlength{\rowlabw}
\setlength{\rowlabw}{1.15cm}

\newlength{\imgw}
\setlength{\imgw}{0.212\textwidth}

\newlength{\rowraise}
\setlength{\rowraise}{0.0\baselineskip}

\newcommand{\rowlabel}[1]{%
	\raisebox{\rowraise}[0pt][0pt]{%
		\makebox[\rowlabw][c]{\rotatebox[origin=c]{90}{\textbf{#1}}}%
	}%
}

\begin{figure*}[!t]
	\centering
	\setlength{\tabcolsep}{1.6pt}
	\renewcommand{\arraystretch}{1.0}
	\captionsetup{font=small,skip=4pt}
	
	\begin{tabular}{@{}C{\rowlabw}@{\hspace{0.03cm}}C{\imgw}@{\hspace{0.03cm}}C{\imgw}@{\hspace{0.03cm}}C{\imgw}@{\hspace{0.03cm}}C{\imgw}@{}}
		
		& \textbf{True Labels} & \textbf{Baseline Model} & \textbf{MPGFRFT-I} & \textbf{MPGFRFT-II} \\[0.10cm]
		
		\rowlabel{SpectralNet}
		& \includegraphics[width=\linewidth]{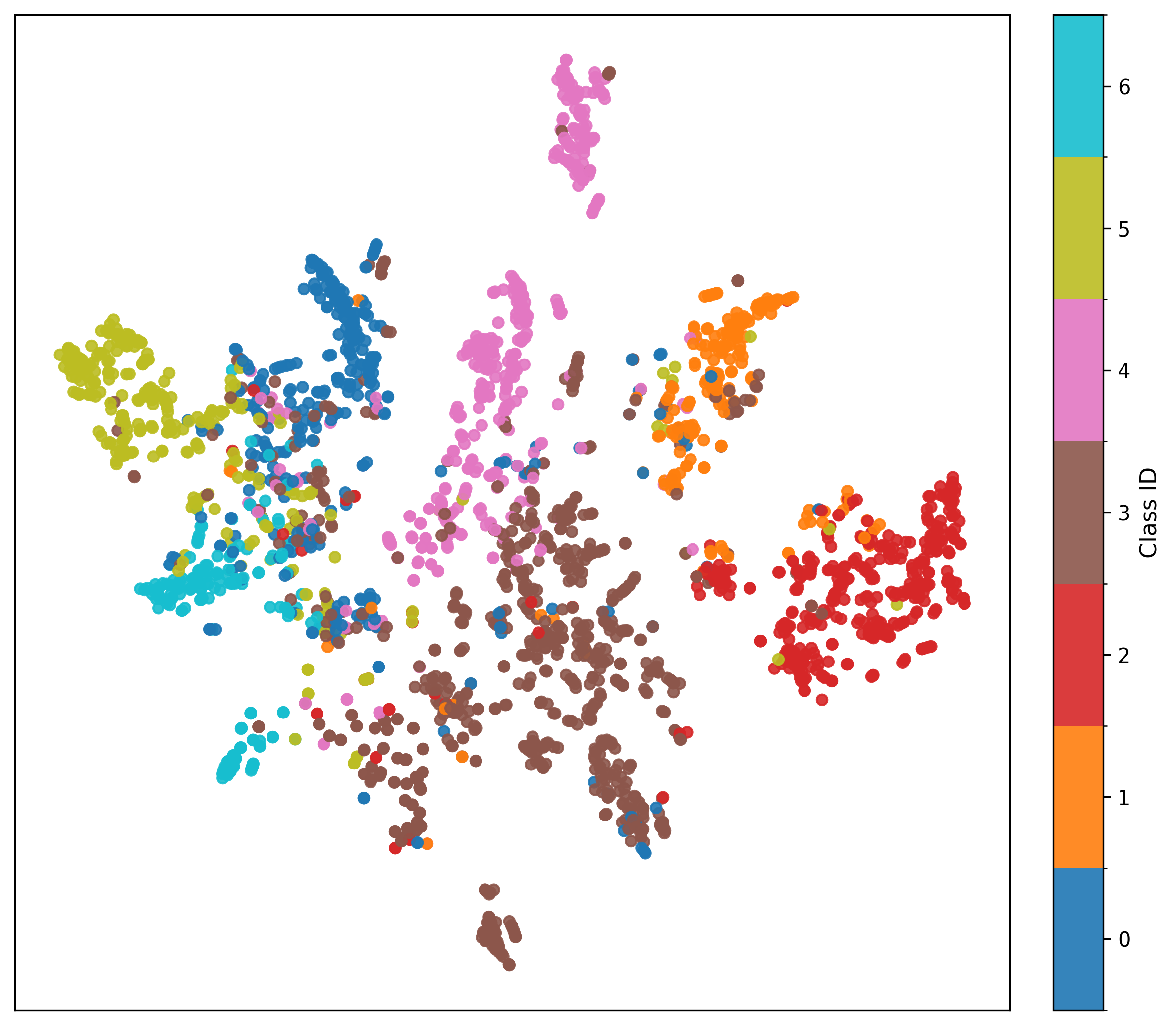}
		& \includegraphics[width=\linewidth]{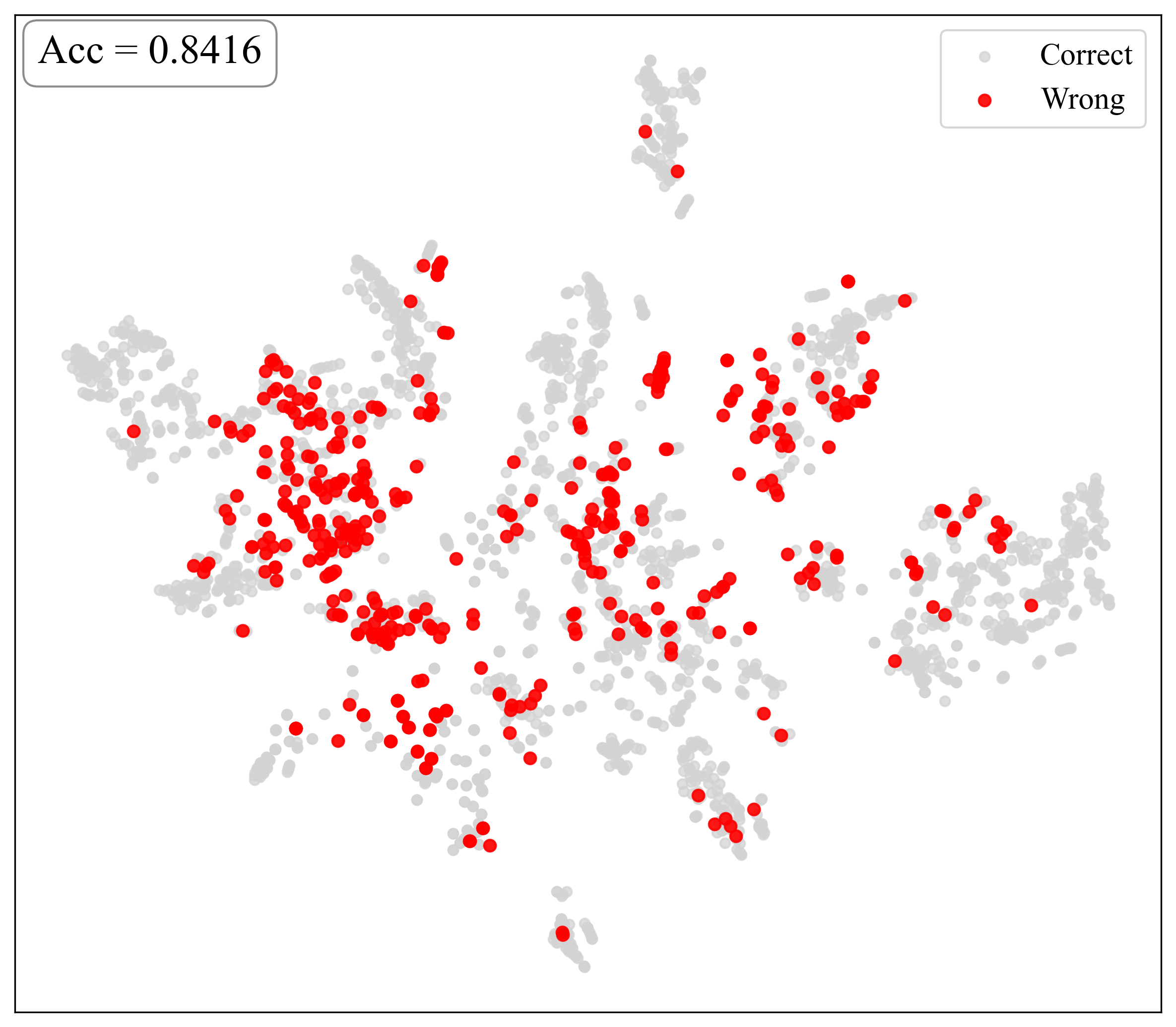}
		& \includegraphics[width=\linewidth]{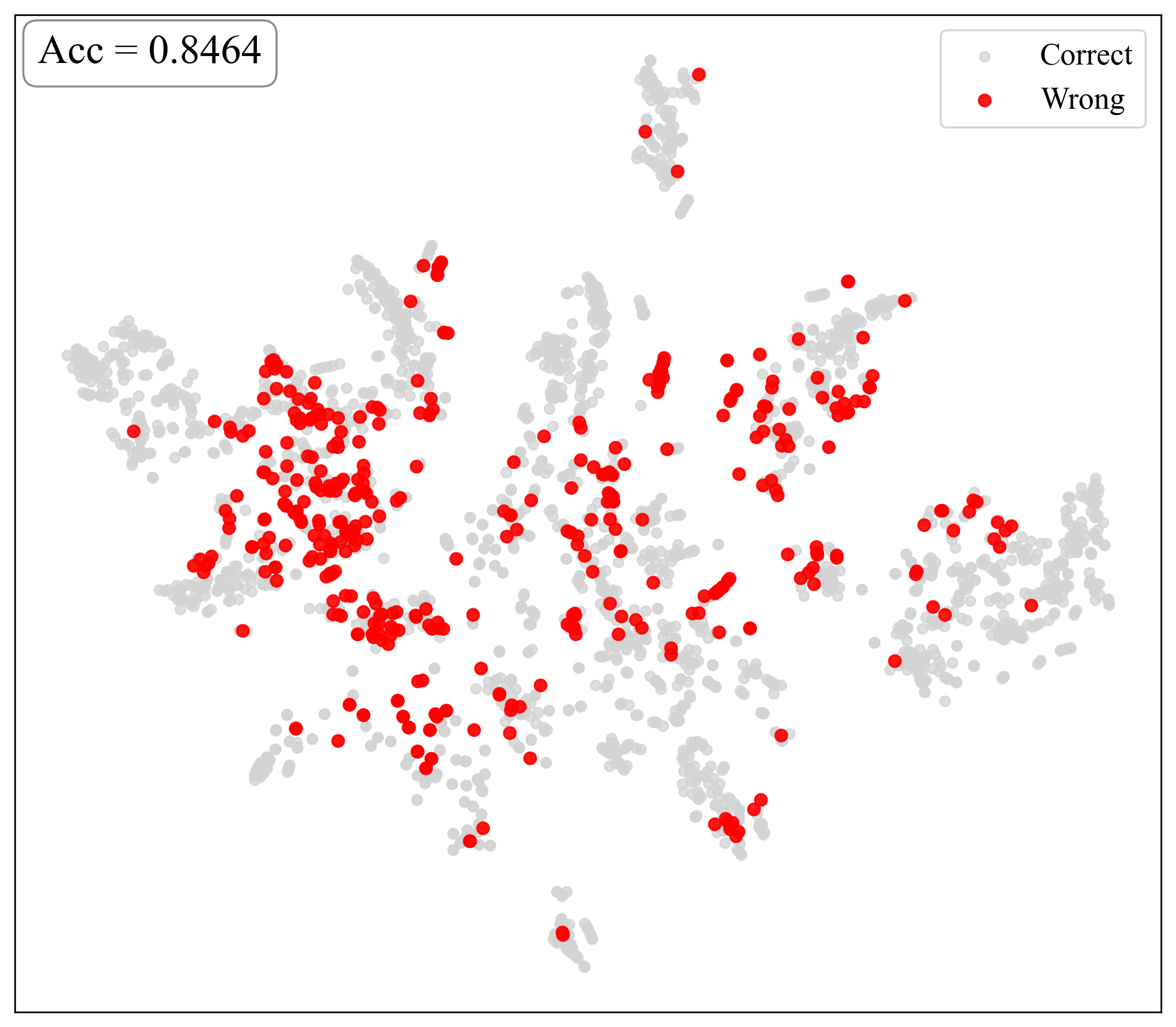}
		& \includegraphics[width=\linewidth]{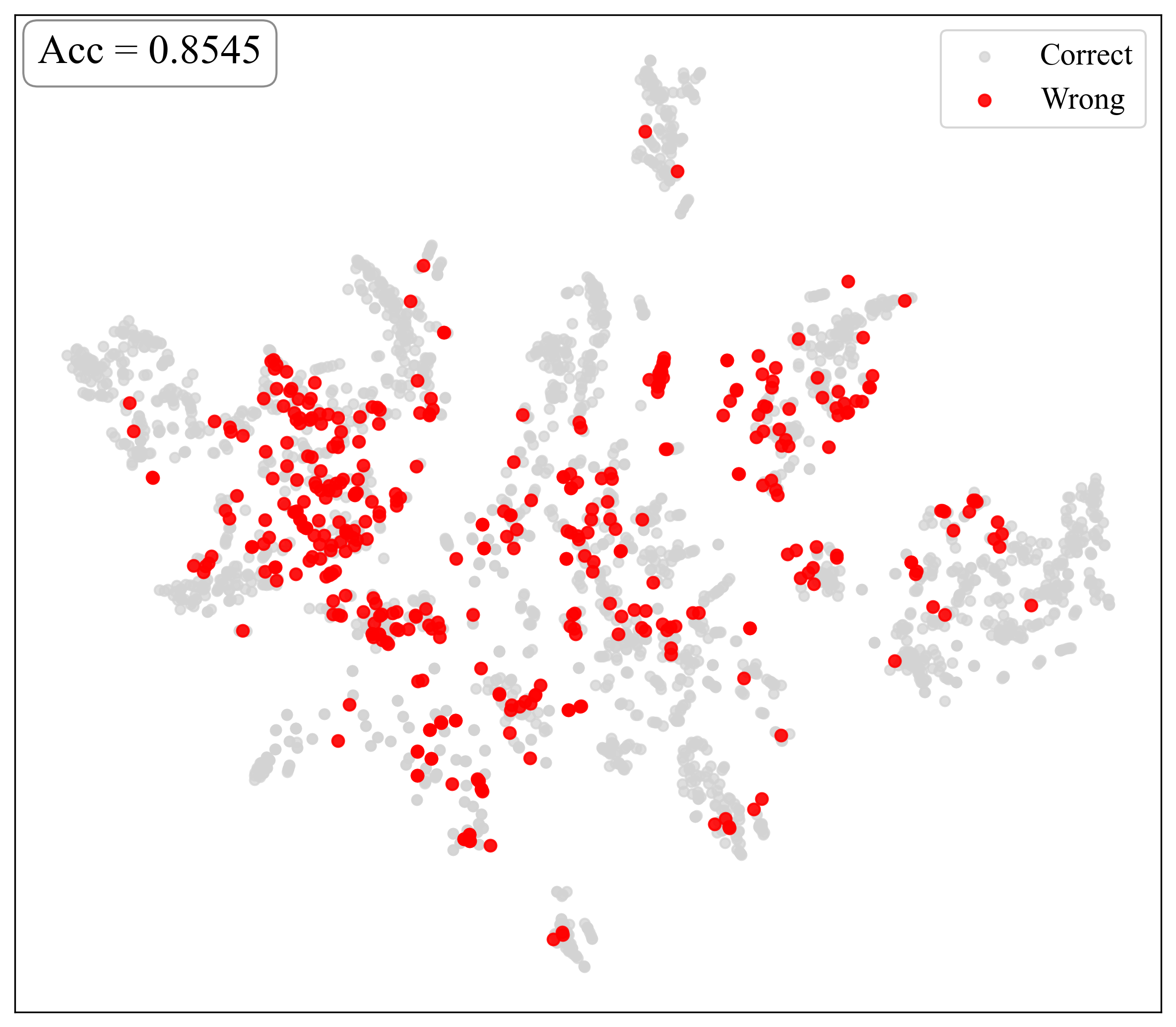} \\[0.08cm]
		
		\rowlabel{LanczosNet}
		& \includegraphics[width=\linewidth]{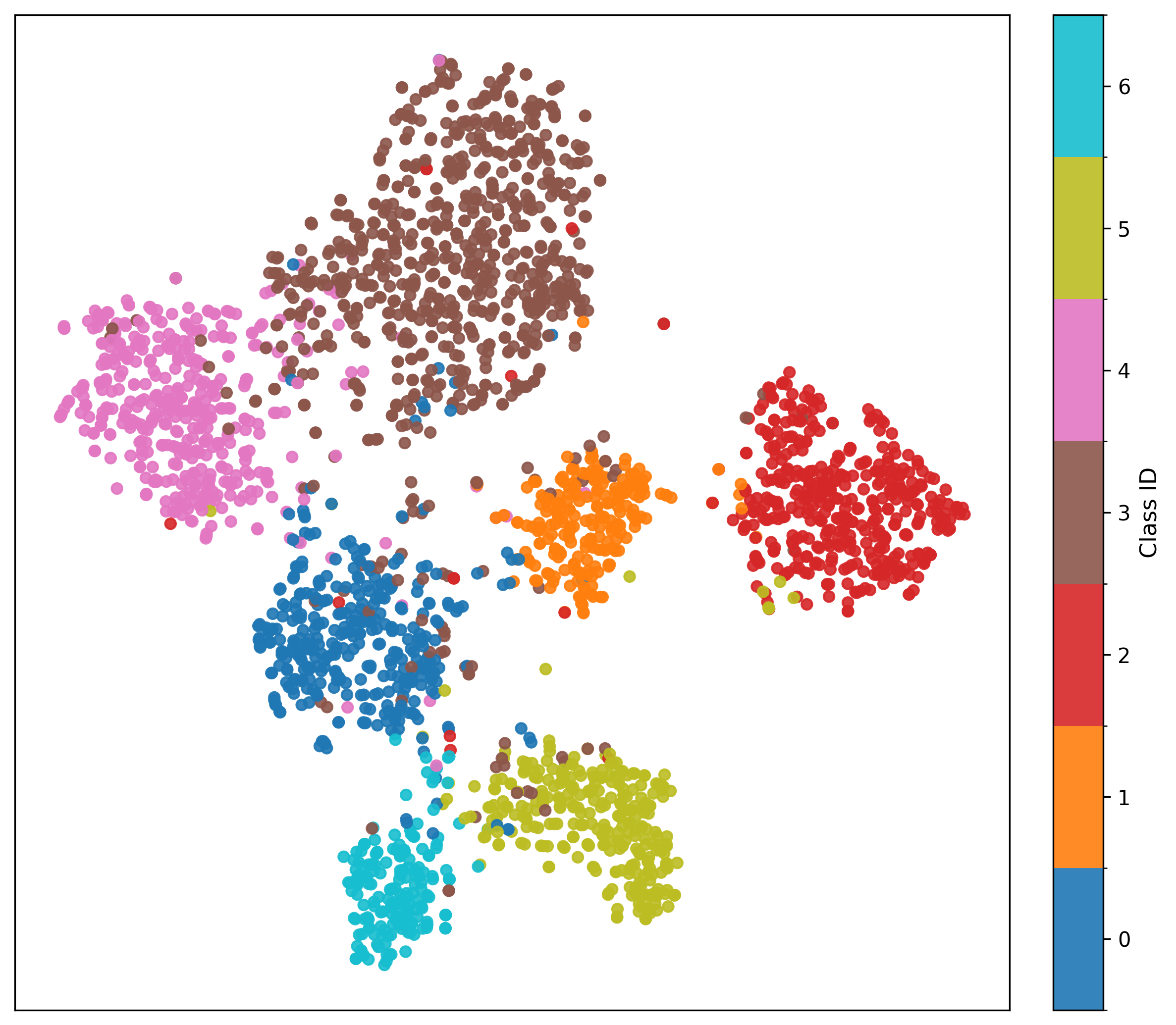}
		& \includegraphics[width=\linewidth]{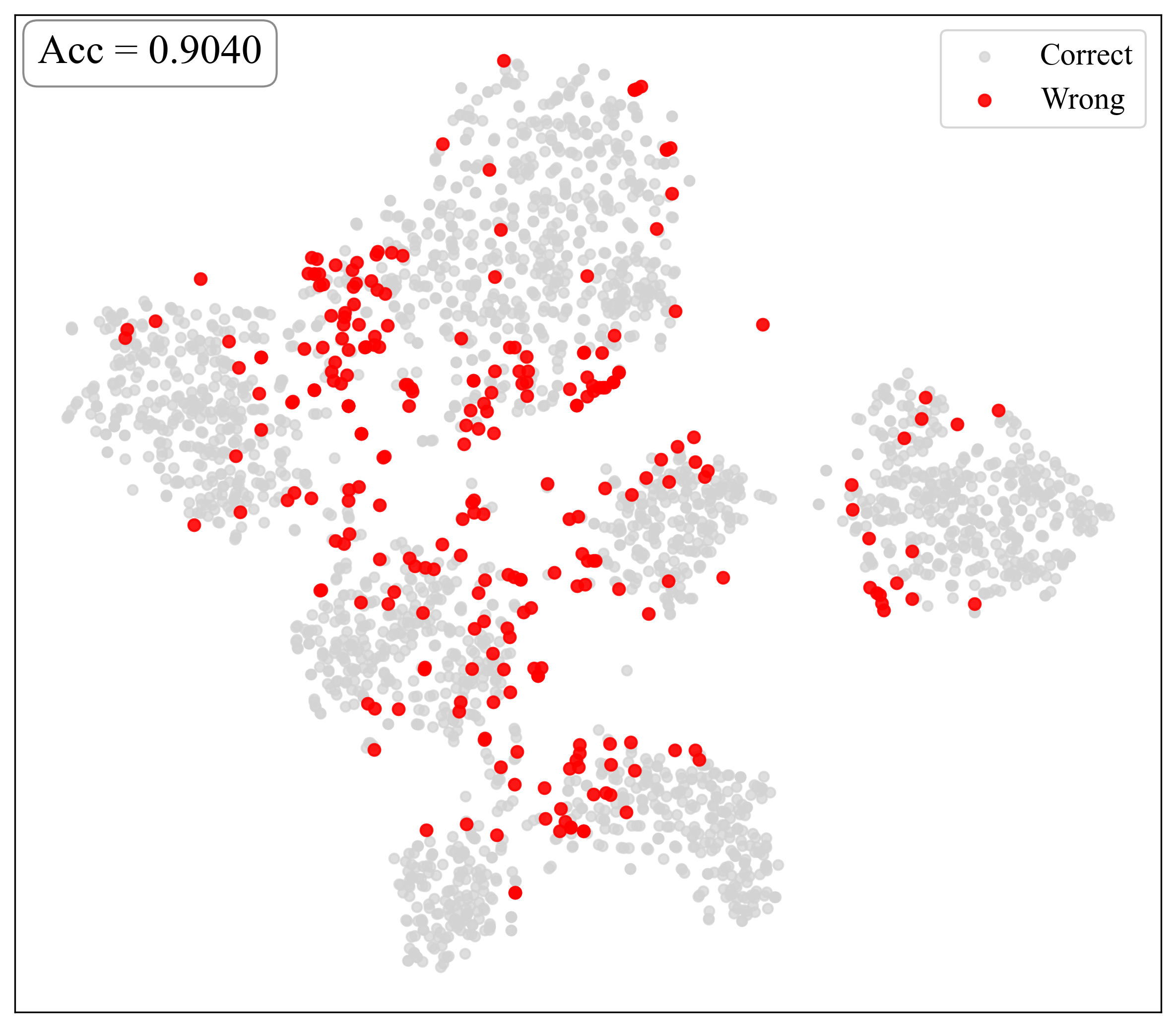}
		& \includegraphics[width=\linewidth]{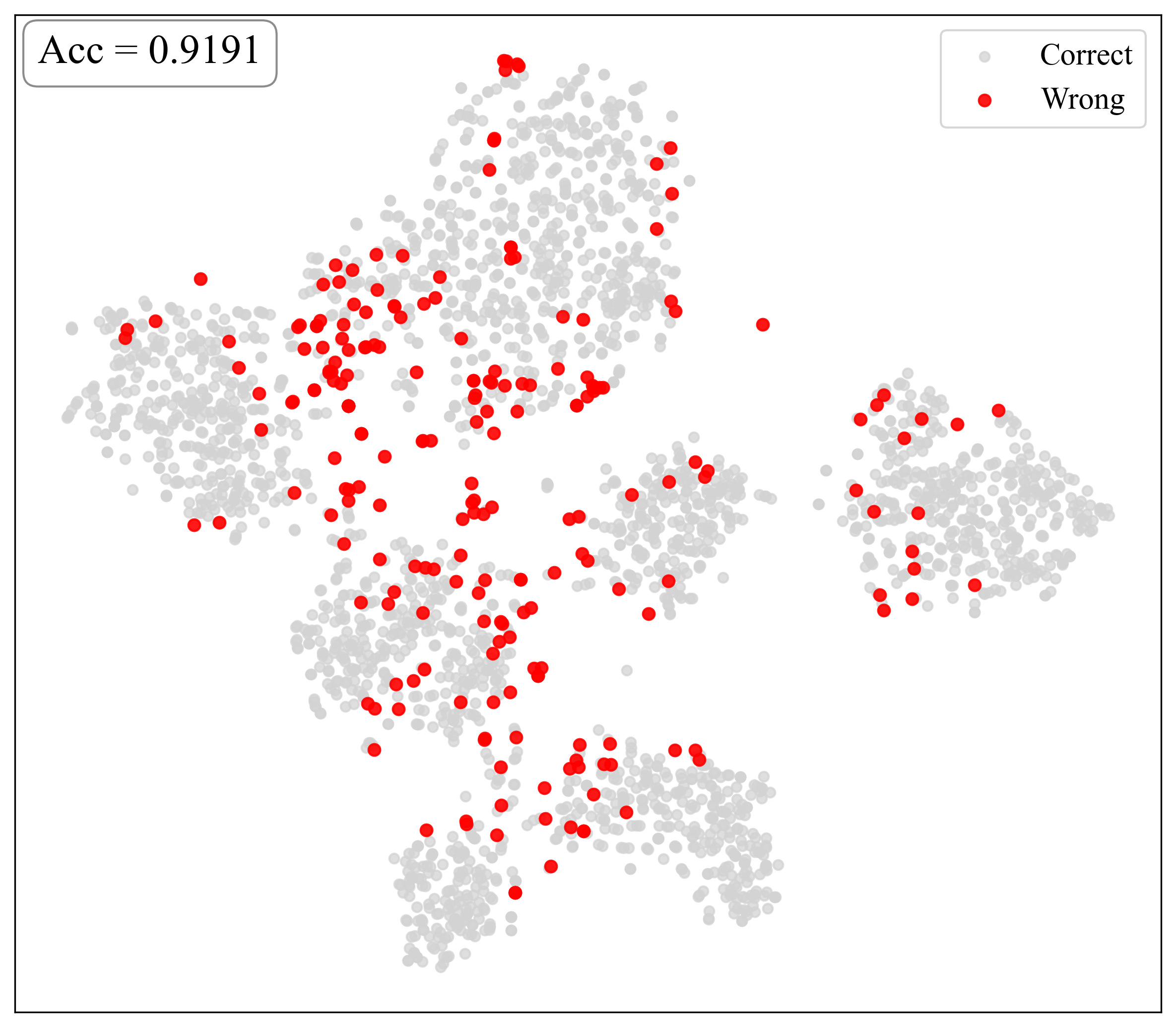}
		& \includegraphics[width=\linewidth]{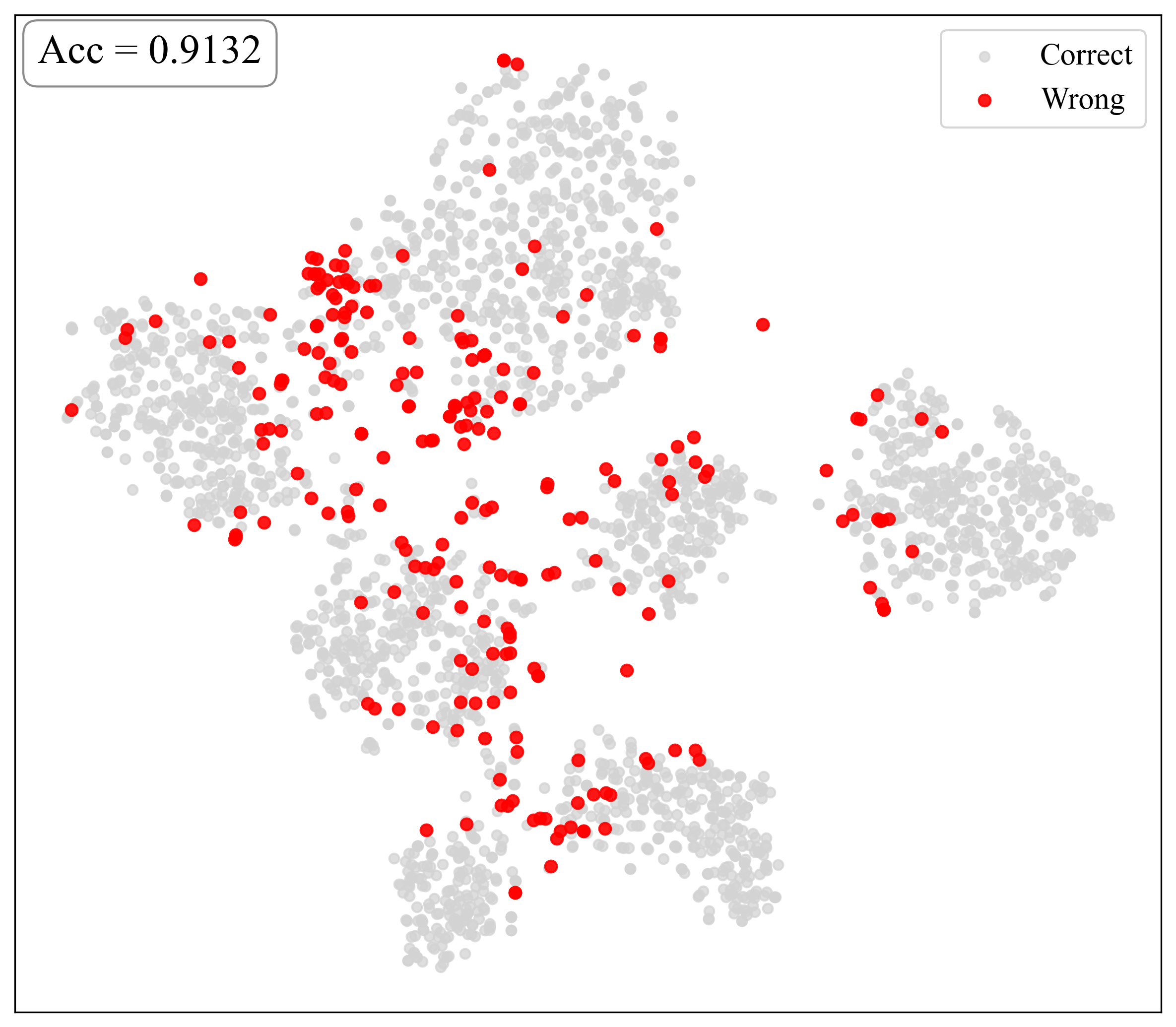} \\[0.08cm]
		
		\rowlabel{Specformer}
		& \includegraphics[width=\linewidth]{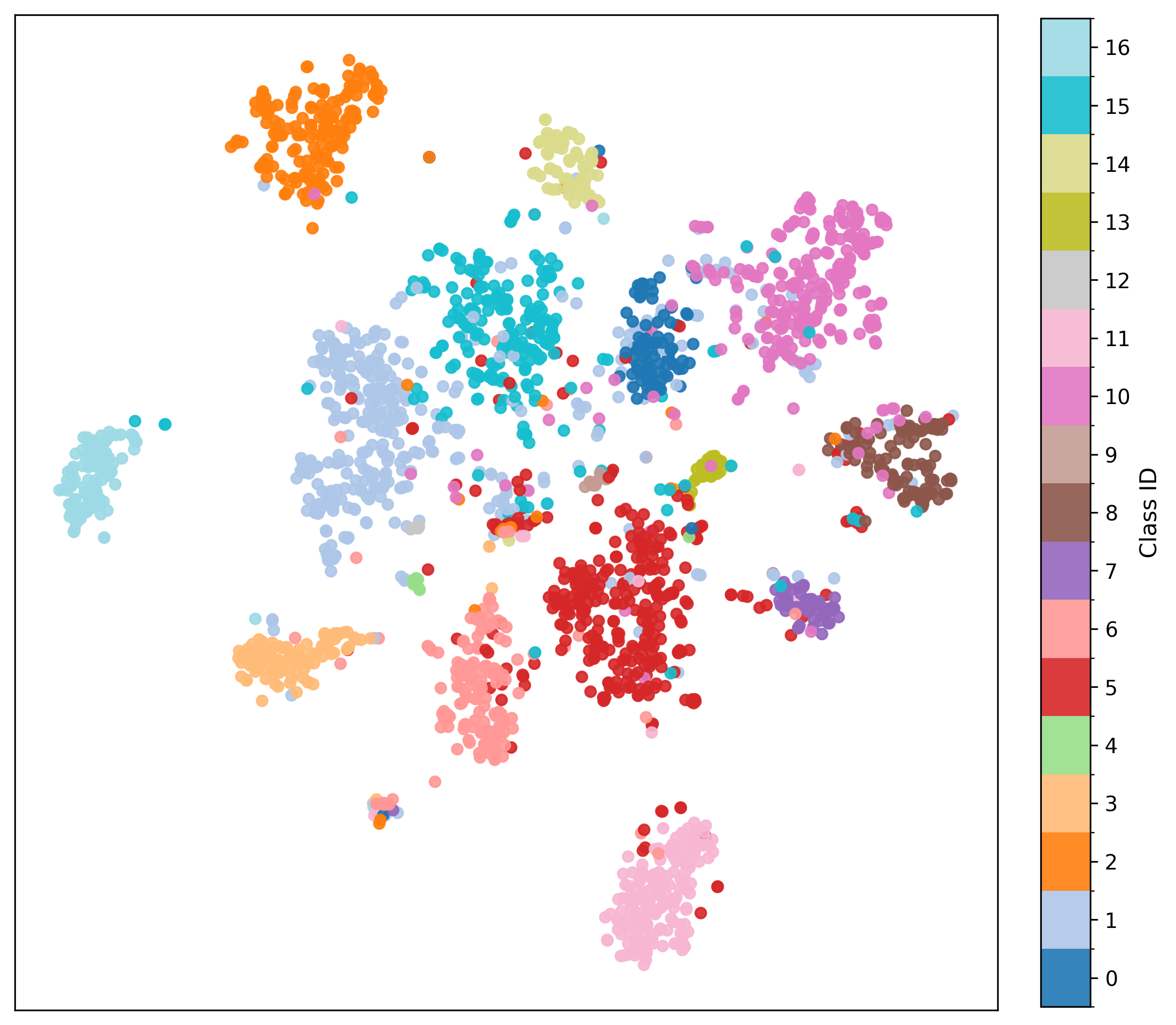}
		& \includegraphics[width=\linewidth]{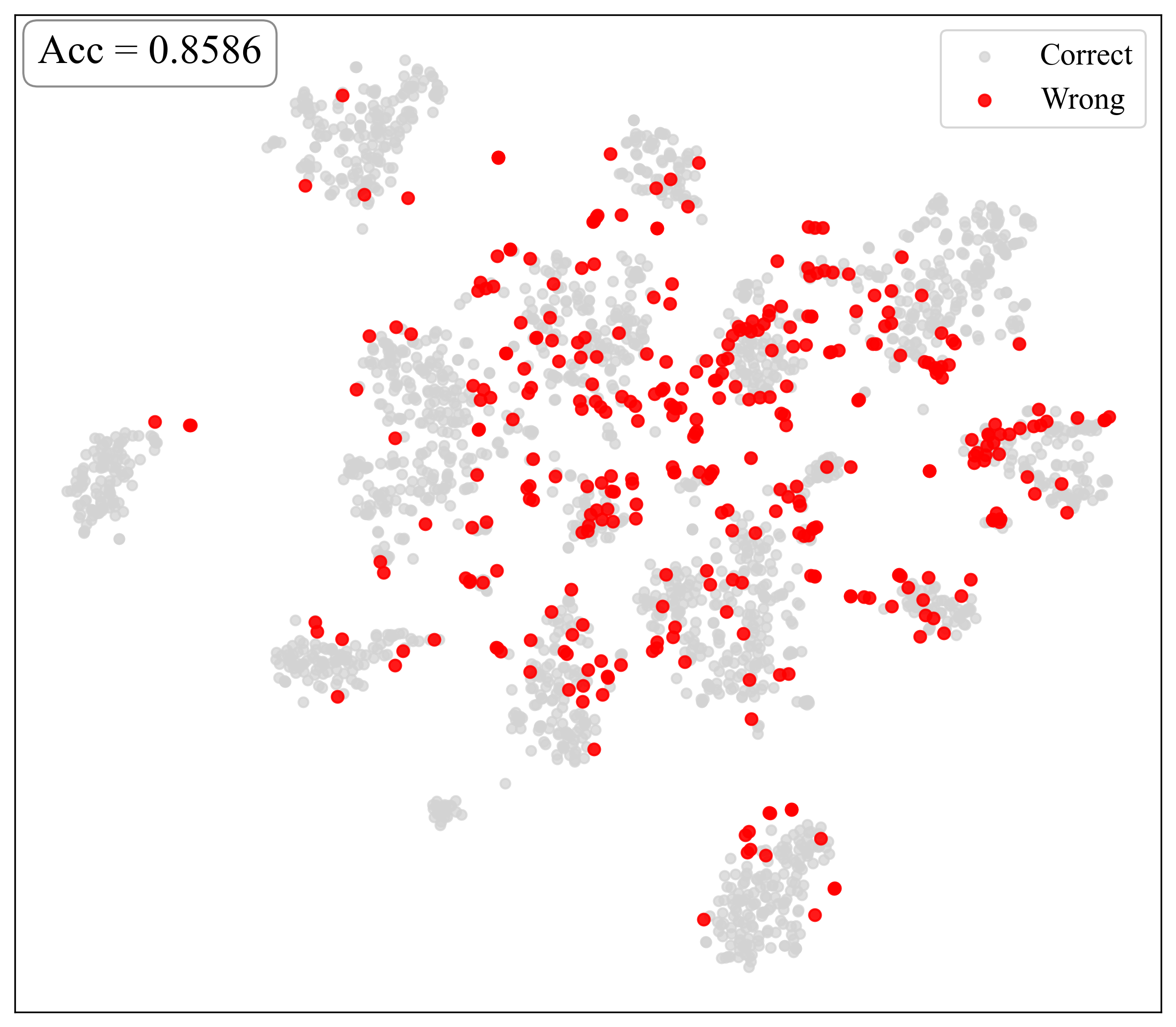}
		& \includegraphics[width=\linewidth]{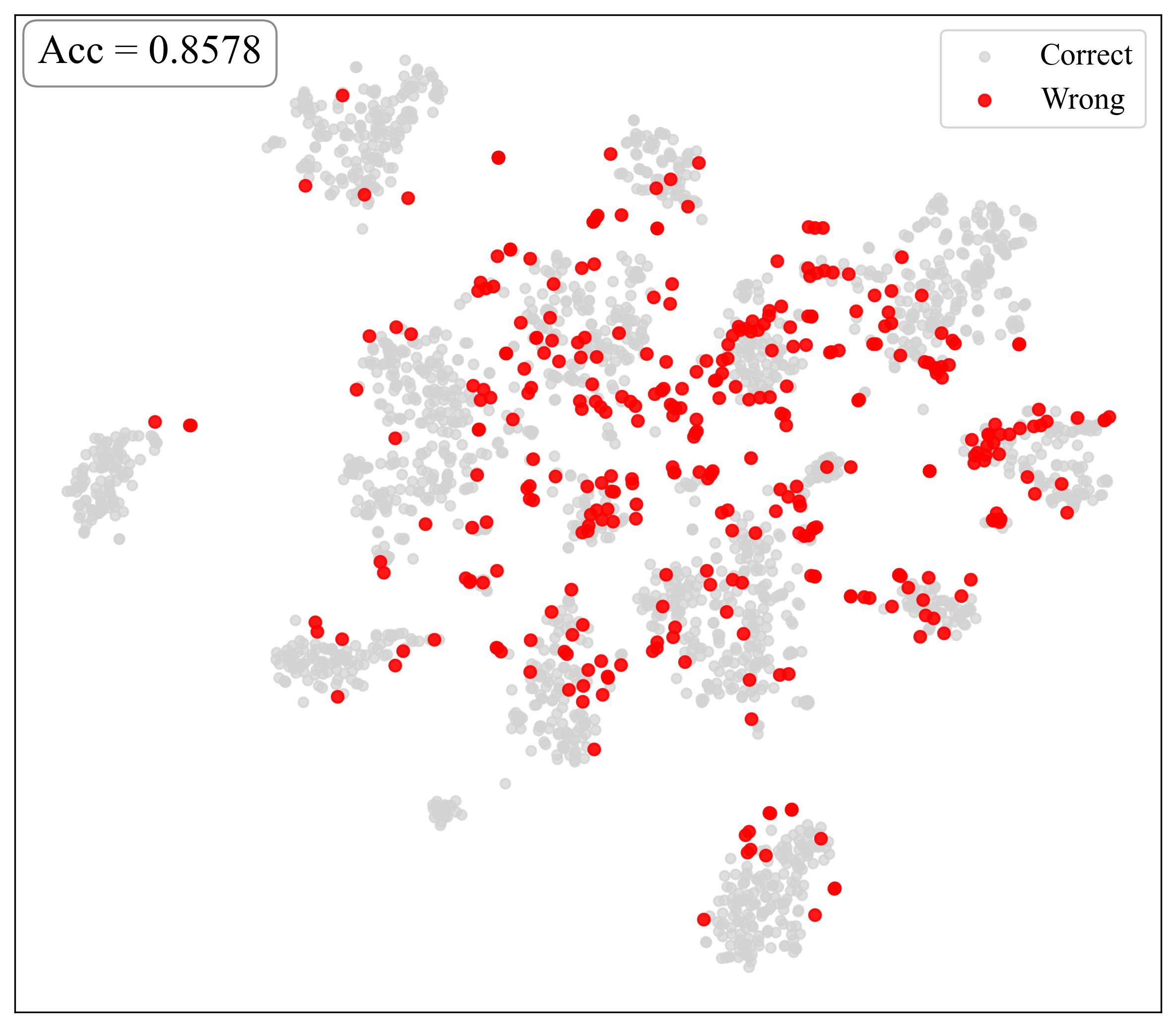}
		& \includegraphics[width=\linewidth]{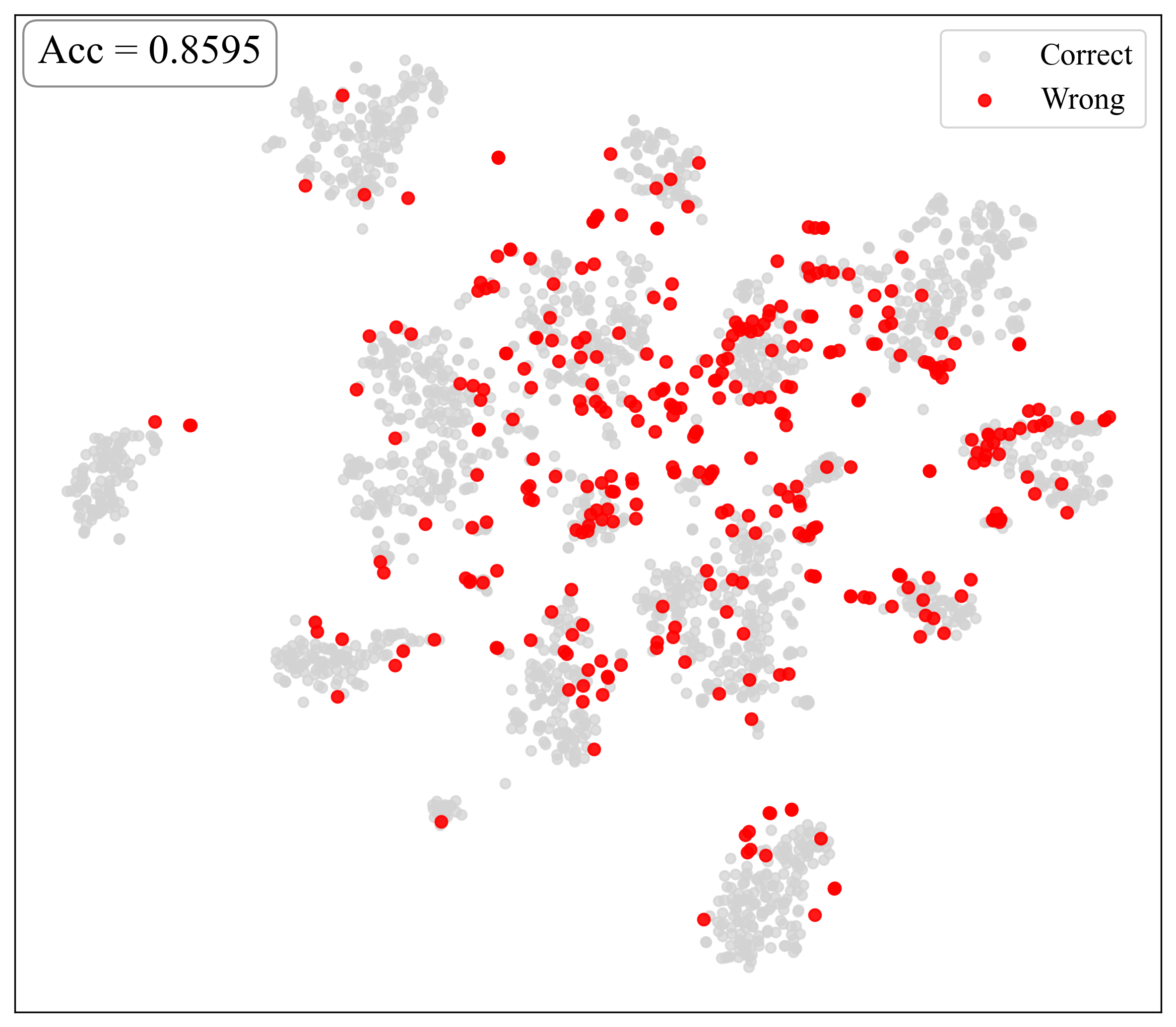} \\[0.08cm]
		
		\rowlabel{GrokFormer}
		& \includegraphics[width=\linewidth]{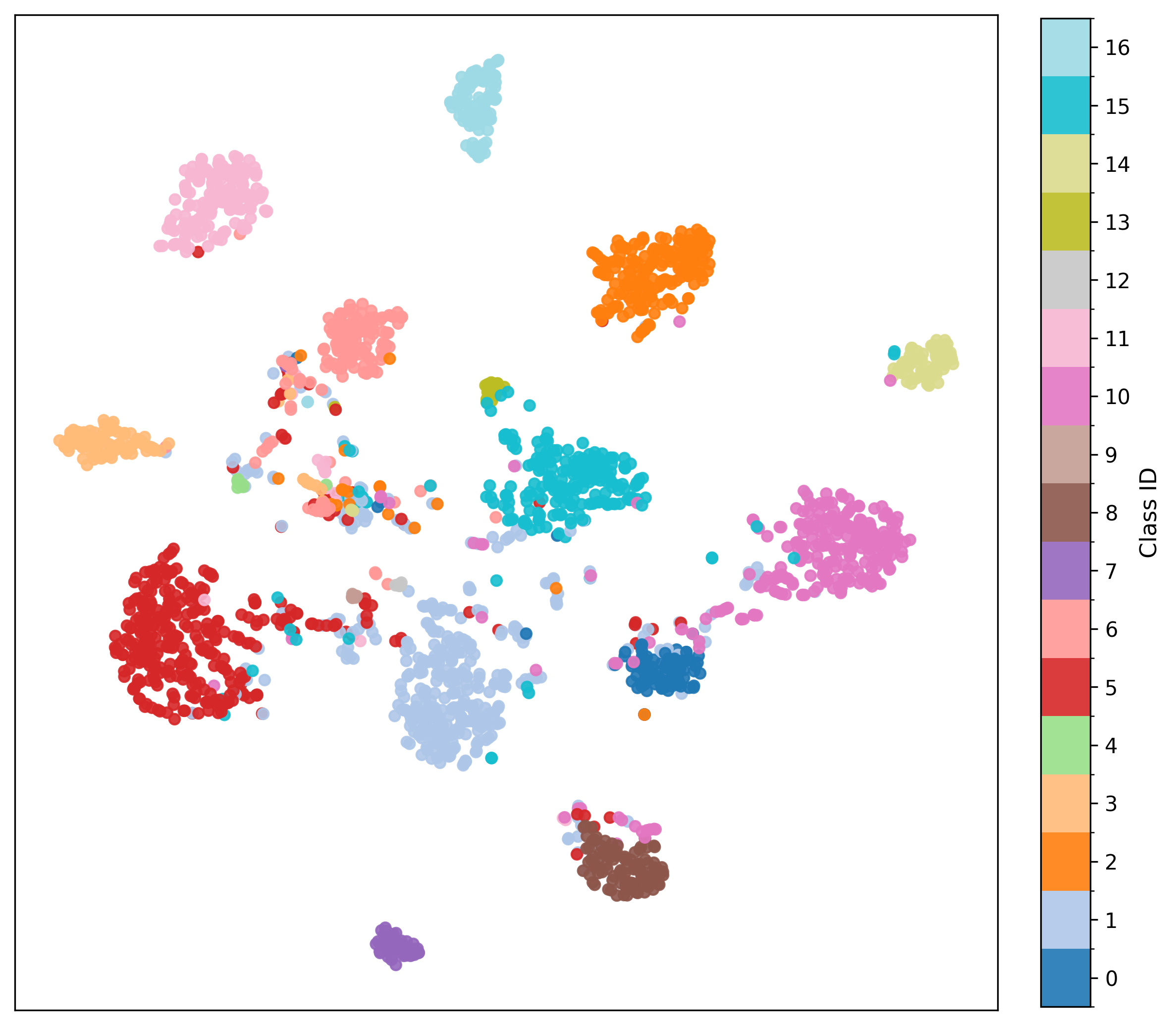}
		& \includegraphics[width=\linewidth]{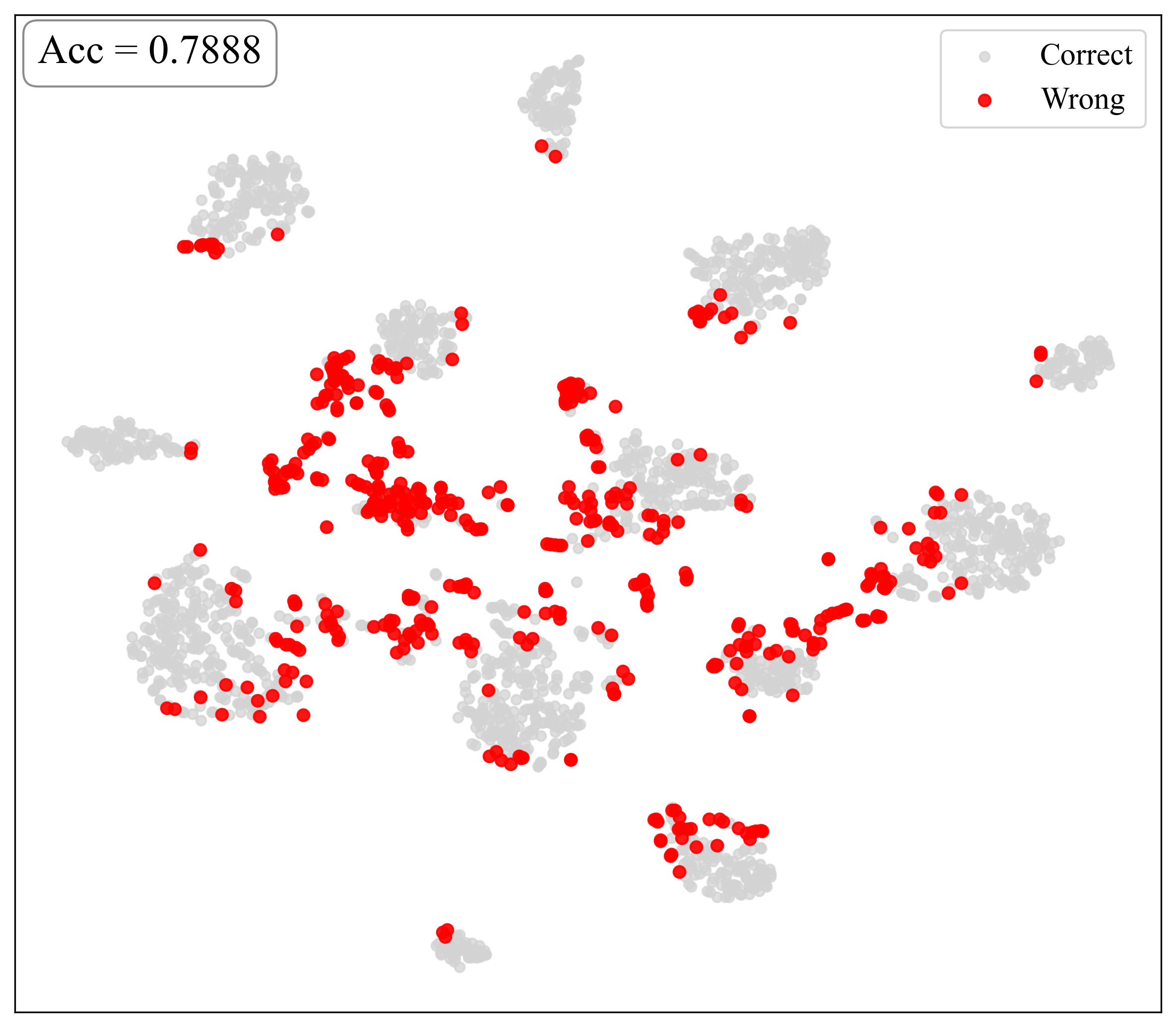}
		& \includegraphics[width=\linewidth]{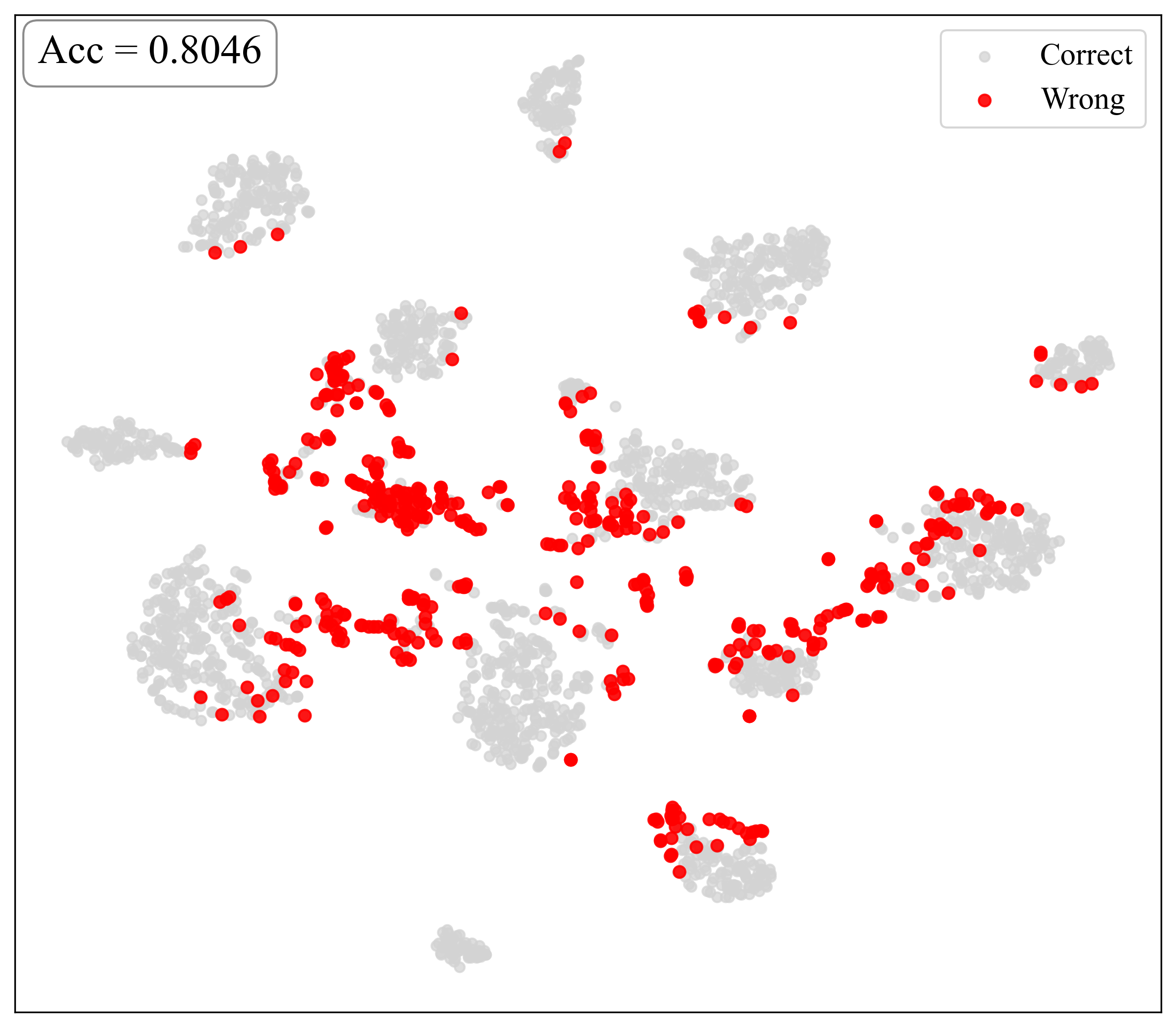}
		& \includegraphics[width=\linewidth]{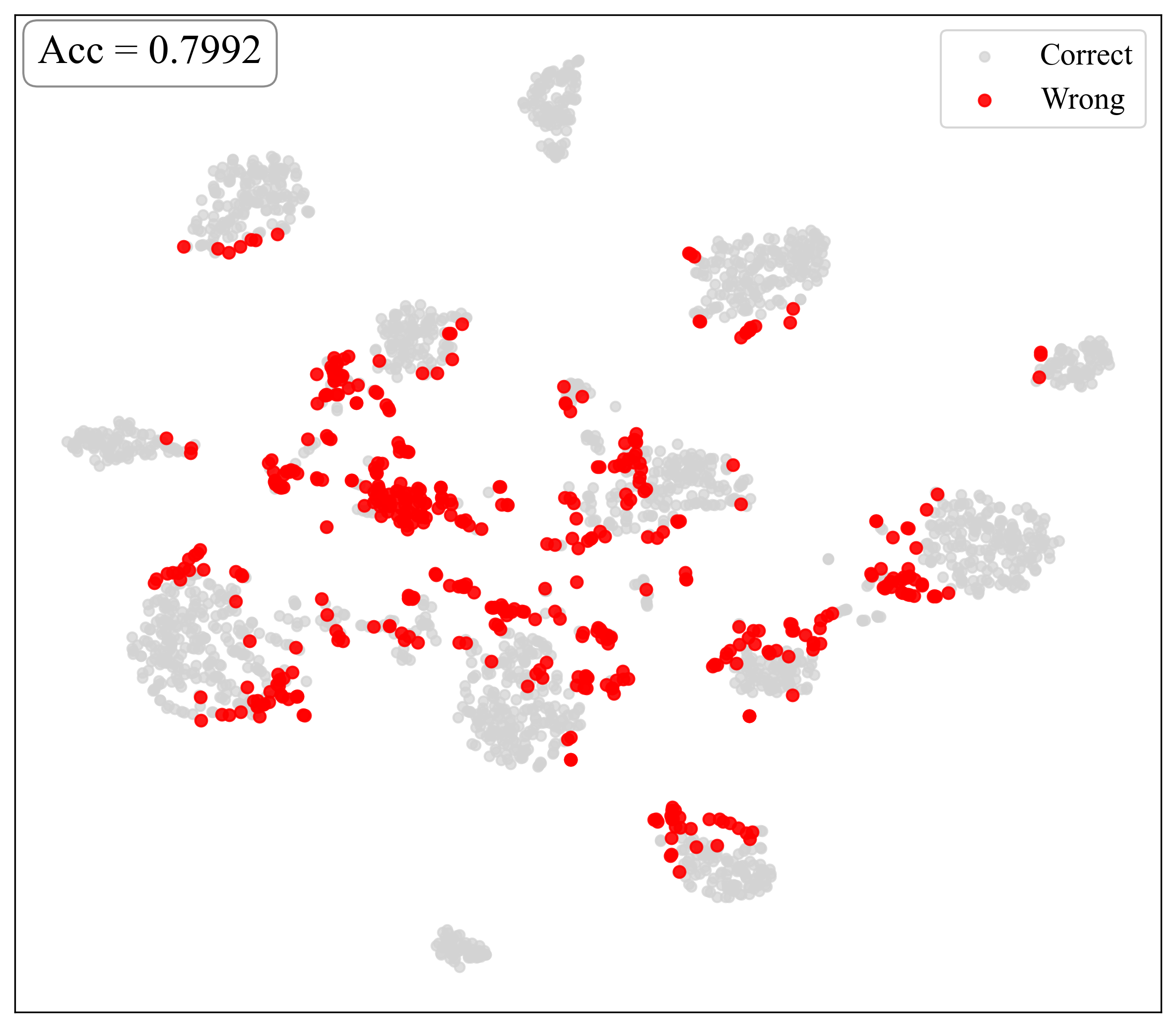} \\
		
	\end{tabular}
	
	\caption{Comparison of t-SNE visualizations of the ground-truth labels and prediction error distributions for SpectralNet, LanczosNet, Specformer, and GrokFormer under three settings: baseline, MPGFRFT-I, and MPGFRFT-II. Red and gray represent misclassified and correctly classified nodes, respectively. The first two rows show results on Cora, while the last two rows show results on Wiki.}
	\label{fig:tsne graphs}
\end{figure*}

\begin{figure*}[!t]
	\centering
	
	\begin{minipage}[b]{0.03\linewidth}~\end{minipage}\hfill
	\begin{minipage}[b]{0.47\linewidth}
		\centering \textbf{MPGFRFT-I}
	\end{minipage}\hfill
	\begin{minipage}[b]{0.47\linewidth}
		\centering \textbf{MPGFRFT-II}
	\end{minipage}
	
	\vspace{0.2cm}
	
	\begin{minipage}[b]{0.06\linewidth}~\end{minipage}\hfill
	\begin{minipage}[b]{0.223\linewidth}
		\centering \textbf{Box Plot of Final}\\[-1pt]\textbf{$\left | \boldsymbol{a} - \mathbf{1}_N \right | $}
	\end{minipage}\hfill
	\begin{minipage}[b]{0.247\linewidth}
		\centering \textbf{Update Norm Curve of}\\[-1pt]\textbf{$\|\boldsymbol{a}^{(t)}-\boldsymbol{a}^{(t-1)}\|_2$}
	\end{minipage}\hfill
	\begin{minipage}[b]{0.223\linewidth}
		\centering \textbf{Box Plot of Final}\\[-1pt]\textbf{$\left | \boldsymbol{a} - \mathbf{1}_N \right | $}
	\end{minipage}\hfill
	\begin{minipage}[b]{0.247\linewidth}
		\centering \textbf{Update Norm Curve of}\\[-1pt]\textbf{$\|\boldsymbol{a}^{(t)}-\boldsymbol{a}^{(t-1)}\|_2$}
	\end{minipage}
	
	\vspace{0.2cm}
	
	
	\begin{minipage}[c]{0.03\linewidth} 
		\centering \rotatebox{90}{\textbf{SpectralNet}}
	\end{minipage}\hfill
	\begin{minipage}[c]{0.95\linewidth} 
		\begin{minipage}[t]{0.235\linewidth} 
			\vspace{0pt}
			\includegraphics[width=\linewidth]{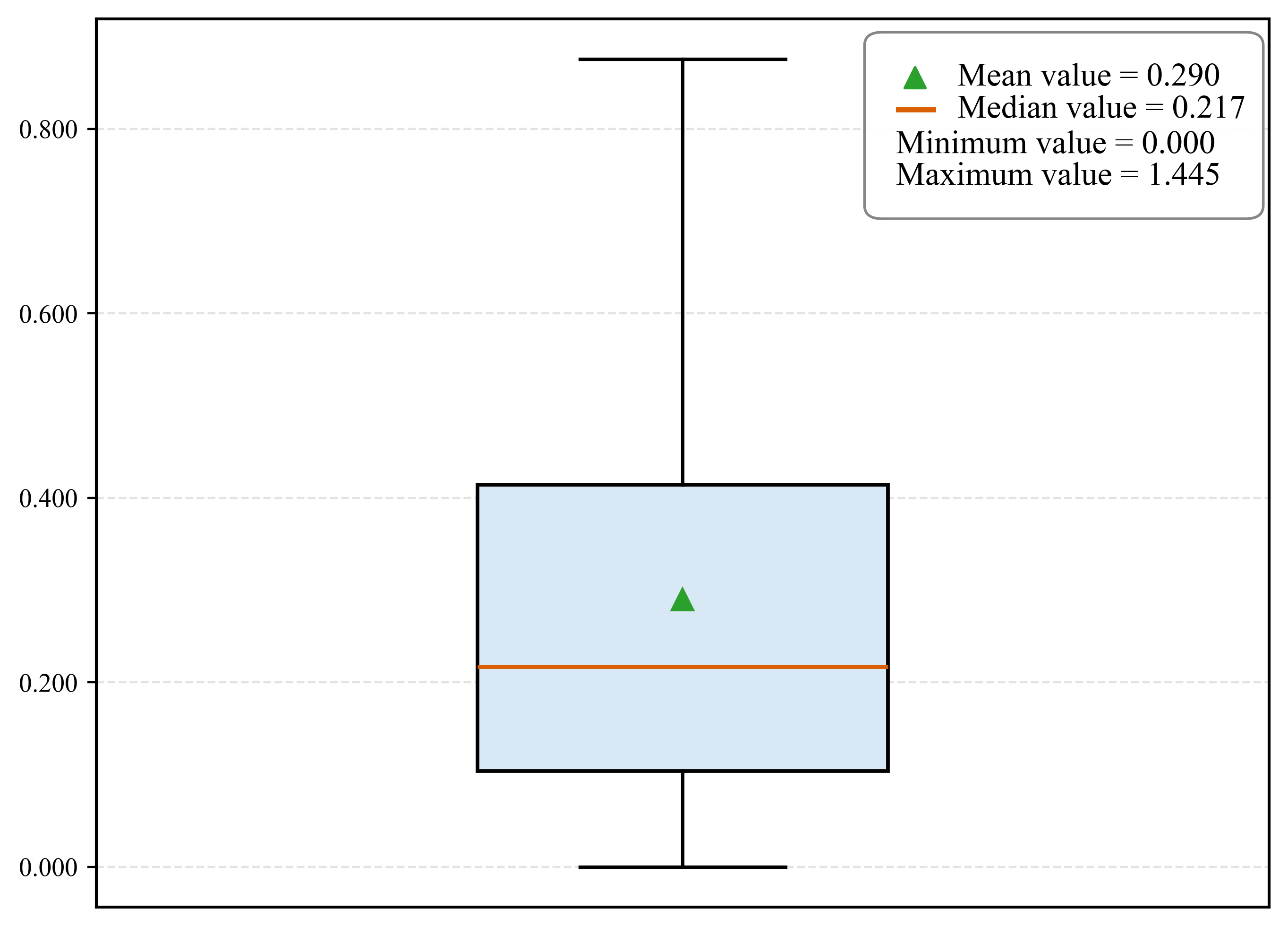}
		\end{minipage}\hfill
		\begin{minipage}[t]{0.260\linewidth} 
			\vspace{0pt}
			\includegraphics[width=\linewidth]{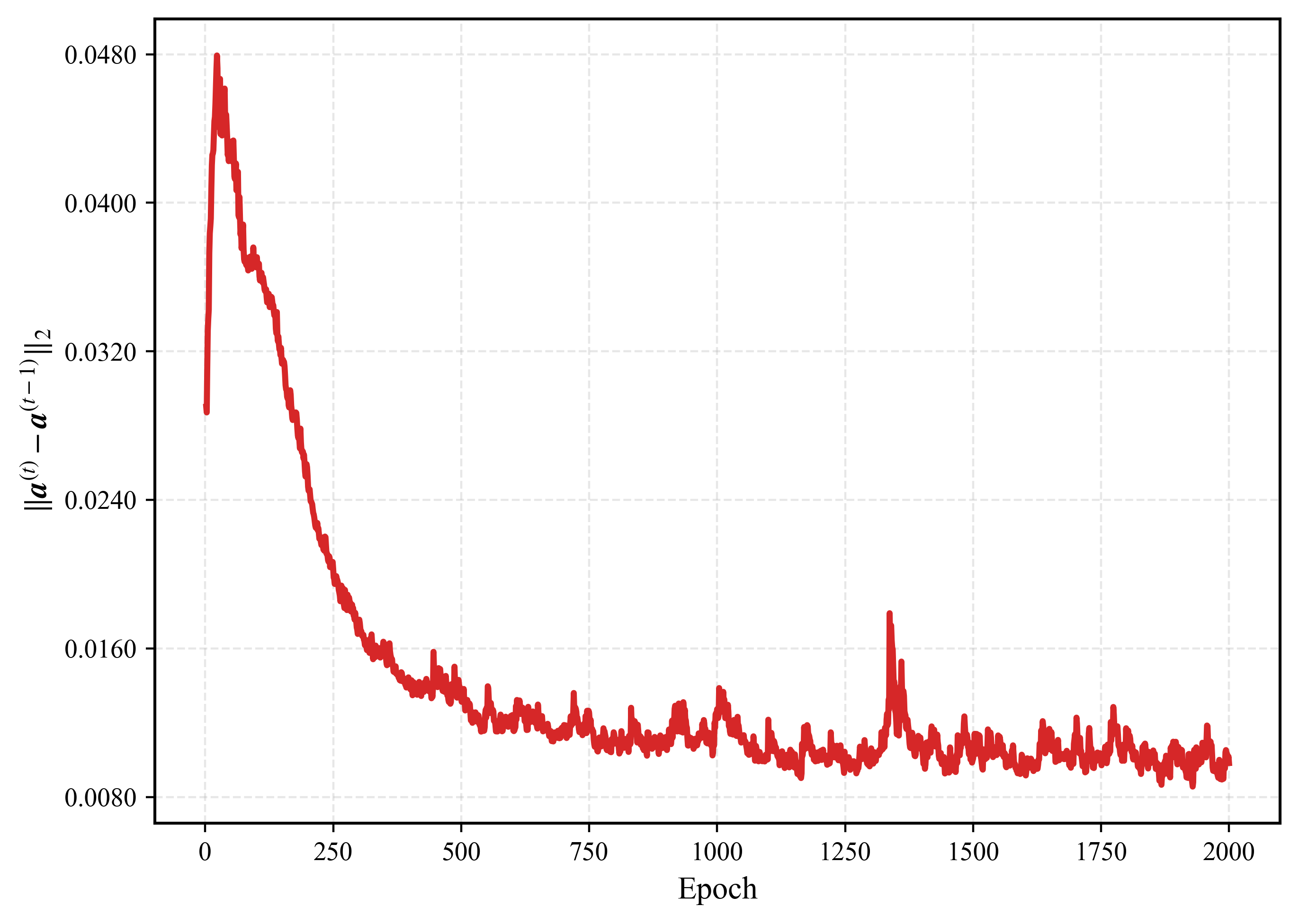}
		\end{minipage}\hfill
		\begin{minipage}[t]{0.235\linewidth}
			\vspace{0pt}
			\includegraphics[width=\linewidth]{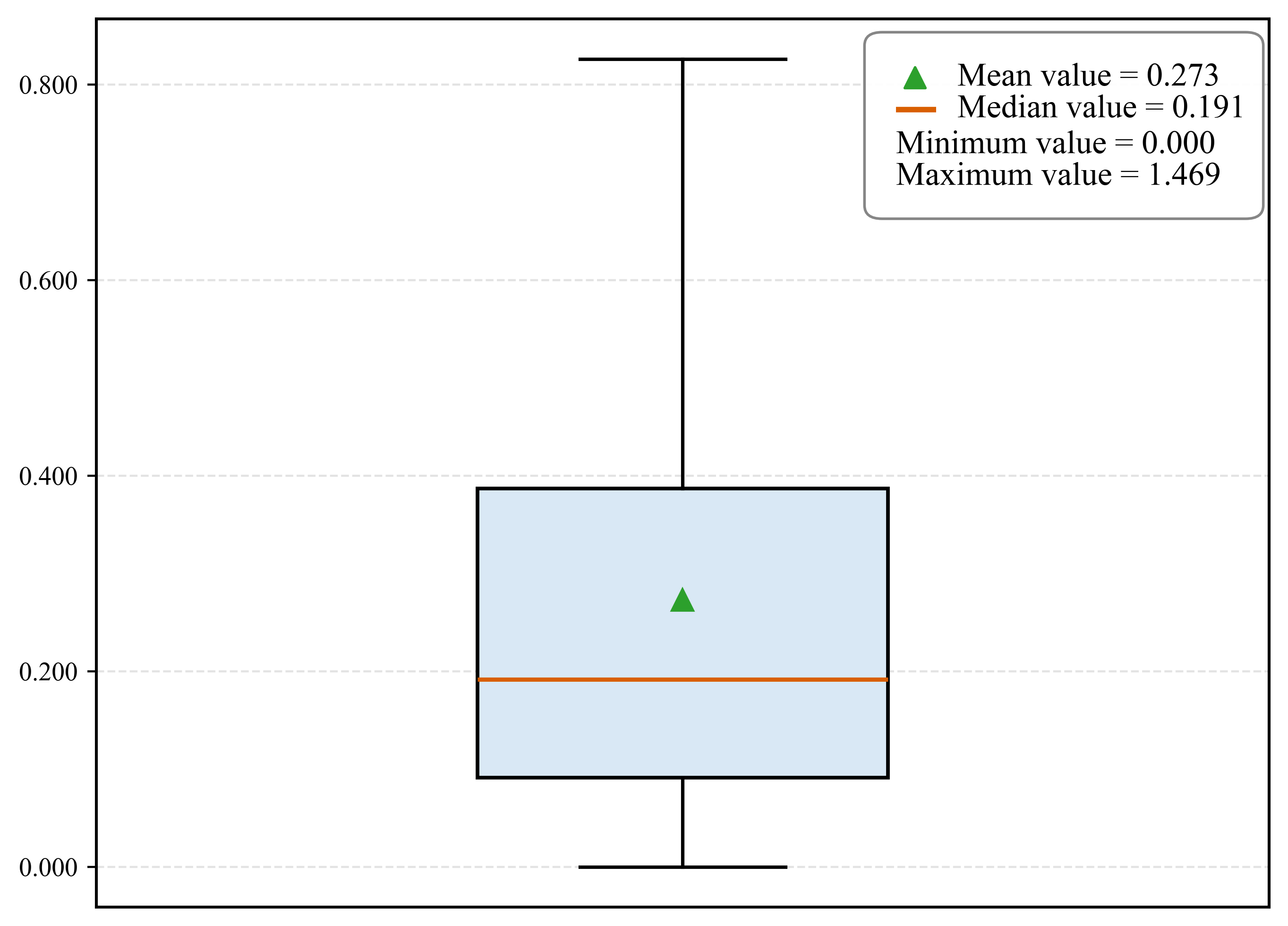}
		\end{minipage}\hfill
		\begin{minipage}[t]{0.260\linewidth}
			\vspace{0pt}
			\includegraphics[width=\linewidth]{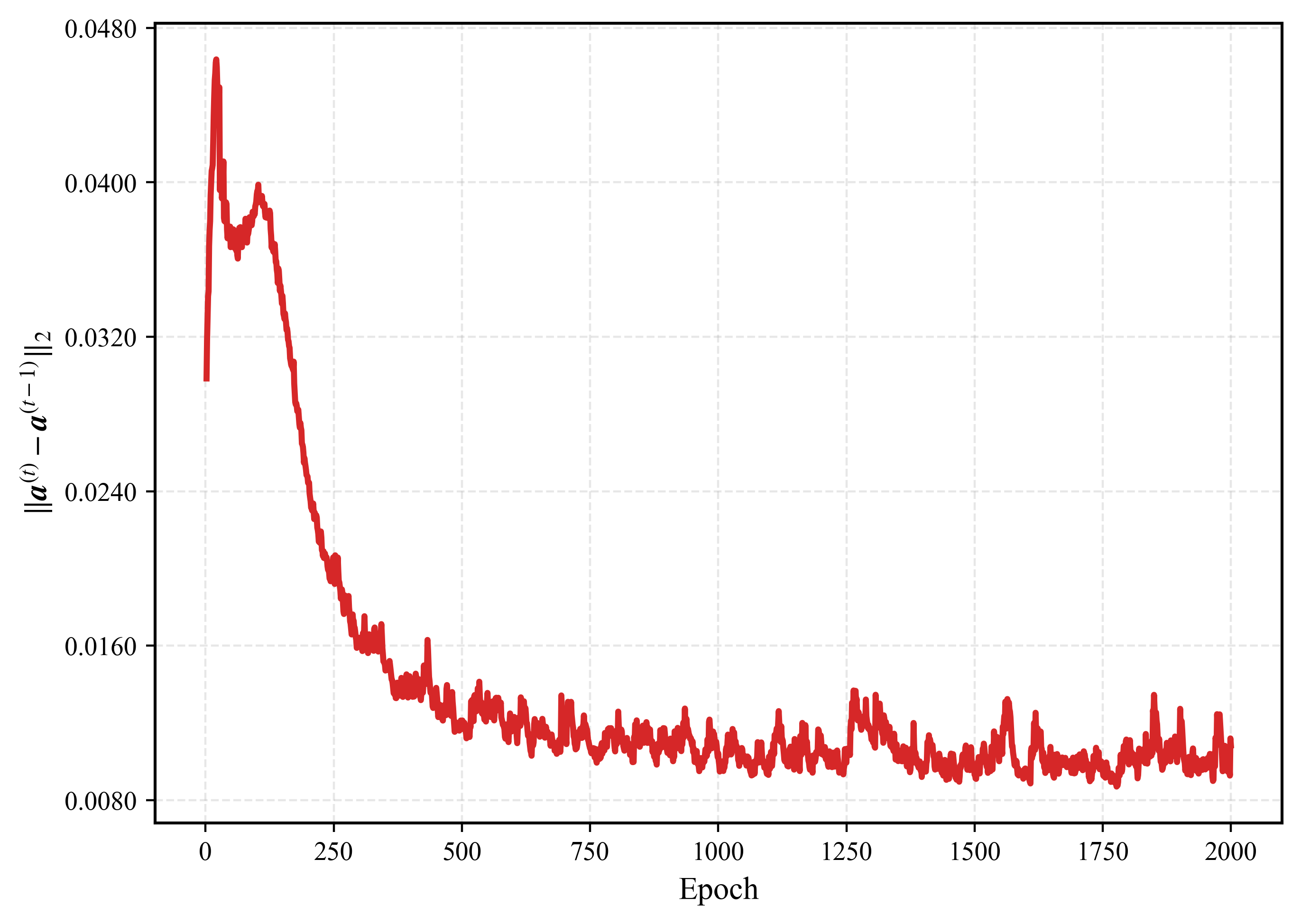}
		\end{minipage}
	\end{minipage}
	
	\vspace{0.15cm}
	
	\begin{minipage}[c]{0.03\linewidth}
		\centering \rotatebox{90}{\textbf{LanczosNet}}
	\end{minipage}\hfill
	\begin{minipage}[c]{0.95\linewidth}
		\begin{minipage}[t]{0.235\linewidth}
			\vspace{0pt}
			\includegraphics[width=\linewidth]{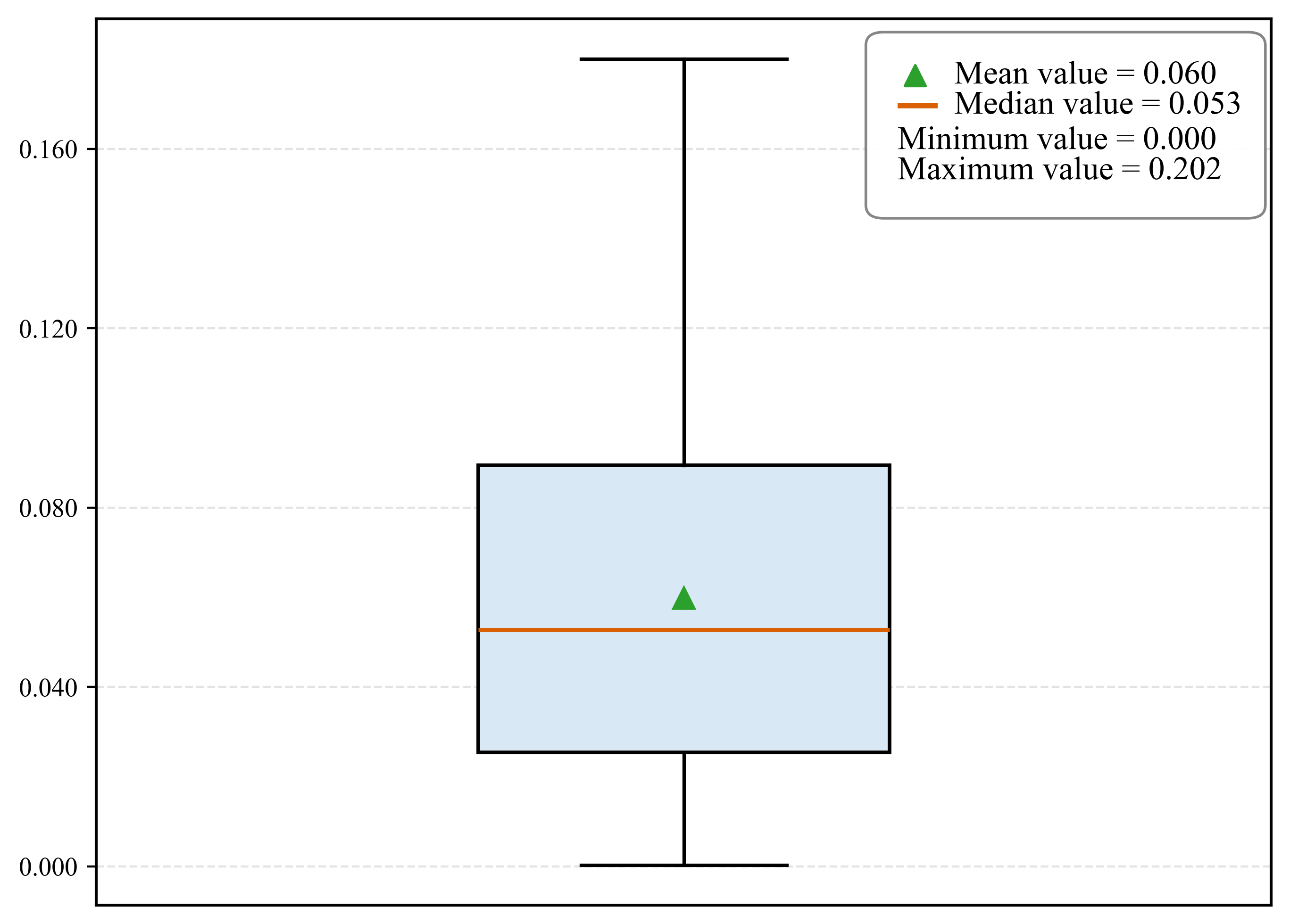}
		\end{minipage}\hfill
		\begin{minipage}[t]{0.260\linewidth}
			\vspace{0pt}
			\includegraphics[width=\linewidth]{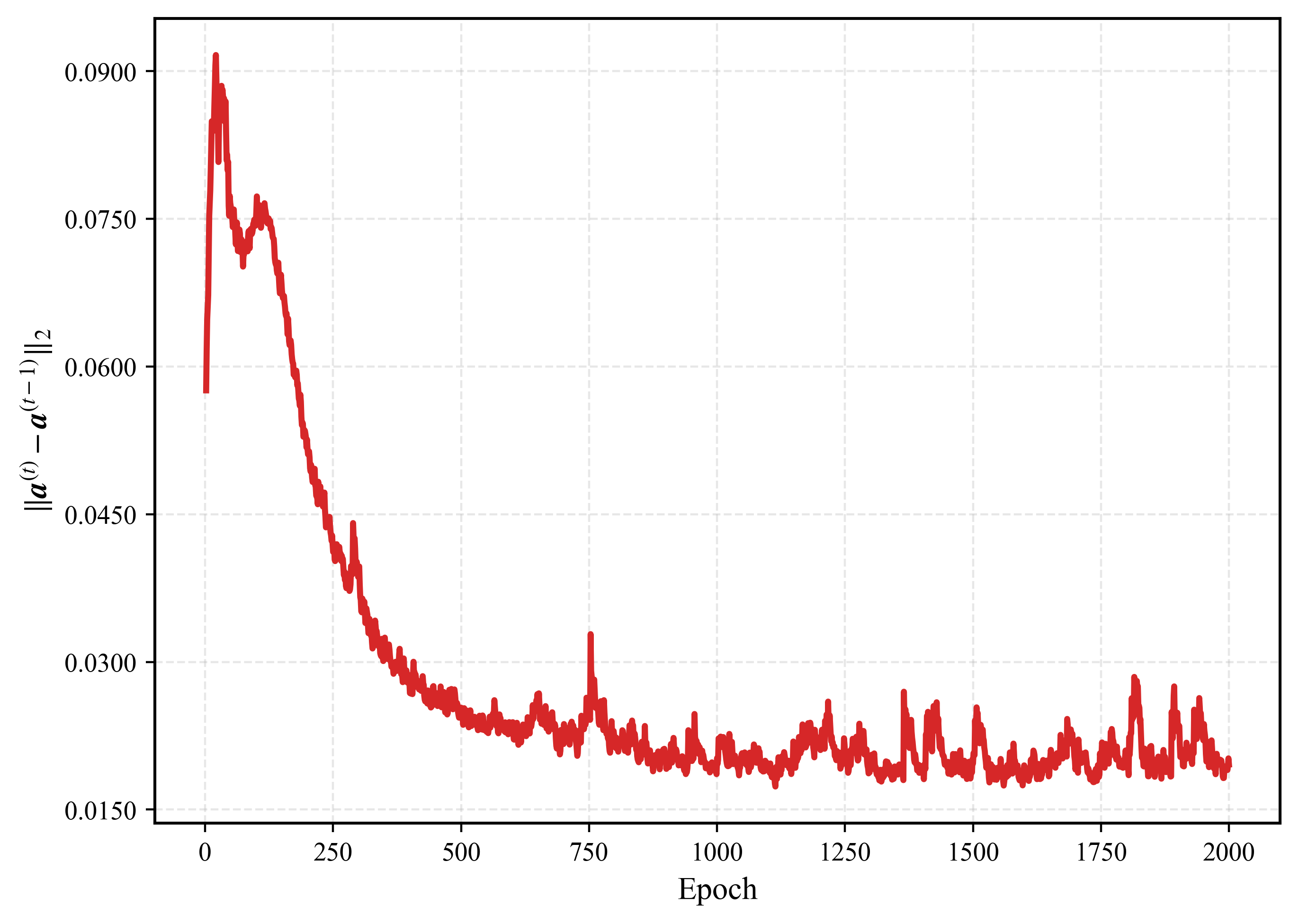}
		\end{minipage}\hfill
		\begin{minipage}[t]{0.235\linewidth}
			\vspace{0pt}
			\includegraphics[width=\linewidth]{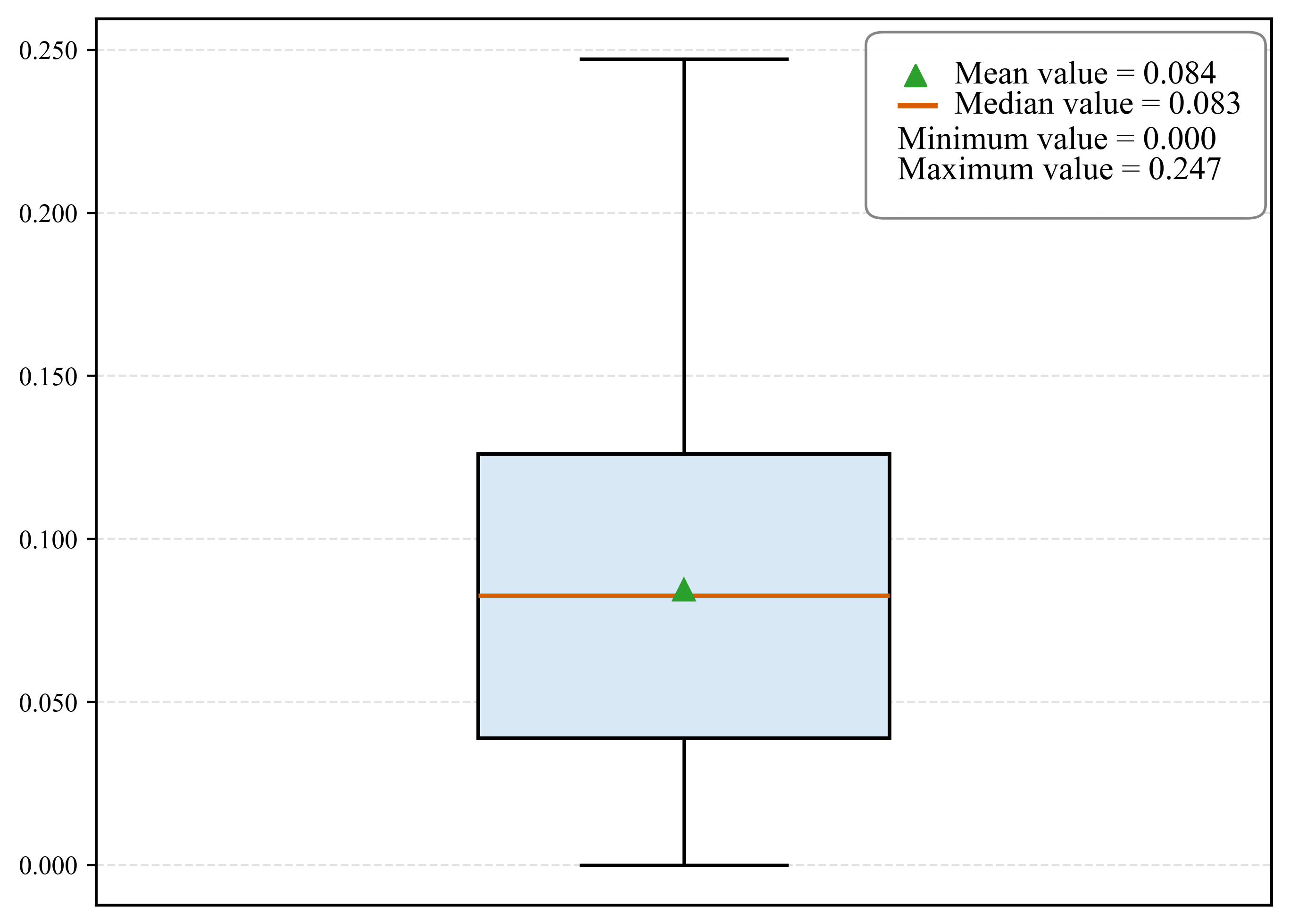}
		\end{minipage}\hfill
		\begin{minipage}[t]{0.260\linewidth}
			\vspace{0pt}
			\includegraphics[width=\linewidth]{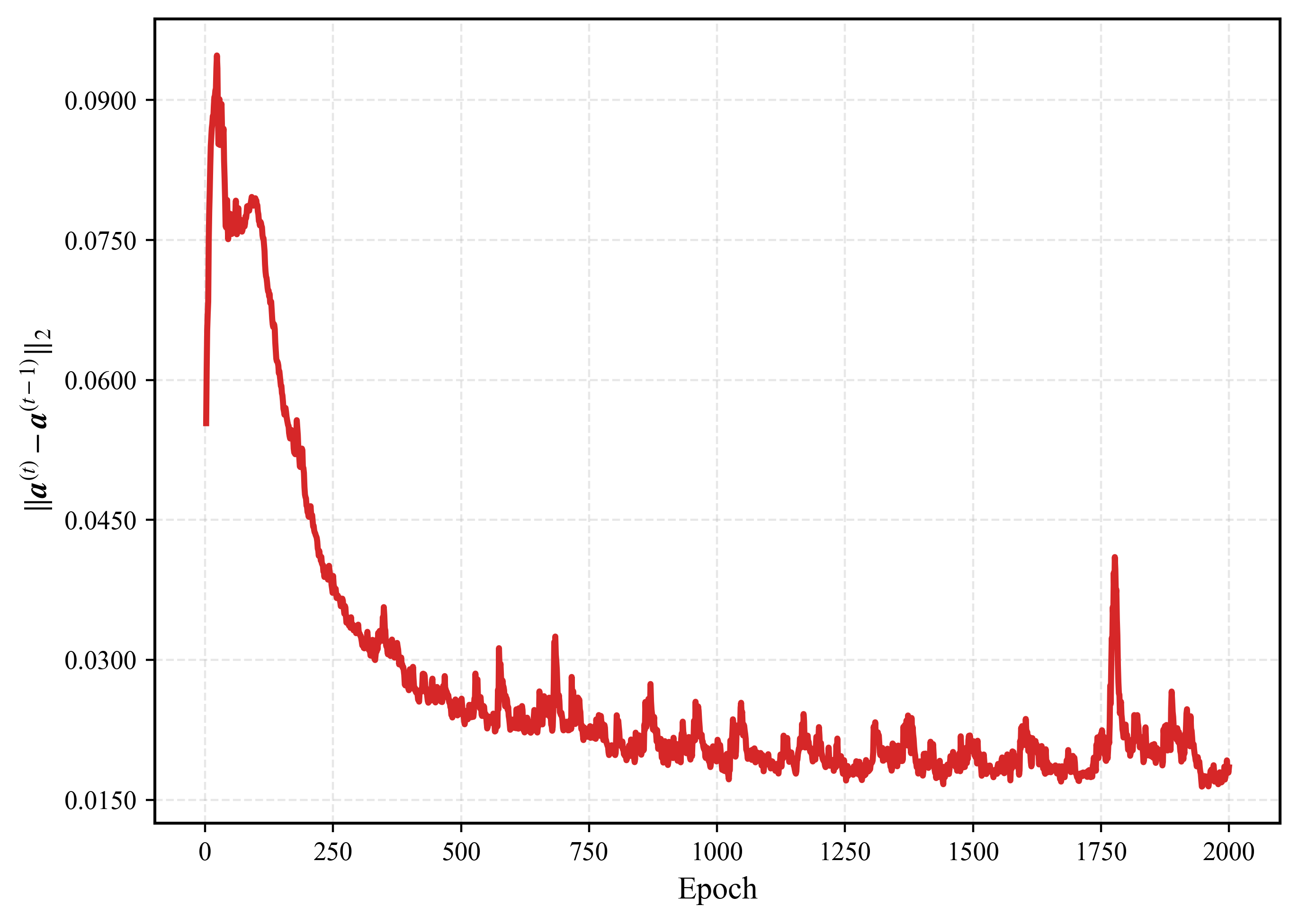}
		\end{minipage}
	\end{minipage}
	
	\vspace{0.15cm}
	
	\begin{minipage}[c]{0.03\linewidth}
		\centering \rotatebox{90}{\textbf{Specformer}}
	\end{minipage}\hfill
	\begin{minipage}[c]{0.95\linewidth}
		\begin{minipage}[t]{0.235\linewidth}
			\vspace{0pt}
			\includegraphics[width=\linewidth]{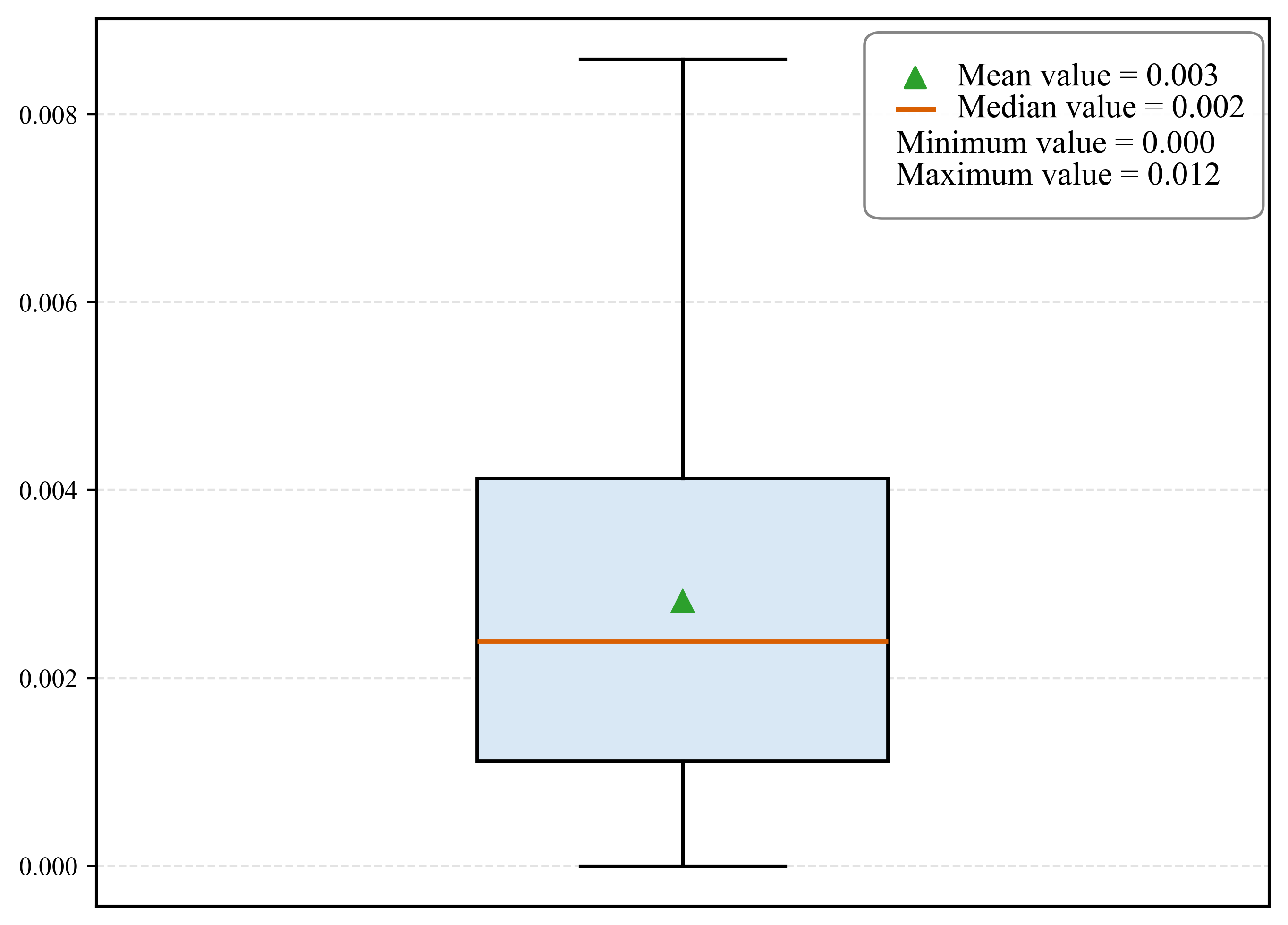}
		\end{minipage}\hfill
		\begin{minipage}[t]{0.260\linewidth}
			\vspace{0pt}
			\includegraphics[width=\linewidth]{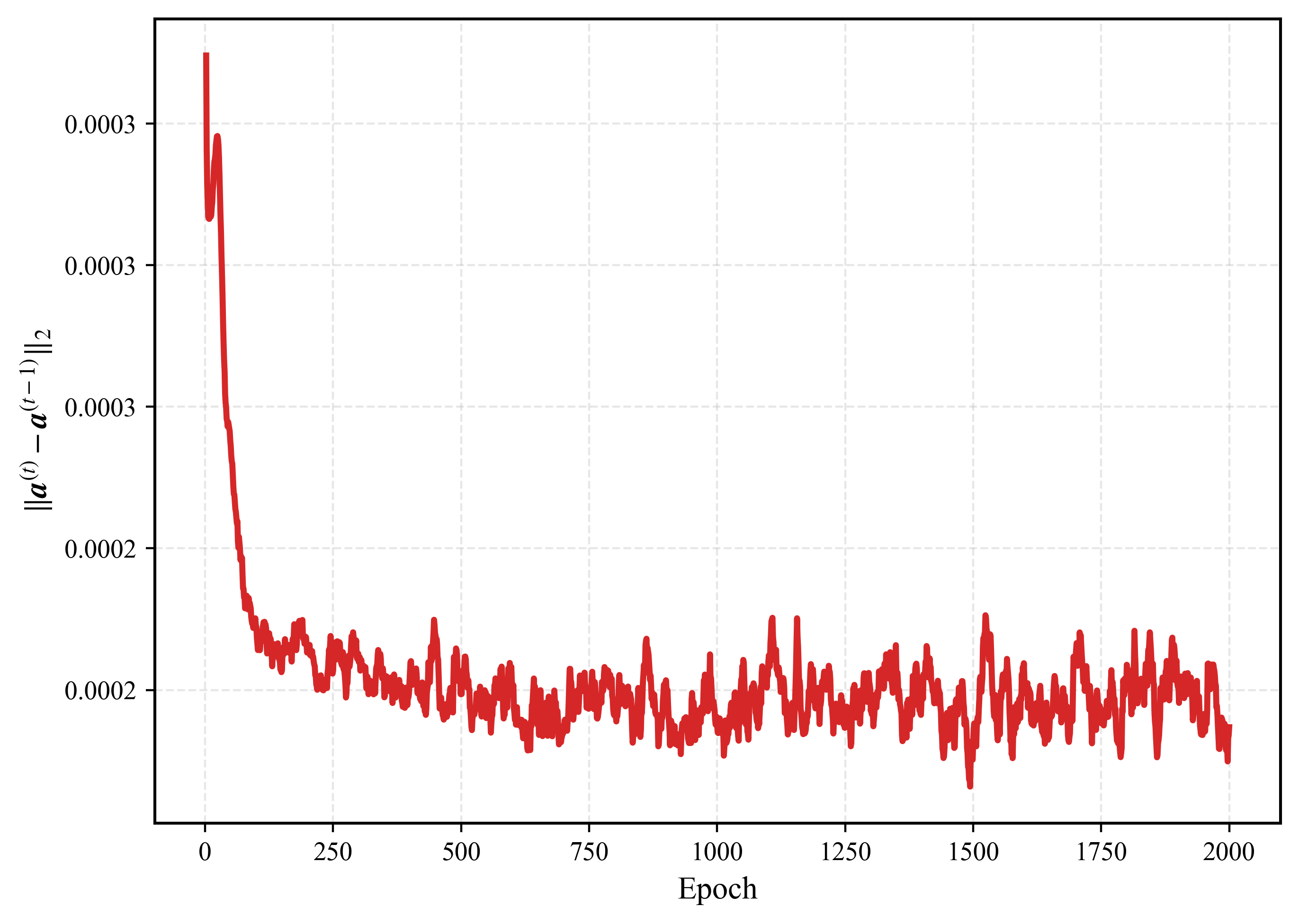}
		\end{minipage}\hfill
		\begin{minipage}[t]{0.235\linewidth}
			\vspace{0pt}
			\includegraphics[width=\linewidth]{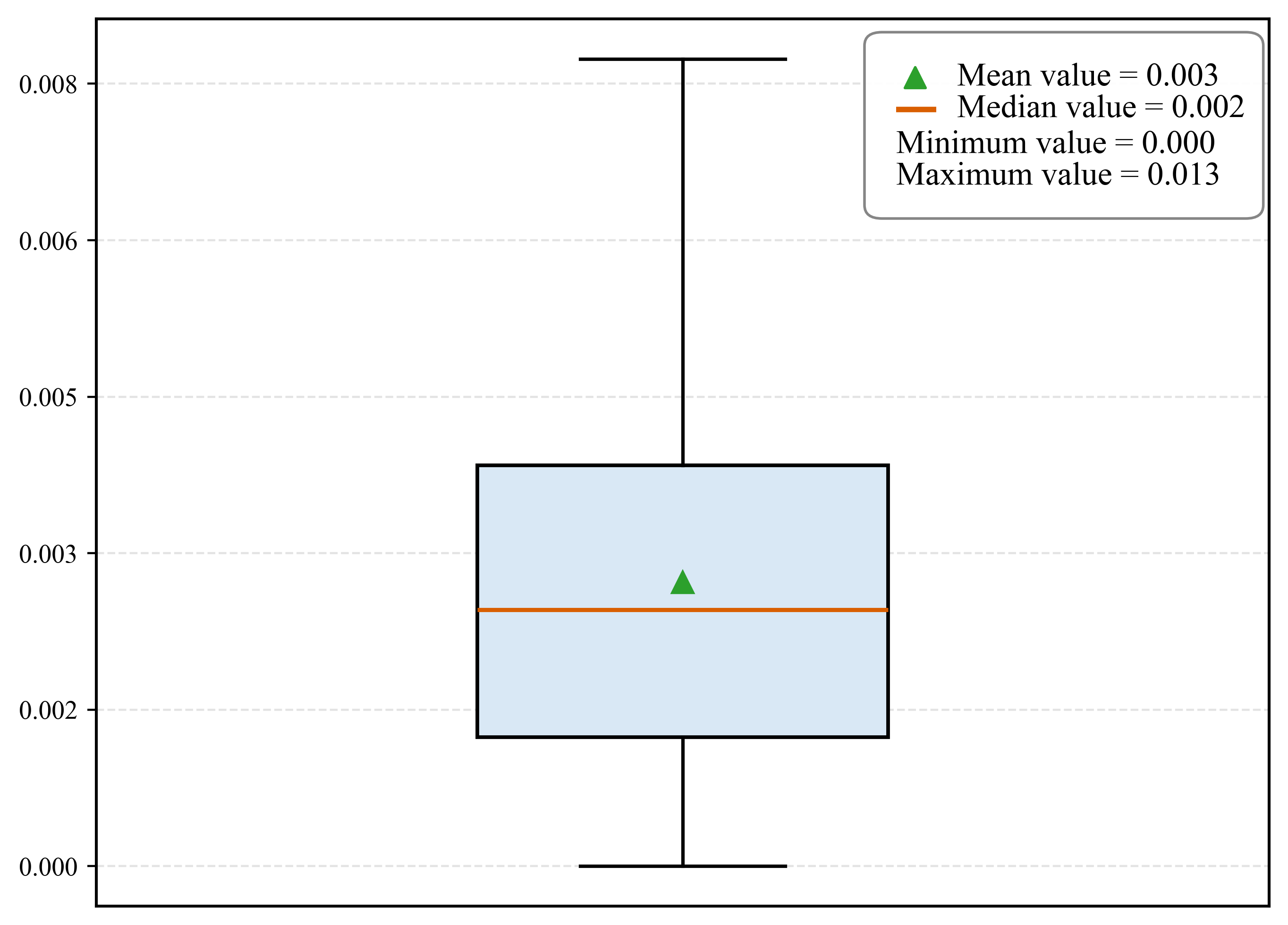}
		\end{minipage}\hfill
		\begin{minipage}[t]{0.260\linewidth}
			\vspace{0pt}
			\includegraphics[width=\linewidth]{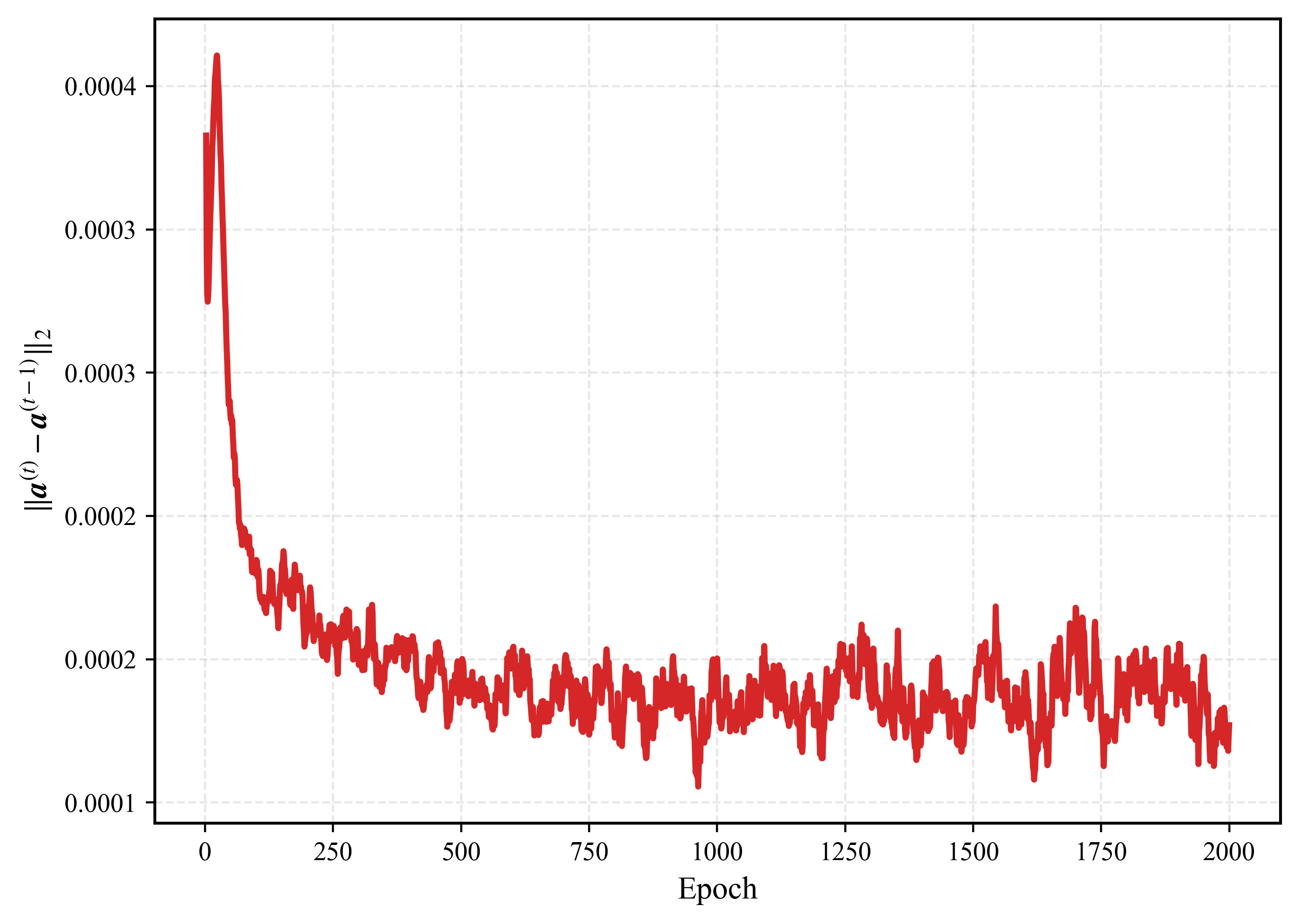}
		\end{minipage}
	\end{minipage}
	
	\vspace{0.15cm}
	
	\begin{minipage}[c]{0.03\linewidth}
		\centering \rotatebox{90}{\textbf{GrokFormer}}
	\end{minipage}\hfill
	\begin{minipage}[c]{0.95\linewidth}
		\begin{minipage}[t]{0.235\linewidth}
			\vspace{0pt}
			\includegraphics[width=\linewidth]{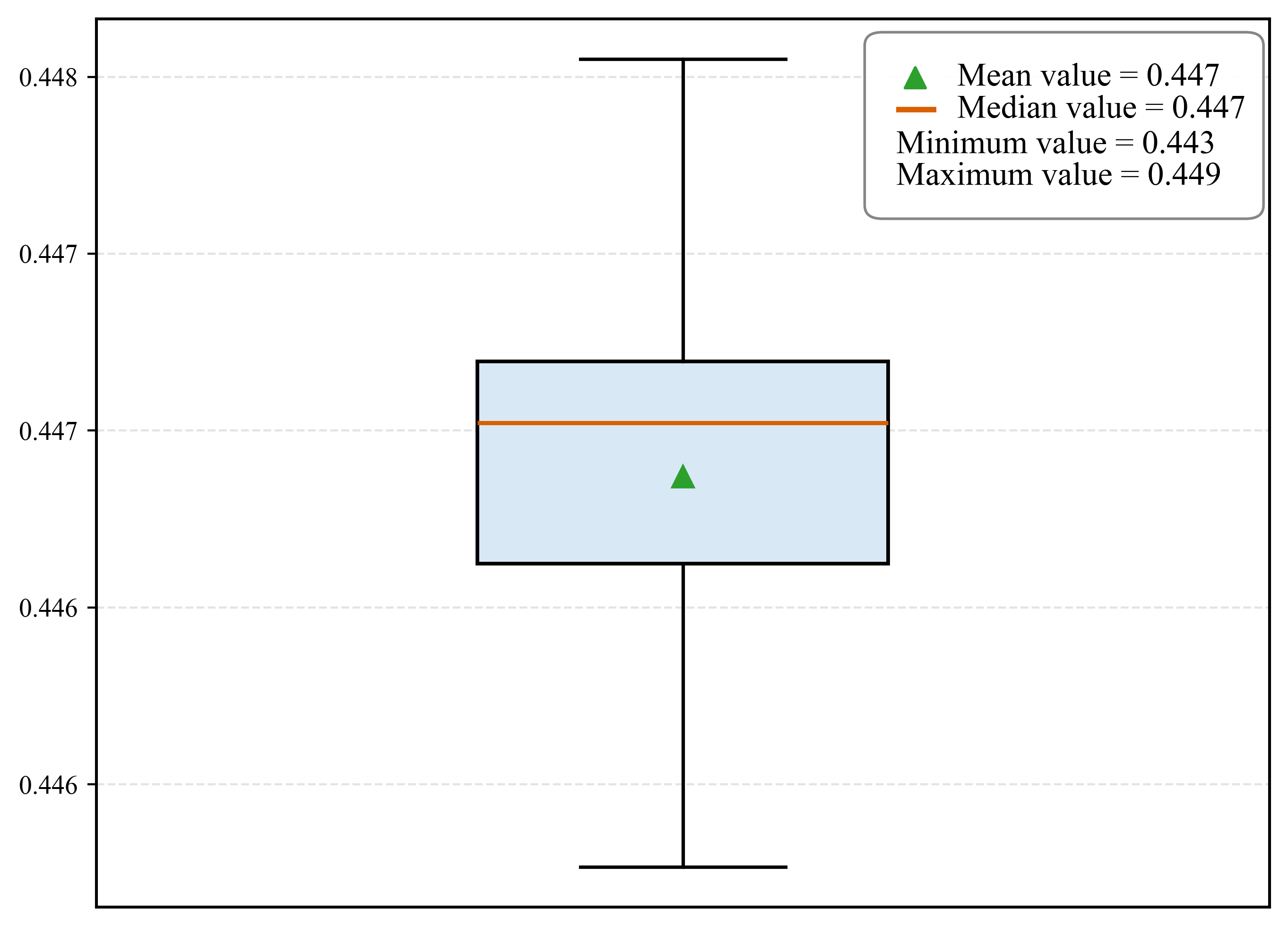}
		\end{minipage}\hfill
		\begin{minipage}[t]{0.260\linewidth}
			\vspace{0pt}
			\includegraphics[width=\linewidth]{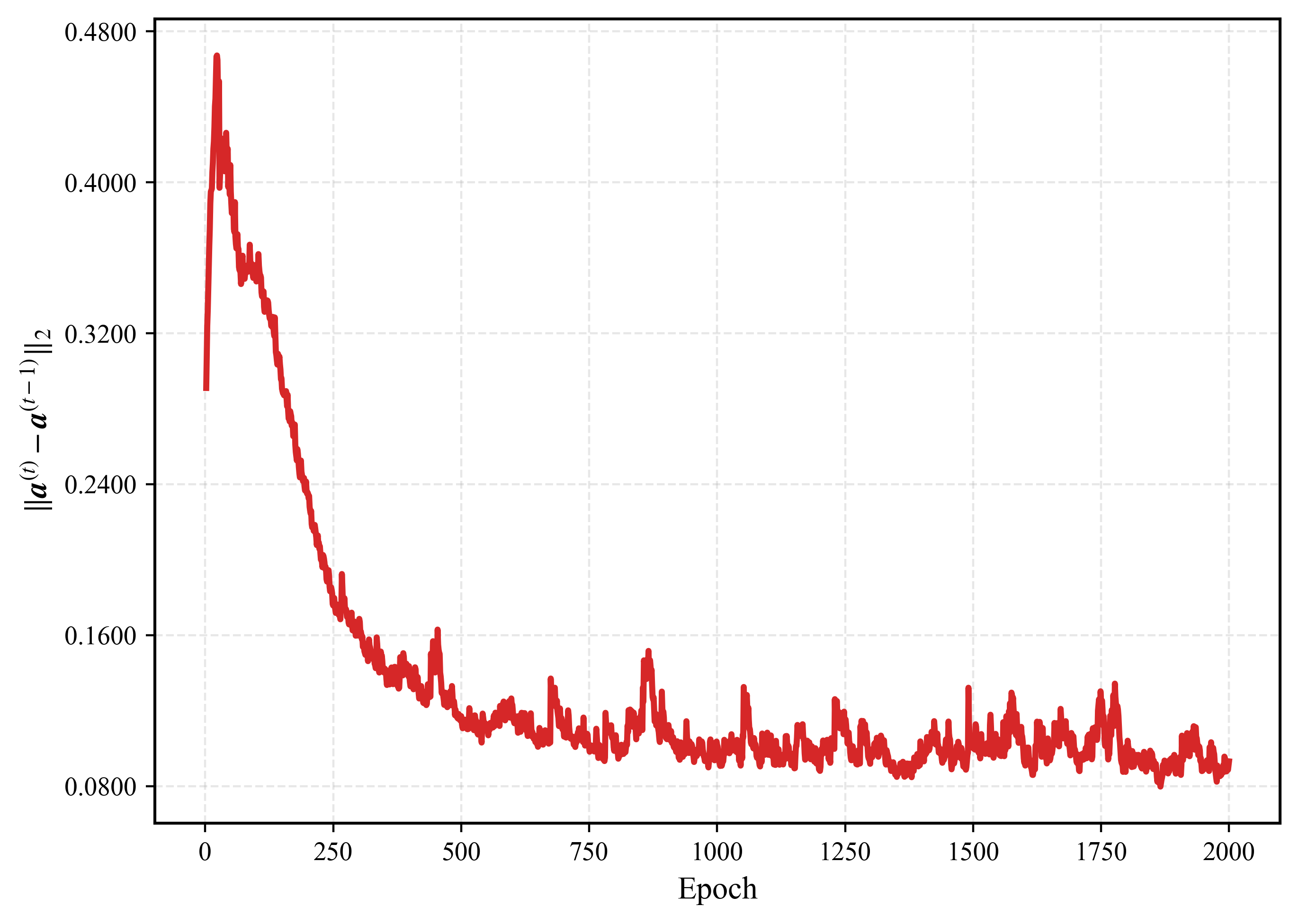}
		\end{minipage}\hfill
		\begin{minipage}[t]{0.235\linewidth}
			\vspace{0pt}
			\includegraphics[width=\linewidth]{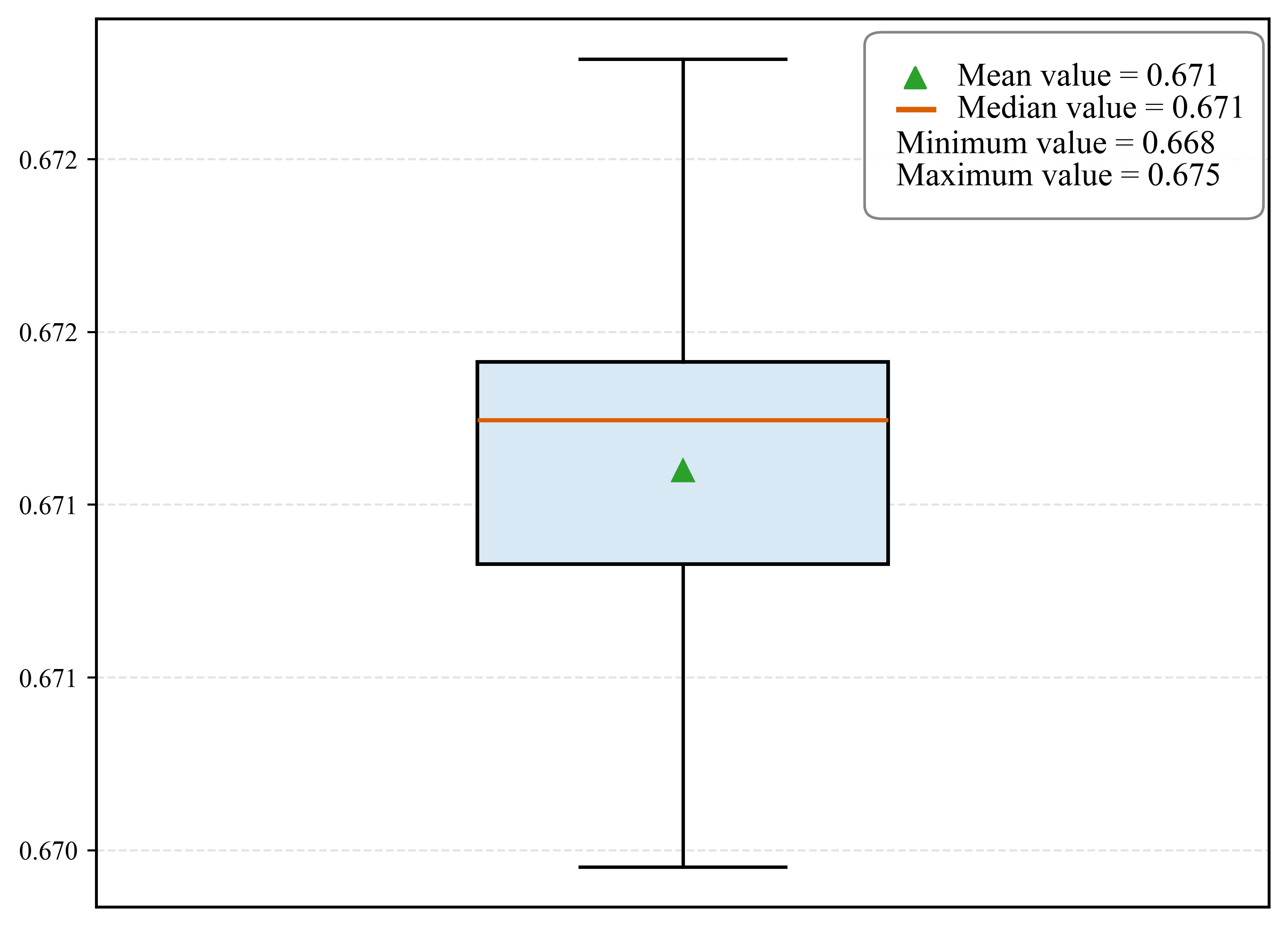}
		\end{minipage}\hfill
		\begin{minipage}[t]{0.260\linewidth}
			\vspace{0pt}
			\includegraphics[width=\linewidth]{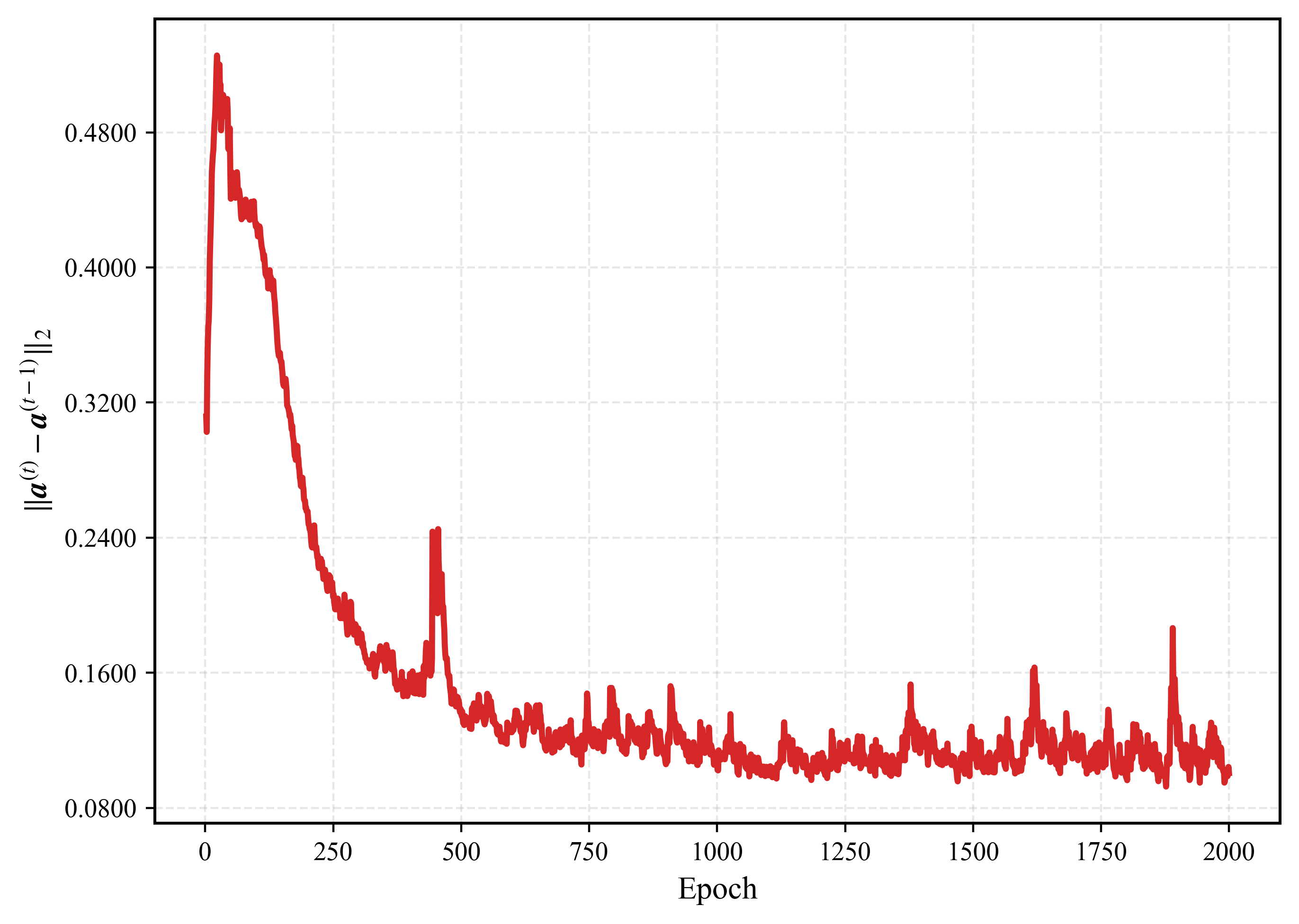}
		\end{minipage}
	\end{minipage}
	
	\vspace{0.2cm}
	
	\caption{Comparison of parameter distributions and update norm curves for SpectralNet, LanczosNet, Specformer, and GrokFormer under the MPGFRFT-I and MPGFRFT-II settings. The first two rows correspond to the Cora dataset, while the last two rows correspond to the Wiki dataset.}
	\label{fig:box graphs and update norm curves graphs}
\end{figure*}

\begin{table*}[htbp]
	\centering
	\caption{Performance gain compared to the original backbone models across 10 datasets: mean accuracy (\%) $\pm$ std.}
	\label{tab:node_classification_results}
	\renewcommand{\arraystretch}{1.08}
	\setlength{\tabcolsep}{4.5pt}
	\resizebox{\textwidth}{!}{
		\begin{tabular}{llcccccccccc}
			\toprule
			& & \multicolumn{4}{c}{Homophilic Datasets} & \multicolumn{6}{c}{Heterophilic Datasets} \\
			\cmidrule(lr){3-6} \cmidrule(lr){7-12}
			& Version & Cora & Citeseer & Photo & Computers & Twitch-PT & Wiki & Actor & Texas & Cornell & Wisconsin \\
			\midrule
			
			\multirow{3}{*}{SpectralCNN}
			& Original & 83.17 $\pm$ 0.87 & 71.06 $\pm$ 1.93 & 91.13 $\pm$ 0.65 & 89.23 $\pm$ 0.39 & 87.64 $\pm$ 3.30 & 65.66 $\pm$ 1.39 & 28.87 $\pm$ 3.77 & 74.75 $\pm$ 7.70 & 64.47 $\pm$ 9.13 & 60.50 $\pm$ 8.46 \\
			& MPFSR-I     & 84.04 $\pm$ 1.34 & \textbf{71.84 $\pm$ 1.23} & \textbf{91.73 $\pm$ 0.61} & \textbf{89.47 $\pm$ 0.33} & \textbf{87.91 $\pm$ 2.74} & 66.07 $\pm$ 1.33 & \textbf{33.82 $\pm$ 0.60} & \textbf{75.90 $\pm$ 7.69} & \textbf{66.17 $\pm$ 8.95} & \textbf{61.63 $\pm$ 7.07} \\
			& MPFSR-II     & \textbf{84.19 $\pm$ 1.23} & 71.16 $\pm$ 1.54 & 91.09 $\pm$ 0.88 & 89.41 $\pm$ 0.35 & 87.75 $\pm$ 3.03 & \textbf{66.18 $\pm$ 1.21} & 33.79 $\pm$ 0.63 & 75.08 $\pm$ 8.19 & 63.19 $\pm$ 14.08 & 60.75 $\pm$ 8.30 \\
			\midrule
			
			\multirow{3}{*}{LanczosNet}
			& Original & 88.03 $\pm$ 1.70 & 80.55 $\pm$ 1.19 & 92.03 $\pm$ 0.60 & 89.57 $\pm$ 0.69 & 88.85 $\pm$ 1.37 & 70.38 $\pm$ 1.05 & 36.07 $\pm$ 1.38 & 89.18 $\pm$ 2.66 & 87.23 $\pm$ 5.21 & 80.50 $\pm$ 4.78 \\
			& MPFSR-I     & \textbf{88.51 $\pm$ 1.42} & \textbf{81.31 $\pm$ 1.90} & \textbf{93.59 $\pm$ 0.79} & \textbf{89.79 $\pm$ 0.71} & 89.06 $\pm$ 1.39 & \textbf{72.08 $\pm$ 1.32} & 36.10 $\pm$ 1.66 & 89.55 $\pm$ 3.93 & \textbf{87.71 $\pm$ 2.80} & \textbf{83.62 $\pm$ 3.64} \\
			& MPFSR-II     & 88.16 $\pm$ 1.03 & 79.10 $\pm$ 1.24 & 91.09 $\pm$ 0.88 & 89.74 $\pm$ 0.66 & \textbf{89.14 $\pm$ 1.30} & 70.56 $\pm$ 0.94 & \textbf{36.25 $\pm$ 0.68} & \textbf{89.98 $\pm$ 4.12} & 87.54 $\pm$ 5.40 & 83.25 $\pm$ 4.51 \\
			\midrule
			
			\multirow{3}{*}{Specformer}
			& Original & 88.57 $\pm$ 1.01 & \textbf{81.49 $\pm$ 0.94} & 95.48 $\pm$ 0.32 & 90.87 $\pm$ 0.66 & 73.40 $\pm$ 3.10 & 79.31 $\pm$ 1.10 & 41.93 $\pm$ 1.04 & 88.23 $\pm$ 0.38 & 88.23 $\pm$ 4.19 & 74.63 $\pm$ 7.52 \\
			& MPFSR-I     & 89.16 $\pm$ 1.06 & 80.49 $\pm$ 1.87 & \textbf{95.98 $\pm$ 0.41} & \textbf{91.21 $\pm$ 0.52} & \textbf{73.43 $\pm$ 3.14} & 79.36 $\pm$ 1.20 & \textbf{42.24 $\pm$ 1.11} & 88.52 $\pm$ 0.43 & \textbf{89.36 $\pm$ 3.14} & 79.36 $\pm$ 1.20 \\
			& MPFSR-II     & \textbf{89.19 $\pm$ 1.22} & 80.69 $\pm$ 1.77 & 95.77 $\pm$ 0.54 & 91.17 $\pm$ 0.61 & 73.32 $\pm$ 2.96 & \textbf{79.86 $\pm$ 1.34} & 42.11 $\pm$ 1.09 & \textbf{89.16 $\pm$ 4.28} & 89.36 $\pm$ 8.51 & \textbf{80.13 $\pm$ 5.05} \\
			\midrule
			
			\multirow{3}{*}{GrokFormer}
			& Original & 89.57 $\pm$ 1.43 & 81.92 $\pm$ 1.25 & 95.52 $\pm$ 0.52 & 90.89 $\pm$ 0.56 & 89.16 $\pm$ 1.34 & 79.89 $\pm$ 1.81 & 42.98 $\pm$ 1.48 & 94.59 $\pm$ 2.08 & 89.17 $\pm$ 3.44 & 90.63 $\pm$ 4.41 \\
			& MPFSR-I     & \textbf{89.82 $\pm$ 1.32} & \textbf{82.13 $\pm$ 0.82} & \textbf{95.75 $\pm$ 0.39} & \textbf{91.29 $\pm$ 0.55} & 89.16 $\pm$ 1.34 & \textbf{80.61 $\pm$ 0.94} & \textbf{44.14 $\pm$ 1.12} & \textbf{96.72 $\pm$ 2.19} & \textbf{91.49 $\pm$ 3.56} & \textbf{94.63 $\pm$ 2.63} \\
			& MPFSR-II     & 89.71 $\pm$ 1.56 & 82.09 $\pm$ 1.79 & 95.71 $\pm$ 0.44 & 91.05 $\pm$ 0.89 & \textbf{89.53 $\pm$ 1.34} & 78.10 $\pm$ 1.58 & 43.97 $\pm$ 1.37 & 93.44 $\pm$ 1.94 & 89.36 $\pm$ 3.20 & 93.75 $\pm$ 2.06 \\
			\bottomrule
			
		\end{tabular}
	}
\end{table*}

\indent To provide an intuitive comparison of the learned spectral representations, Figs.~\ref{fig:tsne graphs} and~\ref{fig:box graphs and update norm curves graphs} provide complementary evidence for the proposed MPGFRFT-based framework from the perspectives of representation quality and optimization behavior. Specifically, Fig.~\ref{fig:tsne graphs} presents the t-SNE visualizations of ground-truth labels and prediction error distributions for four models under three settings, Fig.~\ref{fig:box graphs and update norm curves graphs} shows the box plots of the element-wise absolute deviations of the final parameter vector from the all-ones initialization, i.e., $|\boldsymbol{a}-\mathbf{1}_N|$, together with the update norm curves $|\boldsymbol{a}^{(t)}-\boldsymbol{a}^{(t-1)}|_2$ for the same models under MPGFRFT-I and MPGFRFT-II. 

\indent Fig.~\ref{fig:tsne graphs} shows that misclassified nodes are mainly located in overlapping regions and near class boundaries in the embedding space. Compared with the baseline, the MPGFRFT-based variants exhibit more structured error distributions in several cases, indicating improved discrimination in difficult regions. After introducing MPGFRFT-I and MPGFRFT-II, the overall embedding structure remains stable, while the changes in the error distributions indicate that the spectral representations are adjusted in a way that benefits node classification. Fig.~\ref{fig:box graphs and update norm curves graphs} shows the training behavior of the proposed multi-parameter transform. For both MPGFRFT variants, the update norms decrease rapidly at the beginning of training and then remain at low levels, indicating a stable optimization process. Although local fluctuations appear in some cases, they do not change the overall downward trend. The box plots show that the learned parameter distributions vary across backbone models and between MPGFRFT-I and MPGFRFT-II, suggesting that the proposed multi-parameter transform adapts to different spectral architectures rather than converging to a uniform parameter pattern. These results indicate that the proposed MPGFRFT-based framework improves spectral representations and maintains stable optimization across different baseline models.

\section{Conclusion} \label{sec6}
\indent In this paper, we proposed two types of MPGFRFTs to overcome the fundamental limitations of the conventional GFRFT, which restricts spectral modulation to a single global fractional order. By introducing a multiple-parameter order vector, the MPGFRFT framework achieves superior spectral adaptability and expressive power. Alongside establishing the rigorous theoretical foundations and a thorough computational complexity analysis, we developed a dynamically learnable order vector scheme tailored for task-driven optimization.  Furthermore, we proposed a novel, plug-and-play MPFSR module that intrinsically combines proactive spectral modulation with neural network optimization. Crucially, we mathematically proved the spectral stability of the proposed module by strictly bounding its perturbation relative to the standard GFT, thereby guaranteeing a reliable end-to-end optimization. By seamlessly integrating the MPGFRFT operators into existing spectral GNN architectures, the module enables the dynamic, data-driven tuning of fractional bases. Experiments on graph node classification tasks using representative GNN backbones validate the practical value of this integration. Empirical results consistently demonstrate that the MPFSR-enhanced GNNs achieve steady enhancements in node representation learning, highlighting the broad applicability of our framework. Ultimately, this work not only provides a highly generalizable framework for graph signal processing but also opens new avenues for designing adaptive, theoretically grounded graph neural networks.

\bibliography{mybib}

@ARTICLE{Scarselli08Graph,
	author={Scarselli, Franco and Gori, Marco and Tsoi, Ah Chung and Hagenbuchner, Markus and Monfardini, Gabriele},
	journal={IEEE Trans. Neural Netw. Learn. Syst.}, 
	title={The Graph Neural Network Model}, 
	year={2009},
	month={Dec.},
	volume={20},
	number={1},
	pages={61--80},
	doi={10.1109/TNN.2008.2005605}}

@article{corso2024graph,
	author={Corso, Gabriele and Stark, Hannes and Jegelka, Stefanie and Jaakkola, Tommi and Barzilay, Regina},
	journal={Nat. Rev. Methods Primers},
	title={Graph neural networks},
	year={2024},
	month={Mar.},
	volume={4},
	number={1},
	pages={17},
	doi={https://doi.org/10.1038/s43586-024-00294-7}
}

@article{wu2020comprehensive,
	author={Wu, Zong Han and Pan, Shi Rui and Chen, Feng Wen and Long, Guo Dong and Zhang, Cheng Qi and Yu, Philip S},
	journal={IEEE Trans. Neural Netw. Learn. Syst.},
	title={A comprehensive survey on graph neural networks},
	volume={32},
	number={1},
	pages={4--24},
	year={2020},
	month={Mar.},
	publisher={IEEE},
	doi={10.1109/TNNLS.2020.2978386}
}

@inproceedings{li2021spatial,
	title={Spatial-temporal fusion graph neural networks for traffic flow forecasting},
	author={Li, Meng Zhang and Zhu, Zhan Xing},
	booktitle={Proc. AAAI Conference on Artificial Intelligence},
	volume={35},
	number={5},
	pages={4189--4196},
	year={2021},
	month={May},
	doi={https://doi.org/10.1609/aaai.v35i5.16542}
}

@INPROCEEDINGS{gilmer17a,
	author={Gilmer, Justin and Schoenholz, Samuel S. and Riley, Patrick F. and Vinyals, Oriol and Dahl, George E.},
	booktitle={Proc. 34th International Conference on Machine Learning (ICML)}, 
	title={Neural Message Passing for Quantum Chemistry}, 
	year={2017},
	month={Aug.},
	volume={70},
	number={},
	pages={1263--1272},
	doi={},
}

@INPROCEEDINGS{wang22am,
	author={Wang, Xi Yuan and Zhang, Mu Han},
	booktitle={Proc. 39th International Conference on Machine Learning (ICML)},
	title = {How Powerful are Spectral Graph Neural Networks},
	year={2022},
	volume={162},
	number={},
	month={Jul.},
	pages={23341--23362},
	doi={},
}

@INPROCEEDINGS{mo2025autosgnn,
	title={Auto{SGNN}: automatic propagation mechanism discovery for spectral graph neural networks},
	author={Mo, Shi Bing and Wu, Kai and Gao, Qi Xuan and Teng, Xiang Yi and Liu, Jing},
	booktitle={Proc. AAAI Conference on Artificial Intelligence},
	volume={39},
	number={18},
	pages={19493--19502},
	year={2025},
	month={Mar.},
	doi={https://doi.org/10.1609/aaai.v39i18.34146}
}

@ARTICLE{Ortega18GSP,
	author={Ortega, Antonio and Frossard, Pascal and Kovacevic, Jelena and Moura, Jose M. F. and Vandergheynst, Pierre},
	journal={Proc. IEEE}, 
	title={Graph Signal Processing: Overview, Challenges, and Applications}, 
	year={2018},
	month={May},
	volume={106},
	number={5},
	pages={808--828},
	doi={10.1109/JPROC.2018.2820126}}

@INPROCEEDINGS{wu2019simplifying,
	author = {Wu, Felix and Souza, Amauri and Zhang, Tian Yi and Fifty, Christopher and Yu, Tao and Weinberger, Kilian},
	booktitle = {Proc. 36th International Conference on Machine Learning (ICML)},
	title = {Simplifying Graph Convolutional Networks},
	year={2019},
	volume={97},
	number={},
	month={Jun.},
	pages={6861--6871},
	doi={},
}

@INPROCEEDINGS{bo2021beyond,
	title={Beyond low-frequency information in graph convolutional networks},
	author={Bo, De Yu and Wang, Xiao and Shi, Chuan and Shen, Huawei},
	booktitle={Proc. AAAI Conference on Artificial Intelligence},
	volume={35},
	number={5},
	pages={3950--3957},
	year={2021},
	month={May},
	doi={https://doi.org/10.1609/aaai.v35i5.16514},
}

@INPROCEEDINGS{he2022convolutional,
	title = {Convolutional Neural Networks on Graphs with Chebyshev Approximation, Revisited},
	author = {He, Ming Guo and Wei, Zhe Wei and Wen, Ji Rong},
	booktitle = {Proc. Advances in Neural Information Processing Systems (NeurIPS)},
	pages = {7264--7276},
	volume = {35},
	year = {2022},
	month={Dec.},
}

@INPROCEEDINGS{wang2022powerful,
	author = {Wang, Xi Yuan and Zhang, Mu Han},
	booktitle = {Proc. 39th International Conference on Machine Learning (ICML)},
	title={How Powerful are Spectral Graph Neural Networks},
	year={2022},
	volume={162},
	number={},
	month={Jul.},
	pages={23341--23362},
	doi={},
}

@INPROCEEDINGS{he2021bernnet,
	author = {He, Ming Guo and Wei, Zhe Wei and Huang, Zeng Feng and Xu, Hong Teng},
	booktitle = {Proc. Advances in Neural Information Processing Systems (NeurIPS)},
	year={2021},
	title={Bernnet: learning arbitrary graph spectral filters via bernstein approximation},
	volume={34},
	number={},
	month={Dec.},
	pages={14239--14251},
	doi={},
}

@INPROCEEDINGS{xu2019graph,
	title={Graph convolutional networks using heat kernel for semi-supervised learning},
	author={Xu, Bing Bing and Shen, Hua Wei and Cao, Qi and Cen, Ke Ting and Cheng, Xue Qi},
	booktitle={Proc. 28th International Joint Conference on Artificial Intelligence (IJCAI)},
	pages={1928--1934},
	year={2019},
	month={Aug.},
	volume={},
	number={},
	doi={10.5555/3367243.3367306},
}

@INPROCEEDINGS{dong2021adagnn,
	author={Dong, Yu Shun and Ding, Kaize and Jalaian, Brian and Ji, Shui Wang and Li, Jun Dong},
	booktitle={Proc. 30th ACM International Conference on Information \& Knowledge Management (CIKM)},
	title={Ada{GNN}: Graph Neural Networks with Adaptive Frequency Response Filter},
	year={2021},
	volume={},
	number={Oct.},
	month={},
	pages={392--401},
	doi={10.1145/3459637.3482226},
}

@Book{ortega22introduction,
	title={Introduction to {G}raph {S}ignal {P}rocessing},
	author={Ortega, Antonio},
	year={2022},
	publisher={Cambridge University Press}
}

@ARTICLE{Leus23,
	author={Leus, Geert and Marques, Antonio G. and Moura, Jose M.F. and Ortega, Antonio and Shuman, David I},
	journal={IEEE Signal Process. Mag.}, 
	title={Graph Signal Processing: History, development, impact, and outlook}, 
	year={2023},
	month={Jun.},
	volume={40},
	number={4},
	pages={49--60},
	doi={10.1109/MSP.2023.3262906}}

@ARTICLE{Ortega08,
	author={Ortega, Antonio and Frossard, Pascal and Kovacevic, Jelena and Moura, Jose M. F. and Vandergheynst, Pierre},
	journal={Proc. IEEE}, 
	title={Graph Signal Processing: Overview, Challenges, and Applications}, 
	year={2018},
	month={Apr.},
	volume={106},
	number={5},
	pages={808--828},
	doi={10.1109/JPROC.2018.2820126}}

@ARTICLE{Gavili17,
	author={Gavili, Adnan and Zhang, Xiao Ping},
	journal={IEEE Trans. Signal Process.}, 
	title={On the Shift Operator, Graph Frequency, and Optimal Filtering in Graph Signal Processing}, 
	year={2017},
	mon={Sep.},
	volume={65},
	number={23},
	pages={6303--6318},
	doi={10.1109/TSP.2017.2752689}}

@ARTICLE{Lu19,
	author={Lu, Keng-Shih and Ortega, Antonio},
	journal={IEEE Trans. Signal Process.}, 
	title={Fast Graph {F}ourier Transforms Based on Graph Symmetry and Bipartition}, 
	year={2019},
	month={Aug.},
	volume={67},
	number={18},
	pages={4855--4869},
	doi={10.1109/TSP.2019.2932882}}

@ARTICLE{Domingos20,
	author={Domingos, Joao and Moura, Jose M. F.},
	journal={IEEE Trans. Signal Process.}, 
	title={Graph {F}ourier Transform: A Stable Approximation}, 
	year={2020},
	month={Jul.},
	volume={68},
	number={},
	pages={4422--4437},
	doi={10.1109/TSP.2020.3009645}}

@ARTICLE{Sandryhaila13,
	author={Sandryhaila, Aliaksei and Moura, Jose M. F.},
	journal={IEEE Trans. Signal Process.}, 
	title={Discrete Signal Processing on Graphs}, 
	year={2013},
	month={Jan.},
	volume={61},
	number={7},
	pages={1644--1656},
	doi={10.1109/TSP.2013.2238935}}

@book{bochner1949fourier,
	title={Fourier {T}ransforms},
	author={Bochner, Salomon and Chandrasekharan, Komaravolu},
	year={1949},
	publisher={Princeton University Press},
	address={Princeton}
}

@Book{nussbaumer1982fast,
	title={The {F}ast {F}ourier {T}ransform},
	author={Nussbaumer, Henri J and Nussbaumer, Henri J},
	year={1982},
	publisher={Springer},
	address={Berlin}
}

@ARTICLE{Ji23,
	author={Ji, Feng and Tay, Wee Peng and Ortega, Antonio},
	journal={IEEE Trans. Signal Process.}, 
	title={Graph Signal Processing Over a Probability Space of Shift Operators}, 
	year={2023},
	month={Mar.},
	volume={71},
	number={},
	pages={1159--1174},
	doi={10.1109/TSP.2023.3263675}}

@ARTICLE{Girault15,
	author={Girault, Benjamin and Goncalves, Paulo and Fleury, Eric},
	journal={IEEE Signal Process. Lett.}, 
	title={Translation on Graphs: An Isometric Shift Operator}, 
	year={2015},
	month={Oct.},
	volume={22},
	number={12},
	pages={2416--2420},
	doi={10.1109/LSP.2015.2488279}}

@inproceedings{Wang17,
	author={Wang, Yi Qian and Li, Bing Zhao and Cheng, Qi Yuan},
	booktitle={Proc. Asia-Pacific Signal and Information Processing Association Annual Summit and Conference (APSIPA ASC)}, 
	title={The fractional {F}ourier transform on graphs}, 
	year={2017},
	volume={},
	number={},
	month={Dec.},
	pages={105-110},
	doi={10.1109/APSIPA.2017.8282010}
}

@INPROCEEDINGS{Wang18,
	author={Wang, Yi Qian and Li, Bing Zhao},
	booktitle={Proc. 14th IEEE International Conference on Signal Processing (ICSP)}, 
	title={The Fractional {F}ourier Transform on Graphs: Sampling and Recovery}, 
	year={2018},
	month={Aug.},
	volume={},
	number={},
	pages={1103--1108},
	doi={10.1109/ICSP.2018.8652296}}

@ARTICLE{Ozturk21,
	author={Ozturk, Cuneyd and Ozaktas, Haldun M. and Gezici, Sinan and Koç, Aykut},
	journal={IEEE Trans. Signal Process.}, 
	title={Optimal Fractional {F}ourier Filtering for Graph Signals}, 
	year={2021},
	month={May},
	volume={69},
	number={},
	pages={2902--2912},
	doi={10.1109/TSP.2021.3079804}}

@article{yan2021windowed,
	author={Yan, Fang Jia and Li, Bing Zhao},
	journal={Digit. Signal Process.},
	title={Windowed fractional {F}ourier transform on graphs: Properties and fast algorithm},
	year={2021},
	month={Nov.},
	volume={118},
	number={},
	pages={103210},
	doi={https://doi.org/10.1016/j.dsp.2021.103210}
}

@article{gan2025windowed,
	author={Gan, Yu Chen and Chen, Jian Yi and Li, Bing Zhao},
	journal={Digit. Signal Process.},
	title={The windowed two-dimensional graph fractional {F}ourier transform},
	year={2025},
	month={Jul},
	volume={162},
	number={},
	pages={105191},
	doi={https://doi.org/10.1016/j.dsp.2025.105191}
}

@ARTICLE{Alik24Wiener,
	author={Alikasifoglu, Tuna and Kartal, Bünyamin and Koc, Aykut},
	journal={IEEE Signal Process. Lett.}, 
	title={Wiener Filtering in Joint Time-Vertex Fractional {F}ourier Domains}, 
	year={2024},
	month = {May},
	volume={31},
	number={},
	pages={1319-1323},
	doi={10.1109/LSP.2024.3396664}}

@article{alikacsifouglu2025joint,
	author={Alikasifoglu, Tuna and Kartal, Bunyamin and Ozgunay, Eray and Koc, Aykut},
	journal={Signal Process.},
	title={Joint time-vertex fractional {F}ourier transform},
	year={2025},
	month={Aug.},
	volume={233},
	number={},
	pages={109944},
	doi={https://doi.org/10.1016/j.sigpro.2025.109944}
}

@ARTICLE{MPDFRFT06,
	author={Soo Chang Pei and Wen Liang Hsue},
	journal={IEEE Signal Process. Lett.}, 
	title={The multiple-parameter discrete fractional {F}ourier transform}, 
	year={2006},
	volume={13},
	month={Jun.},
	number={6},
	pages={329--332},
	doi={10.1109/LSP.2006.871721}}

@ARTICLE{MPDFRFT16,
	author={Kang, Xue Jing and Tao, Ran and Zhang, Feng},
	journal={IEEE Trans. Signal Process.}, 
	title={Multiple-Parameter Discrete Fractional Transform and its Applications}, 
	year={2016},
	month={Mar.},
	volume={64},
	number={13},
	pages={3402--3417},
	doi={10.1109/TSP.2016.2544740}}

@ARTICLE{Alikasifoglu24,
	author={Alikasifoglu, Tuna and Kartal, Bunyamin and Koc, Aykut},
	journal={IEEE Trans. Signal Process.}, 
	title={Graph Fractional {F}ourier Transform: A Unified Theory}, 
	year={2024},
	month={Aug.},
	volume={72},
	pages={3834--3850},
	doi={10.1109/TSP.2024.3439211}}

@ARTICLE{Song22,
	author={Song, Xiao Ying and Chai, Li and Zhang, Jing Xin},
	journal={IEEE Trans. Pattern Anal. Mach. Intell.}, 
	title={Graph Signal Processing Approach to {QSAR/QSPR} Model Learning of Compounds}, 
	year={2022},
	month={Oct.},
	volume={44},
	number={4},
	pages={1963--1973},
	doi={10.1109/TPAMI.2020.3032718}}

@ARTICLE{Patane23,
	author={Patane, Giuseppe},
	journal={IEEE Trans. Pattern Anal. Mach. Intell.}, 
	title={Fourier-Based and Rational Graph Filters for Spectral Processing}, 
	year={2023},
	month={May},
	volume={45},
	number={6},
	pages={7063--7074},
	doi={10.1109/TPAMI.2022.3177075}}

@ARTICLE{Qi22,
	author={Qi, Wen Fa and Guo, Si Rui and Hu, Wei},
	journal={IEEE Trans. Image Process.}, 
	title={Generic Reversible Visible Watermarking via Regularized Graph {F}ourier Transform Coding}, 
	year={2022},
	month={Dec.},
	volume={31},
	number={},
	pages={691--705},
	doi={10.1109/TIP.2021.3134466}}

@ARTICLE{Pan09,
	author={Pan, Wei and Qin, Kai Huai and Chen, Yao},
	journal={IEEE Trans. Pattern Anal. Mach. Intell.}, 
	title={An Adaptable-Multilayer Fractional {F}ourier Transform Approach for Image Registration}, 
	year={2009},
	month={Apr.},
	volume={31},
	number={3},
	pages={400--414},
	doi={10.1109/TPAMI.2008.83}}

@ARTICLE{Zhao24,
	author={Zhao, Xu Dong and Zhang, Meng Meng and Tao, Ran and Li, Wei and Liao, Wen Zhi and Tian, Lian Fang and Philips, Wilfried},
	journal={IEEE Trans. Neural Netw. Learn. Syst.}, 
	title={Fractional {F}ourier Image Transformer for Multimodal Remote Sensing Data Classification}, 
	year={2024},
	month={Feb.},
	volume={35},
	number={2},
	pages={2314--2326},
	doi={10.1109/TNNLS.2022.3189994}}

@ARTICLE{Isufi22,
	author={Isufi, Elvin and Gama, Fernando and Ribeiro, Alejandro},
	journal={IEEE Trans. Pattern Anal. Mach. Intell.}, 
	title={EdgeNets: Edge Varying Graph Neural Networks}, 
	year={2022},
	month={Sep.},
	volume={44},
	number={11},
	pages={7457--7473},
	doi={10.1109/TPAMI.2021.3111054}}

@ARTICLE{Liu25,
	author={Liu, You Fa and Du, Bo},
	journal={IEEE Trans. Neural Netw. Learn. Syst.}, 
	title={Frequency Domain-Oriented Complex Graph Neural Networks for Graph Classification}, 
	year={2025},
	month={Feb.},
	volume={36},
	number={2},
	pages={2733--2746},
	doi={10.1109/TNNLS.2024.3351762}}

@ARTICLE{Xia21,
	author={Xia, Feng and Sun, Ke and Yu, Shuo and Aziz, Abdul and Wan, Liangtian and Pan, Shirui and Liu, Huan},
	journal={IEEE Trans. Artif. Intell.}, 
	title={Graph Learning: A Survey}, 
	year={2021},
	month={Apr.},
	volume={2},
	number={2},
	pages={109--127},
	doi={10.1109/TAI.2021.3076021}}

@ARTICLE{Li24,
	author={Li, Nai Qi and Li, Wen Jie and Gao, Ying Hua and Li, Yi Ming and Bao, Ji Gang and Kuruoglu, Ercan E. and Jiang, Yong and Xia, Shu Tao},
	journal={IEEE Trans. Artif. Intell.}, 
	title={Node-Level Graph Regression With Deep {G}aussian Process Models}, 
	year={2024},
	month={Jun.},
	volume={5},
	number={6},
	pages={3257--3269},
	doi={10.1109/TAI.2023.3347177}}

@inproceedings{bruna2014spectral,
	title={Spectral networks and locally connected networks on graphs},
	author={Bruna, Joan and Zaremba, Wojciech and Szlam, Arthur and LeCun, Yann},
	booktitle={Proc. International Conference on Learning Representations (ICLR)},
	year={2013},
	volume={},
	number={},
	month={Dec.},
	pages={},
	doi={}
}

@inproceedings{liao2019lanczosnet,
	title={Lanczos{N}et: Multi-scale deep graph convo-lutional networks},
	author={Liao, Ren Jie and Zhao, Zhi Zhen and Urtasun, Raquel and Zemel, Richard S},
	booktitle={Proc. International Conference on Learning Representations (ICLR)},
	year={2019},
	volume={},
	number={},
	month={Jan.},
	pages={},
	doi={}
}

@inproceedings{Bo2023,
	author={Bo, De Yu and Shi, Chuan and Wang, Le Le and Liao, Ren Jie},
	booktitle={Proc. International Conference on Learning Representations (ICLR)}, 
	title={Specformer: Spectral Graph Neural Networks Meet Transformers}, 
	year={2023},
	volume={},
	number={},
	month={Mar.},
	pages={},
	doi={}
}

@inproceedings{Ai2025,
	author={Ai, Guo Guo and Pang, Guan Song and Qiao, He Zhe and Gao, Yuan and Yan, Hui},
	booktitle={Proc. International Conference on Machine Learning (ICML)}, 
	title={GrokFormer: Graph {F}ourier Kolmogorov-Arnold Transformers}, 
	year={2025},
	volume={},
	number={},
	month={Jul.},
	pages={13--19},
	doi={}
}

@article{sen2008collective,
	title={Collective classification in network data},
	author={Sen, Prithviraj and Namata, Galileo and Bilgic, Mustafa and Getoor, Lise and Galligher, Brian and Eliassi-Rad, Tina},
	journal={AI Mag.},
	volume={29},
	number={3},
	pages={93--93},
	year={2008},
	month={Sep.},
	doi={https://doi.org/10.1609/aimag.v29i3.2157},
	
}

@article{shchur2018pitfalls,
	title={Pitfalls of graph neural network evaluation},
	author={Shchur, Oleksandr and Mumme, Maximilian and Bojchevski, Aleksandar and Gunnemann, Stephan},
	journal={arXiv preprint arXiv:1811.05868},
	year={2018}
}

@inproceedings{pei2020geomgcn,
	title     = {Geom-{GCN}: Geometric Graph Convolutional Networks},
	author    = {Pei, Hongbin and Wei, Bingzhe and Chang, Kevin Chen-Chuan and Lei, Yu and Yang, Bo},
	booktitle = {Proc. International Conference on Learning Representations (ICLR)},
	year      = {2020},
	volume={},
	number={},
	month={Apr.},
	pages={},
	doi={}
}

@techreport{craven1998learning,
	title={Learning to extract symbolic knowledge from the world wide web},
	author={Craven, Mark and McCallum, Andrew and PiPasquo, Dan and Mitchell, Tom and Freitag, Dayne},
	year={1998},
	number={CMU-CS-98-122},    
	institution={Carnegie Mellon University, School of Computer Science}, 
	type={Technical Report},   
	address={Pittsburgh, PA, USA},
	note={Accession Number: ADA356047} 
}

@article{rozemberczki2021multi,
	title={Multi-scale attributed node embedding},
	author={Rozemberczki, Benedek and Allen, Carl and Sarkar, Rik},
	journal={J. Complex Netw.},
	volume={9},
	number={2},
	pages={cnab014},
	year={2021},
	month={Apr.},
	publisher={Oxford University Press},
	doi={https://doi.org/10.1093/comnet/cnab014}
}

@article{yang2020scaling,
	title={Scaling attributed network embedding to massive graphs},
	author={Yang, Ren Chi and Shi, Jie Ming and Xiao, Xiao Kui and Yang, Yin and Liu, Jun Cheng and Bhowmick, Sourav S},
	journal={Proc. VLDB Endow.},
	volume={14},
	number={1},
	pages={37--49},
	year={2020},
	month={Sep.},
	publisher={Association for Computing Machinery}
}

@inproceedings{velivckovic2018graph,
	author={Velickovic, Petar and Cucurull, Guillem and Casanova, Arantxa and Romero, Adriana and Lio, Pietro and Bengio, Yoshua},
	booktitle={Proc. International Conference on Learning Representations (ICLR)}, 
	title={Graph Attention Networks}, 
	year={2018},
	volume={},
	number={},
	month={Apr.},
	pages={},
	doi={10.17863/CAM.48429}
}

@inproceedings{zhu2020beyond,
	author={Zhu, Jiong and Yan, Yu Jun and Zhao, Ling Xiao and Heimann, Mark and Akoglu, Leman and Koutra, Danai},
	booktitle={Proc. Conference on Neural Information Processing Systems (NeurIPS)}, 
	title={Beyond Homophily in Graph Neural Networks: Current Limitations and Effective Designs}, 
	year={2020},
	volume={33},
	number={},
	month={Dec.},
	pages={7793--7804},
	doi={}
}

@inproceedings{chen2023node,
	title={From node interaction to hop interaction: New effective and scalable graph learning paradigm},
	author={Chen, Jie and Li, Zi Long and Zhu, Yin and Zhang, Jun Ping and Pu, Jian},
	booktitle={Proc. IEEE/CVF Conference on Computer Vision and Pattern Recognition (CVPR)},
	pages={7876--7885},
	year={2023},
	month={Jun.},
}

@inproceedings{chien2021adaptive,
	author={Chien, Eli and Peng, Jian Hao and Li, Pan and Milenkovic, Olgica},
	booktitle={Proc. International Conference on Learning Representations (ICLR)}, 
	title={Adaptive Universal Generalized {P}age{R}ank Graph Neural Network}, 
	year={2021},
	volume={},
	number={},
	month={May},
	pages={},
	doi={}
}

@inproceedings{huang2024higher,
	title={Higher-order graph convolutional network with flower-petals laplacians on simplicial complexes},
	author={Huang, Yi Ming and Zeng, Yu Jie and Wu, Qiang and Lu, Lin Yuan},
	booktitle={Proceedings of the AAAI conference on artificial intelligence},
	volume={38},
	number={11},
	pages={12653--12661},
	year={2024}
}

@article{dwivedi2020generalization,
	title={A generalization of transformer networks to graphs},
	author={Dwivedi, Vijay Prakash and Bresson, Xavier},
	journal={arXiv preprint arXiv:2012.09699},
	year={2020}
}

@inproceedings{rampavsek2022recipe,
	author={Rampasek, Ladislav and Galkin, Michael and Dwivedi, Vijay Prakash and Luu, Anh Tuan and Wolf, Guy and Beaini, Dominique},
	booktitle={Proc. Conference on Neural Information Processing Systems (NeurIPS)},
	title={Recipe for a General, Powerful, Scalable Graph Transformer},
	year={2022},
	volume={35},
	number={},
	month={Dec.},
	pages={14501--14515},
	doi={}
}

@inproceedings{wu2022nodeformer,
	author={Wu, Qi Tian and Zhao, Wen Tao and Li, Ze Nan and Wipf, David P and Yan, Jun Chi},
	booktitle={Proc. Conference on Neural Information Processing Systems (NeurIPS)},
	title={NodeFormer: A Scalable Graph Structure Learning Transformer for Node Classification},
	year={2022},
	volume={35},
	number={},
	month={Nov.},
	pages={27387--27401},
	doi={}
}

@inproceedings{wu2023sgformer,
	author={Wu, Qi Tian and Zhao, Wen Tao and Yang, Chen Xiao and Zhang, Heng Rui and Nie, Fan and Jiang, Hai Tian and Bian, Ya Tao and Yan, Jun Cchi},
	booktitle={Proc. Conference on Neural Information Processing Systems (NeurIPS)},
	title={SGFormer: Simplifying and Empowering Transformers for Large-Graph Representations},
	year={2023},
	volume={36},
	number={},
	month={Dec.},
	pages={64753--64773},
	doi={}
}

@inproceedings{chen2023nagphormer,
	author={Chen, Jin Song and Gao, Kai Yuan and Li, Gai Chao and He, Kun},
	booktitle={Proc. International Conference on Learning Representations (ICLR)},
	title={NAGphormer: A Tokenized Graph Transformer for Node Classification in Large Graphs},
	year={2022},
	volume={},
	number={},
	month={Apr.},
	pages={},
	doi={}
}

@inproceedings{ma2024polyformer,
	author={Ma, Jia Hong and He, Ming Guo and Wei, Zhe Wei},
	booktitle={Proc. ACM SIGKDD Conference on Knowledge Discovery and Data Mining (KDD)},
	title={PolyFormer: Scalable Node-wise Filters via Polynomial Graph Transformer},
	year={2024},
	volume={},
	number={},
	month={Aug.},
	pages={2118--2129},
	doi={10.1145/3637528.3671849}
}
\bibliographystyle{IEEEtran}

\begin{IEEEbiographynophoto}{Manjun~Cui}
	received the B.S. degree in Mathematics and Applied Mathematics from Yancheng Teachers University, Yancheng, Jiangsu, China, in 2022. She is currently pursuing the Ph.D. degree in mathematics with the School of Mathematics and Statistics, Nanjing University of Information Science and Technology, Nanjing, Jiangsu, China.
\end{IEEEbiographynophoto}

\begin{IEEEbiographynophoto}{Xiaopeng~Cheng}
	received the B.S. degree in Information and Computing Science from Anhui University, Hefei, Anhui, China, in 2023. He is currently pursuing the M.S. degree in mathematics with the School of Mathematics and Statistics, Nanjing University of Information Science and Technology, Nanjing, Jiangsu, China.
\end{IEEEbiographynophoto}

\begin{IEEEbiographynophoto}{Yangfan~He}
	received the Ph.D. degree in space physics from Wuhan University, Wuhan, China, in 2022. Since 2022, she has been with the School of Communication and Artificial Intelligence, School of Integrated Circuits, Nanjing Institute of Technology, Nanjing, Jiangsu, China. She is currently a Part-time Postdoc with the School of Atmospheric Physics, Nanjing University of Information Science and Technology, Nanjing, Jiangsu, China. She has published more than 10 journal articles in IEEE TRANSACTIONS ON INFORMATION THEORY, IEEE TRANSACTIONS ON SIGNAL PROCESSING, SIGNAL PROCESSING, JOURNAL OF GEOPHYSICAL RESEARCH: SPACE PHYSICS, etc. Her current research interests include signal processing, plasma waves, and magnetosphere-ionosphere coupling.
\end{IEEEbiographynophoto}

\begin{IEEEbiographynophoto}{Zhichao~Zhang}
	(Member, IEEE) received the B.S. degree in mathematics and applied mathematics from Gannan Normal University, Ganzhou, Jiangxi, China, in 2012, and the Ph.D. degree in mathematics of uncertainty processing from Sichuan University, Chengdu, Sichuan, China, in 2018. From September 2017 to September 2018, he was a Visiting Student Researcher with the Department of Electrical and Computer Engineering, Tandon School of Engineering, New York University, Brooklyn, NY, USA, where he was awarded a grant from the China Scholarship Council. Since 2019, he has been with the School of Mathematics and Statistics, Nanjing University of Information Science and Technology, Nanjing, Jiangsu, China, where he is currently a Full Professor and a Ph.D Supervisor. From January 2021 to January 2023, he was a Macau Young Scholars Post-Doctoral Fellow of information and communication engineering with the School of Computer Science and Engineering, Macau University of Science and Technology, Macau, SAR, China. He has published more than 100 journal articles in IEEE TRANSACTIONS ON INFORMATION THEORY, IEEE TRANSACTIONS ON SIGNAL PROCESSING, IEEE TRANSACTIONS ON SIGNAL AND INFORMATION PROCESSING OVER NETWORKS, IEEE SIGNAL PROCESSING LETTERS, IEEE COMMUNICATIONS LETTERS, SIGNAL PROCESSING, JOURNAL OF FOURIER ANALYSIS AND APPLICATIONS, etc. His current research interests include mathematical theories, methods, and applications in fractional domain signal and information processing, including fractional domain time-frequency analysis, sparse information processing, graph signal processing and graph neural network. He was a member of the International Association of Engineers, the China Society for Industrial and Applied Mathematics, the Chinese Institute of Electronics, and the Beijing Society for Interdisciplinary Science. He was the Vice President of the Jiangsu Society for Computational Mathematics and the Director of the Jiangsu Society for Industrial and Applied Mathematics. He was listed among world’s top 2\% scientists recognized by Stanford University from 2021 to 2025.
\end{IEEEbiographynophoto}
%

\end{document}